\documentclass[10pt]{iopart}

\DeclareMathAlphabet{\mathpzc}{OT1}{pzc}{m}{it}
\usepackage{graphicx}
\usepackage{cancel}  %
\usepackage{booktabs}
\usepackage{cite}  %
\usepackage{xcolor}
\usepackage{epstopdf}
\usepackage{hyperref}
\usepackage{soul} %
\usepackage{xifthen}
\usepackage{ulem} 
\usepackage{units} %
\usepackage{upgreek}  %
\usepackage{tabularx}
\usepackage{titlesec}  %
\usepackage{etoolbox}
\usepackage{hyphenat} %

\expandafter\let\csname equation*\endcsname\relax
\expandafter\let\csname endequation*\endcsname\relax
\usepackage{amsmath}
\usepackage{amssymb}

\definecolor{myblue}{rgb}{0, 0, 0.5}
\hypersetup{colorlinks = true, citecolor=myblue, linkcolor=myblue}

\usepackage{bbold}

\DeclareFontFamily{U}{mathx}{\hyphenchar\font45}
\DeclareFontShape{U}{mathx}{m}{n}{<-> mathx10}{}
\DeclareSymbolFont{mathx}{U}{mathx}{m}{n}
\DeclareMathAccent{\widebar}{0}{mathx}{"73}

\newcommand{\Eqref}[1]{\mbox{equation\hspace{0.25em}\eqref{#1}}}
\newcommand{\Eqsref}[1]{\mbox{equations\hspace{0.25em}\eqref{#1}}}
\newcommand{\EQref}[1]{\mbox{Equation\hspace{0.25em}\eqref{#1}}}
\newcommand{\EQsref}[1]{\mbox{Equations\hspace{0.25em}\eqref{#1}}}
\newcommand{\figref}[1]{\mbox{figure\hspace{0.25em}\ref{#1}}}
\newcommand{\Figref}[1]{\mbox{Figure\hspace{0.25em}\ref{#1}}}
\newcommand{\figsref}[1]{\mbox{figures\hspace{0.25em}\ref{#1}}}

\newcommand{\secref}[1]{\mbox{section\hspace{0.25em}\ref{#1}}}
\newcommand{\Secref}[1]{\mbox{Section\hspace{0.25em}\ref{#1}}}

\newcommand{\appref}[1]{\mbox{appendix\hspace{0.25em}\ref{#1}}}

\newcommand{\refcite}[1]{\mbox{ref.\hspace{0.25em}\cite{#1}}}

\definecolor{gray}{RGB}{200,200,200}

\renewcommand{\emph}{\textit}

\newcommand{\identity}{\mathbb{1}}

\newcommand{\vect}{\boldsymbol}
\newcommand{\tensor}{\boldsymbol}

\newcommand{\diff}{\text{d}}
\newcommand{\difffrac}[2]{\frac{\diff #1}{\diff #2}}
\newcommand{\pfrac}[2]{\frac{\partial #1}{\partial #2}}

\newcommand{\mean}[1]{\langle #1 \rangle}

\newcommand{\ab}{{\alpha\beta}}
\newcommand{\hessian}{H}
\newcommand{\jacobian}{\mathcal J}
\newcommand{\normal}{\mathcal{N}}

\newcommand{\Fbulk}{F_\mathrm{bulk}}
\newcommand{\fBulk}{f_\mathrm{bulk}}
\newcommand{\fSurf}{f_\mathrm{surf}}
\newcommand{\fWall}{\fSurf}
\newcommand{\Df}{\Delta\! f}  %
\newcommand{\DF}{\Delta\! F}
\newcommand{\Dmu}{\Delta\!\mu}
\newcommand{\Ftot}{F_\mathrm{tot}}
\newcommand{\Fint}{F_\mathrm{int}}
\newcommand{\Nc}{{N_\mathrm{c}}} %
\newcommand{\Np}{{N_\mathrm{p}}}
\newcommand{\NbarP}{\widebar N_\mathrm{p}}
\newcommand{\Nu}{N_\mathrm{u}}
\newcommand{\NbarU}{\widebar N_\mathrm{u}}
\newcommand{\Nr}{{N_\mathrm{r}}}
\newcommand{\size}{\Omega}
\newcommand{\Rcrit}{R_\mathrm{crit}}

\newcommand{\Rnuc}{R_\mathrm{nucl}}
\newcommand{\Fnuc}{\DF_\mathrm{nucl}}
\newcommand{\Nd}{{N_\mathrm{d}}}
\newcommand{\Vsys}{V_\mathrm{sys}}
\newcommand{\kBT}{k_\mathrm{B}T}
\newcommand{\capLen}{\ell_\gamma}
\newcommand{\Lelasto}{\ell_{\textrm{cap}}}

\newcommand{\width}{\ell}
\newcommand{\cIn}{c_\mathrm{in}}

\newcommand{\DOut}{D_\mathrm{out}}
\newcommand{\Ddrop}{D_\mathrm{drop}}
\newcommand{\supSat}{\mathcal S}
\newcommand{\phiInt}{\phi_\mathrm{eq}}
\newcommand{\phiSol}{\phi_0}
\newcommand{\phiSup}{\phi_\infty}
\newcommand{\phiIn}{\phi_\mathrm{in}}
\newcommand{\phiOut}{\phi_\mathrm{out}}
\newcommand{\phiBase}{\phi^{(0)}}
\newcommand{\phiBaseIn}{\phiBase_\mathrm{in}}
\newcommand{\phiBaseOut}{\phiBase_\mathrm{out}}
\newcommand{\phiEq}{\phi^\mathrm{eq}}
\newcommand{\phiEqIn}{\phiEq_\mathrm{in}}
\newcommand{\phiEqOut}{\phiEq_\mathrm{out}}

\newcommand{\muSup}{\bar\mu_\infty}
\newcommand{\muEq}{\bar\mu_\mathrm{eq}}

\newcommand{\PEq}{\Pi_\mathrm{eq}}
\newcommand{\Plaplace}{P_\gamma}

\newcommand{\gammaIn}{\gamma_\mathrm{in}}
\newcommand{\gammaOut}{\gamma_\mathrm{out}}
\newcommand{\jIn}{\vect{j}_\mathrm{in}}
\newcommand{\jOut}{\vect{j}_\mathrm{out}}
\newcommand{\vNorm}{v_\perp} %

\newcommand{\Veq}{V_\mathrm{eq}}
\newcommand{\stress}{\sigma}

\newcommand{\stressEq}{\stress^\mathrm{eq}}
\newcommand{\chibar}{\widebar\chi}
\newcommand{\sigmaA}{\sigma^{(\alpha)}}
\newcommand{\sA}{s^{(\alpha)}}
\newcommand{\ka}{k_\mathrm{a}}
\newcommand{\kp}{k_\mathrm{p}}

\newcommand{\kOut}{k_\mathrm{out}}
\newcommand{\sBase}{\Gamma}
\newcommand{\sBaseIn}{\sBase_\mathrm{in}}
\newcommand{\sBaseOut}{\sBase_\mathrm{out}}
\newcommand{\pot}{\varphi}
\newcommand{\LRD}{\ell}
\newcommand{\LRDin}{\LRD_\mathrm{in}}
\newcommand{\LRDout}{\LRD_\mathrm{out}}

\numberwithin{equation}{section} %
\numberwithin{figure}{section} %
\numberwithin{table}{section} %
\titleformat{\subsubsection}[runin]{\normalfont\it}{\thesubsubsection}{5pt}{}[:]  %

\makeatletter
\def\@mkboth#1#2{}
\newlength\appendixwidth
\preto\appendix{\addtocontents{toc}{\protect\patchl@section}}
\newcommand{\patchl@section}{%
  \settowidth{\appendixwidth}{\textbf{Appendix }}%
  \addtolength{\appendixwidth}{1.5em}%
  \patchcmd{\l@section}{1.5em}{\appendixwidth}{}{\ddt}%
}
\makeatother

\providecommand{\DIFdel}[1]{}  %

\begin{document}

\title[Physics of droplet regulation in biological cells]{Physics of droplet regulation in biological cells}

\author{David Zwicker$^{1,*}$, Oliver W. Paulin$^{1,*}$, Cathelijne ter Burg$^{1,*}$}

\address{$^1$ Max Planck Institute for Dynamics and Self-Organization, Am Fa{\ss}berg 17, 37077 G\"{o}ttingen, Germany}

\address{$^*$ equal contribution}

\ead{david.zwicker@ds.mpg.de}
\vspace{10pt}

\begin{indented}
\item[] Version of draft:  \today 
\end{indented}

\begin{abstract}
Droplet formation has emerged as an essential concept for the spatiotemporal organisation of biomolecules in cells.
However, classical descriptions of droplet dynamics based on passive liquid-liquid phase separation cannot capture the complex situation inside cells.
This review discusses three distinct aspects that are crucial in cells:
(i) biomolecules are diverse and individually complex, implying that cellular droplets possess complex internal behaviour, e.g., in terms of their material properties;
(ii) the cellular environment contains many solid-like structures that droplets can wet;
(iii) cells are alive and use fuel to drive processes out of equilibrium.
We illustrate how these principles control droplet nucleation, growth, position, and count to unveil possible regulatory mechanisms in biological cells and other applications of phase separation.
\end{abstract}

\vspace{1pc}

\ioptwocol

\tableofcontents

\newcommand{\sectionseparation}{}

\sectionseparation

\section{Introduction}

\begin{figure*}%
	\centering
	\includegraphics[width=\linewidth]{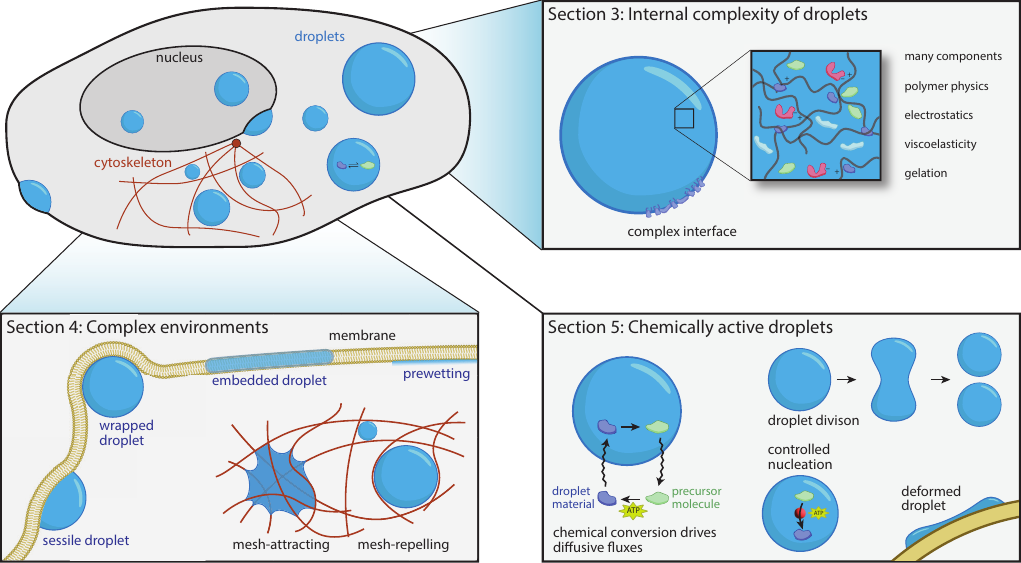}
	\caption{
	\textbf{Physical phenomena relevant to intracellular droplets discussed in this review.}
	Droplets inside cells (upper left) display multiple physical phenomena discussed in separate sections of this review.
	\hyperref[sec:basic_LLPS]{Section \ref{sec:basic_LLPS}} introduces the basic physics of phase separation leading to droplet formation.
	\hyperref[sec:internal_complexity]{\Secref{sec:internal_complexity}} focuses on the internal complexity of droplets originating from the many different interacting molecules that droplets comprise (upper right).
	\hyperref[sec:complex_environment]{\Secref{sec:complex_environment}} discusses the interaction of droplets with complex environments (lower left).
	\hyperref[sec:chemical_reactions]{\Secref{sec:chemical_reactions}} introduces driven chemical reactions and their effects on droplets (lower right).
	} 
	\label{fig:intro}
\end{figure*}

Biological cells contain droplets, known as \emph{biomolecular condensates}, which partition the cells' interior without using lipid membranes~\cite{Banani2017}.
Besides simply structuring cells, these droplets also provide distinct chemical environments~\cite{Kilgore2023},  affect reactions~\cite{Papp2025,Oflynn2021}, sequester molecules~\cite{Glauninger2024}, exert forces~\cite{Wiegand2020}, and can induce phenotypic switching~\cite{Hong2024}.
Droplets are implicated in homeostasis of healthy cells (e.g., in development~\cite{So2021}, gene regulation~\cite{Pei2024,Hirose2022}, protein quality control~\cite{Rajendran2025}, and signaling~\cite{Jaqaman2021}), but also in diseased states~\cite{Alberti2021,Shin2017a} (e.g., neurodegenerative diseases~\cite{Mathieu2020}, and cancer~\cite{Cai2021}).
It is thus not surprising that droplets appear everywhere in cells, from the nucleus~\cite{Lafontaine2020}, to the cytosol~\cite{Banani2017}, to membranes~\cite{Jaqaman2021}, and in all forms of life, encompassing bacteria~\cite{Azaldegui2020}, plants~\cite{Kim2021}, and animals~\cite{Banani2017}.
While it is clear that droplets are crucial for cells, it is less clear how they form and how cells control them.

The physics of droplet formation, known as \emph{phase separation}, has been studied for more than a century~\cite{Gibbs1876}, including in biological contexts~\cite{Wilson1899}.
Phase separation generally describes how a balance between entropy, favouring mixing, and enthalpy, due to interactions, leads to stable droplets in thermodynamic equilibrium. %
While this basic principle likely also governs droplet formation in cells~\cite{Hyman2014}, phase separation in a biological context is significantly more complex for at least three major reasons (\figref{fig:intro}).
First, cells comprise an enormous number of different biomolecules, which can each be incredibly complex~\cite{Holehouse2023}.
Consequently, droplets typically comprise many components and posses complex material properties.
Second, the cellular environment is heterogeneous, filled with membranes, rigid polymers, and other macromolecular structures.
Droplets interact with this environment through wetting, which affects both the droplets and the environment itself.
Third, cells are alive, implying that processes can be driven away from equilibrium, which can profoundly affect droplet formation~\cite{Weber2019}.
One challenge in describing cellular droplets theoretically  lies in the many different length scales involved: microscopic interactions (e.g., \mbox{$\pi$-$\pi$}-stacking, \mbox{cation-$\pi$} interactions, hydrogen bonding, and hydrophobic interactions~\cite{Dignon2020}) induce phase separation; mesoscopic interactions (e.g., complex coacervation~\cite{Sing2020}, entanglement, and gelation) determine mechanical properties of droplets; and macroscopic interactions (e.g., gradients, flow, and wetting) control droplets on the cellular scale.
Theory is necessary to bridge these scales and to arrive at a holistic description of the physical processes that shape droplets in cells.

This review summarises recent work on intracellular phase separation, focusing on the physical processes that allow cells to regulate droplets, i.e., processes that determine when and where droplets form, how large they get, and when they dissolve.
We focus less on how droplets affect cellular processes \textit{in vivo} (reviewed in~\cite{Rippe2025, Rajendran2025, Pei2024, Visser2024, Buchan2024, Mohapatra2023, Millar2023, Holehouse2023, Gormal2023, Lee2022, Hirose2022, Wang2021, Su2021, So2021, Schisa2021, Oflynn2021, Kim2021, Jaqaman2021, Ghosh2021a, Currie2021, Cai2021, Bhat2021, Alberti2021, Swain2020, Adekunle2020, Wiegand2020, Sabari2020a, Mathieu2020, Ong2020, Hondele2020, Cohan2020, Zhao2020a, Greening2020, Wu2020, Wheeler2020, Levental2020, Dignon2020, Chen2020, Zhang2020, Lyon2020, Lafontaine2020, Azaldegui2020, Snead2019, Youn2019, Spannl2019, Alberti2019a, Shin2017a, Alberti2017, Banani2017, Simons2010,Liang2023,Alberti2025, Sabari2024}) or in related \textit{in vitro} studies (reviewed in~\cite{Alberti2025,Sabari2024, Pei2024, Zhou2024, Mangiarotti2024, Cao2024, Romero-Perez2023, Donau2023, Michelin2023, Holt2023, Harrington2023, Abyzov2022, Villegas2022, Scholl2021, Sing2020, Lohse2020, Andreotti2020, Andre2020, Bracha2019, Nakashima2019, Cates2015a, Xu2014, Landfester2006}).
We also neglect external influences, such as electric and magnetic fields, or gravity, which is only relevant in extremely large cells~\cite{Feric2013}, and do not discuss the detailed (bio-)chemistry that leads to phase separation (see~\cite{Hess2025,Holehouse2025,Holehouse2023,Pappu2023,Choi2020a} for details).
Instead, we assume that such interactions can be captured by a suitable coarse-grained free energy~\cite{Sherck2021}, so that we can describe droplets using continuous concentration fields.
Our approach is similar to that taken in other reviews of the theory of phase separation in cells~\cite{Julicher2024, Cates2024, Jacobs2023, Saar2023, Pappu2023, Zwicker2022a, Li2022b, Ghosh2022, Choi2020a, Weber2019, Berry2018}, but with a particular focus on droplet regulation.
In essence, we ask what physical processes (mostly inspired by soft matter physics~\cite{Saarloos2024,Doi2013,Onuki2007}) can cells use to control droplets.

This review is organised into four distinct sections, beginning with a summary of the basic physics of phase separation in \secref{sec:basic_LLPS}.
The three subsequent sections are intended to be independent of each other and can be read in any order.
These sections cover the core aspects of cellular complexity identified above (\figref{fig:intro}): internal complexity of droplets (\secref{sec:internal_complexity}), complexity of the surrounding environment (\secref{sec:complex_environment}), and activity (\secref{sec:chemical_reactions}).

\sectionseparation

\section{Passive liquid-liquid phase separation} \label{sec:basic_LLPS}
In this section, we review the basic physics of phase separation for a multi-component mixture. First, we consider a continuous theory of spatially extended fields with smooth (diffuse) interfaces between phases.  Next, we explore a simplified effective description of well-defined droplets separated from a dilute phase via a thin (sharp) interface. Finally, we discuss regulatory mechanisms that can be utilised by cells to control droplets even without the addition of further complexity.

\subsection{Continuous description}\label{sec:continuous_desc}

To construct a theory of phase separation, we start by considering a coarse-grained picture in which the spatial distributions of different molecules are described by continuous density fields.

\subsubsection{Thermodynamics of liquids}
\label{sec:thermodynamics_multi}
We first consider an isothermal, homogeneous system of volume~$\Vsys$ filled with a liquid comprised of a solvent and $\Nc$ additional components.
For simplicity, we focus on incompressible systems, where the molecular volumes~$\nu_i$ of each component are constant. The generalisation to compressible systems is discussed in \appref{sec:appendix_compressible}.
The composition of the homogeneous system is fully described by  particle counts $N_i$ for $i=1,\ldots,\Nc$ and all remaining space is filled with solvent ($i=0$).
Since the system is isothermal and has constant volume, the relevant thermodynamic potential is the Helmholtz free energy $F = \Vsys f(\phi_1, \ldots, \phi_\Nc)$, where $f$ is a free energy density, which is a function of the $\Nc$ volume fractions $\phi_i = \nu_i N_i / \Vsys$. %
We consider a generic form that combines translational entropies with pairwise interactions between components (known as the \emph{Flory--Huggins free energy}~\cite{Flory1942, Huggins1941}),
\begin{multline}
	\label{eqn:free_energy_many}
	\!\!\!\! f(\phi_1, \ldots, \phi_{\Nc}) = 
\\
		\frac{\kBT}{\nu_0}\!\Biggl[		
		\sum_{i=0}^{\Nc} \frac{\phi_i}{\size_i} \ln(\phi_i)
		+ \!\sum_{i=1}^\Nc w_i \phi_i
		+ \!\!\sum_{i,j=1}^\Nc \!\! \frac{\chi_{ij}}{2} \phi_i\phi_j
	\!\Biggr]
	\,,\!\!
\end{multline}
where the first term accounts for the mixing entropy with $\phiSol = 1  - \sum_{i=1}^\Nc \phi_i$; see \appref{sec:appendix_compressible}.
We also use the molecular volume~$\nu_0$ of the solvent to define non-dimensional molecular sizes $\size_i = \nu_i/ \nu_0$.
The last two terms in \Eqref{eqn:free_energy_many} specify enthalpic interactions using non-dimensional quantities $w_i$ and $\chi_{ij}$, 
which include solvent interactions (see \Eqref{eqn:enthalpy_coefficients_with_solvent}) and whose particular values can be determined from detailed theories~\cite{Pappu2023}.
Here, $w_i$ captures the difference of internal energy between species~$i$ and the solvent, but it can also account for external potentials.
Such potentials can in principle depend on space, due to, for example, gravity or gradients in a regulatory species~\cite{Weber2019}, but we do not discuss this aspect further in this review.
In contrast, the symmetric Flory parameters $\chi_{ij}=\chi_{ji}$ capture interactions:
positive values raise the free energy and thus relate to repulsion between species $i$ and $j$, while negative values imply effective attraction.
Since the interactions $\chi_{ij}$ are measured relative to solvent molecules, the diagonal terms need not vanish; see \appref{sec:appendix_compressible}.
The Flory--Huggins free energy given by \Eqref{eqn:free_energy_many} only describes the basic behavior of phase separating mixtures, and we discuss more complex interactions in \secref{sec:complex_molecules}.

To study thermodynamic equilibria, we derive intensive thermodynamic variables using derivatives of the free energy $F$ with respect to the extensive variables~$N_i$ and $\Vsys$.
Note that these variables are not independent, since $\Vsys = \sum_{i=0}^\Nc \nu_i N_i$ in the incompressible limit.
Consequently, the relevant chemical potentials $\bar\mu_i = (\partial F/\partial N_i)_{N_{j \neq i}, \Vsys}$,
\begin{multline}
	\label{eqn:chemical_potential_many}
	\bar\mu_i = \kBT \biggl[ 
		1 + \ln(\phi_i) - \size_i\bigl(1 + \ln(\phi_0)\bigr)
\\
		+ \size_iw_i + \size_i \sum_{j=1}^\Nc \chi_{ij} \phi_j
	   \biggr]
	  \;,
\end{multline}
 are exchange chemical potentials, quantifying the change of free energy when $\size_i$ solvent particles are replaced by a single particle of type~$i$.
Similarly, the pressure~$\Pi=-(\partial F/\partial \Vsys)_{N_i} = \sum_i\bar\mu_i \phi_i \nu_i^{-1} - f$,
\begin{align}
	\label{eqn:pressure_many}
	\Pi &=  \frac{\kBT}{\nu_0}\Biggl[
		 \sum_{i=0}^{\Nc} \frac{\phi_i}{\size_i}  
		- 1  - \ln(\phi_0)
		+ \sum_{i, j=1}^\Nc \frac{\chi_{ij}}{2} \phi_i\phi_j	   
		\Biggr]
	\;,
\end{align}
is related to the osmotic pressure of solvent exchange; see \appref{sec:appendix_compressible}.
The intensive quantities~$\bar\mu_i$ and $\Pi$ are linked by the Gibbs--Duhem relation~\cite{Julicher2018},
\begin{align}
	\label{eqn:gibbs_duhem}
	\diff \Pi = \sum_{i=1}^\Nc \frac{\phi_i}{\nu_i} \diff \bar\mu_i
	\;,
\end{align}
and their balance determines the coexisting phases.
To show this, we consider thermodynamically large systems, where interfacial and surface energies are negligible, so that the free energy of a system of $\Np$ homogeneous phases reads $F\approx \sum_{n=1}^\Np V_n f\bigl(\Phi^{(n)}\bigr)$, where $\Phi^{(n)} = (\phi^{(n)}_1, \phi^{(n)}_2, \ldots, \phi^{(n)}_{\Nc})$ denotes the composition of the $n$-th phase.
This free energy is minimal when the coexistence conditions~\cite{Weber2019,Zwicker2022},
\begin{subequations}
\label{eqn:coexistence_many}
\begin{align}
	\bar\mu_i \bigl(\Phi^{(1)}\bigr) &
		= \bar\mu_i \bigl(\Phi^{(2)}\bigr) 
		= \cdots 
		= \bar\mu_i \bigl(\Phi^{(\Np)}\bigr)
	\;,
	\label{eqn:coexistence_many_mu}
\\
	\Pi \bigl(\Phi^{(1)}\bigr) &
	= \Pi \bigl(\Phi^{(2)}\bigr)
	= \cdots
	= \Pi \bigl(\Phi^{(\Np)}\bigr)
	\label{eqn:coexistence_many_P}
	\;,
\end{align}
\end{subequations}
are met for components~$i=1,\ldots,\Nc$.
These equations represent chemical and mechanical equilibrium between phases, respectively.
\EQsref{eqn:coexistence_many} give $(\Np-1)(\Nc+1)$ conditions for $\Np \Nc$ independent fractions~$\phi^{(n)}_i$.
The system is thus overdetermined when $\Np>\Nc+1$, illustrating that at most $\Nc+1$ phases of different composition can coexist, consistent with \emph{Gibbs phase rule} for fixed temperature and system volume~\cite{Gibbs1876}.

\subsubsection{Kinetics of heterogeneous systems}
\label{sec:kinetics_multi-component}
To describe droplet dynamics, we must generalise the theory presented in the previous section to heterogeneous systems.
To do this, we assume local equilibrium, neglecting the heat fluctuations which can manifest on long length and time scales in typically active biological systems~\cite{Mabillard2023}.
We can thus assume an isothermal system and define fields for the thermodynamic quantities introduced above.
In particular, the system state is now specified by volume fraction fields $\Phi(\vect r)=\bigl(\phi_1(\vect r), \ldots, \phi_{\Nc}(\vect r)\bigr)$ and its free energy is given by the functional~\cite{Cahn1958,Hoyt1990, Tang1991}
\begin{align}
	\label{eqn:free_energy_functional}
	\Fbulk[\Phi(\vect r)]
		&= \! \int\biggl[f\bigl(\Phi\bigr)
			+ \!\sum_{i,j=1}^{\Nc}\frac{\kappa_{ij}}{2} (\nabla \phi_i) \!\cdot\! (\nabla \phi_j)
		\biggr] \diff V \;,
\end{align}
without interactions with the system's boundary, which we will include in \Eqref{eqn:free_energy_surface_interaction}.
Here, the first term accounts for the local free energy described by \Eqref{eqn:free_energy_many}, and the second term results from a gradient expansion~\cite{Cahn1958}.
The symmetric matrix $\kappa_{ij}$ quantifies how strongly compositional gradients are penalised and generally depends on composition~\cite{Li2022, Mao2018, deGennes1980}.
In the simple case where $\kappa_{ij}$ does not depend on composition, the exchange chemical potentials, $\bar\mu_i = \nu_0 \size_i \delta \Fbulk/\delta \phi_i$,  read
\begin{align}
	\label{eqn:chemical_potential_many_kappa}
	\bar\mu_i &= \nu_0 \size_i \biggl(
		\pfrac{f}{\phi_i}
		- \sum_{j=1}^\Nc \kappa_{ij} \nabla^2 \phi_j
	\biggr)
	\;.
\end{align}
Here, the first term  captures  contributions relevant in homogeneous systems, e.g., those given by \Eqref{eqn:chemical_potential_many}, whereas the second term describes spatial couplings originating from heterogeneities; see \appref{sec:appendix_variation}.

We now derive kinetic equations based on linear non-equilibrium thermodynamics~\cite{DeGroot2013,Julicher2018}, starting from the continuity equation, $\partial_t \phi_i  + \nabla\cdot \vect J_i = s_i$, where $\vect J_i$ denotes spatial fluxes and the source term $s_i(\vect{r})$ accounts for chemical reactions as described in \secref{sec:chemical_reactions}.
To separate fluxes due to hydrodynamic flows (associated with momentum transport) from diffusive fluxes driven by gradients in chemical potentials, we split $\vect J_i$ into a centre-of-mass velocity field~$\vect{v} = \sum_{i=0}^\Nc \vect J_i$ and relative fluxes $\vect j_i=\phi_i(\vect v_i - \vect v)$,  where $\vect v_i = \vect J_i/\phi_i$ is the velocity of component $i$ in the lab frame. 
Since many biomolecules have mass densities close to water, we here assume equal mass density for all components, and refer the reader to \refcite{Julicher2018} for the general case.
Hence,
\begin{equation}
	\label{eqn:continuity_many}
	\partial_t \phi_i +  \nabla \!\cdot\! (\vect{v} \phi_i) +\nabla \!\cdot\! \vect{j}_i = s_i
	\;.
\end{equation}
The velocity field $\vect v$ obeys a Navier--Stokes equation,
\begin{align}
	\label{eqn:navier_stokes}
	\partial_t \vect v + \rho \vect v \!\cdot\! \nabla \vect v = \nabla\cdot \tensor\sigma
	\;,
\end{align}
with constant mass density $\rho$, such that incompressibility implies $\nabla \cdot \vect v=0$.
The stress tensor  $\tensor\sigma$ %
incorporates viscous dissipation, equilibrium stresses originating from phase separation, elastic effects, and potential active stresses that drive flows in cells; see \secref{sec:ch3_internal_stresses} for details.
In contrast, chemical potential gradients drive the relative fluxes~\cite{Julicher2018}, %
\begin{equation}
	\label{eqn:diffusive_fluxes_many}
	\vect{j}_i = - \size_i\sum_{j=1}^{\Nc}  \Lambda_{ij} \nabla \bar\mu_j - \size_i\vect{\xi}_i
\end{equation}
for $i=1,\ldots,\Nc$, and the solvent flux is given by $\vect j_0 =  -\sum_{i=1}^\Nc \vect{j}_i$, so that all relative fluxes sum to zero.
The thermal fluctuations described by $\vect{\xi}_i$ have zero mean and their correlations are governed by the fluctuation--dissipation theorem~\cite{Julicher2018},
\begin{align}
	\label{eqn:fluctuation_dissipation_relation}
	\mean{\vect{\xi}_i(\vect r, t)\vect{\xi}_j(\vect r', t')}
		= 2 \kBT \nu_0 \Lambda_{ij} \identity \delta(\vect r-\vect r')\delta(t-t')
	\;,
\end{align}
where $\identity$ denotes the identity matrix, indicating that spatial components of $\vect{\xi}_i$ are independent.
Onsager's symmetry principle implies $\Lambda_{ij} = \Lambda_{ji}$, but to derive expressions for the $\Nc$-dimensional matrix $\Lambda_{ij}$ it is instructive to express it by the more general $(\Nc+1)$-dimensional mobility matrix $L_{ij}$, which captures the interplay of all components.
We show in \appref{sec:appendix_compressible} that the two matrices are connected by
\begin{equation}
	\label{eqn:mobility_matrix}
	\Lambda_{ij} =  L_{ij} - \frac{\sum_{k,l=0}^\Nc  \size_l \size_k L_{il} L_{kj}}{\sum_{k,l=0}^\Nc \size_l \size_k L_{kl}}
	\;.
\end{equation}
The rescaled mobility matrix~$\Lambda_{ij}$ is identical to $L_{ij}$ in the simple case where the solvent is fast ($L_{00} \rightarrow \infty$).
In contrast, even in the simple case where all components have equal molecular sizes ($\size_i=1$) and mobilities scale with density ($L_{ij} = \lambda \phi_i \delta_{ij}$), \Eqref{eqn:mobility_matrix} implies effective cross-diffusion~\cite{deGennes1980,Kramer1984,Mao2018},
\begin{equation}
	\label{eqn:kramers_mobility}
	\Lambda_{ij} = \lambda\bigl(\phi_i \delta_{ij} - \phi_i\phi_j\bigr)
	\;,
\end{equation}
which captures the effects of the kinetics of the solvent component not described explicitly.
More realistic kinetic models capture additional concentration-dependences~\cite{Kramer1984, Konig2021, Bo2021}.
In all these cases, the thermal fluctuations described by \Eqref{eqn:fluctuation_dissipation_relation} are then correlated, which can be simulated using Cholesky decomposition~\cite{VanVlimmeren1996}.
Taken together, we find   %
\begin{equation}
	\label{eqn:kinetics_many}
	\partial_t \phi_i +  \nabla \!\cdot\! (\vect{v} \phi_i)=
		\nabla \!\cdot\! \biggl(\! \size_i\!\sum_{j=1}^{\Nc}\Lambda_{ij} \nabla \bar\mu_j \!\biggr)
		+ s_i
		+ \size_i \nabla \!\cdot\! \vect{\xi}_i
	\;,
\end{equation}
which is augmented by \Eqref{eqn:navier_stokes} for the flow field~$\vect v$.
However,  flows can be neglected ($\vect{v}=\vect{0}$) when droplet growth is dominated by diffusion.
For example, \textit{C. elegans} centrosomes and P granules of typical radii $R\approx\unit[1]{\upmu m}$~\cite{Decker2011,Brangwynne2009} exhibit velocities of $v\approx\unitfrac[10]{\upmu m}{min}$~\cite{Hird1993,Keating1998}, but the molecular diffusivity of $D\approx\unitfrac[5]{\upmu m^2}{s}$~\cite{Mahen2011,Griffin2011} implies small P\'{e}clet numbers ($vR/D \approx 0.03$), and so hydrodynamic transport is less relevant than diffusive transport. %
Consequently, we also neglect the stochastic component of the hydrodynamic flow~\cite{Landau1959_6}.  %

The coupled partial differential equations~\eqref{eqn:kinetics_many} for the fields $\{\phi_i\}$ typically permit stationary homogeneous solutions.
For instance, without chemical reactions ($s_i=0$), each homogeneous state  is stationary because of material conservation.
The stability of such states can be assessed using linear stability analysis, which involves linearizing \Eqsref{eqn:kinetics_many} and \eqref{eqn:chemical_potential_many_kappa}.
In particular, the homogeneous state $\phi_i(\vect r) = \bar\phi_i$ is stable if the associated Jacobian%
\begin{align}
	\label{eqn:jacobian_multicomp}
	\jacobian_{ij} &= -\nu_0\size_i^2\vect q^2 \sum_{l=1}^\Nc \Lambda_{il}\left[\pfrac{^2f}{\phi_l\partial \phi_j} + \kappa_{lj}\vect q^2\right] + \pfrac{s_i}{\phi_j}
\end{align}
has only negative eigenvalues for all wavevectors~$\vect q$.
In contrast, the homogeneous state is unstable if $\jacobian_{ij}$ has positive eigenvalues and the largest eigenvalue will typically determine the dominant length scale $2\pi/|\vect q|$ from the associated wavevector~$\vect q$.
However, since \Eqref{eqn:kinetics_many} is nonlinear, such a stability analysis only provides information about the initial dynamics.

The late-stage dynamics can in principle be obtained by simulating  equations~\eqref{eqn:kinetics_many}, which requires two boundary conditions at the system's boundary because of the fourth-order spatial derivatives.
Before we introduce such boundary conditions in \secref{sec:complex_environment}, we first consider the simpler case of isolated droplets, where we can employ periodic boundary conditions on all fields.
This situation is typically considered in numerical investigations, e.g., using finite differences~\cite{Zhou2021a}, spectral methods~\cite{Zhu1999,Shrinivas2021}, Lattice-Boltzmann methods~\cite{Thampi2011,Ledesma-Aguilar2014,Semprebon2016,Gsell2022}, or stochastic solvers~\cite{GarciaOjalvo2012, Lord2014}.
Since these numerical simulations are  costly~\cite{Wodo2011}, contemporary computers are not able to simulate multi-component dynamics for the time and length scales relevant in cells~\cite{Zhou2021a}.
However, to discuss dynamics of individual droplets, it often suffices to consider binary phase separation and sharp interfaces (see \secref{sec:eff_model}), which will be the main focus of this review. %

\subsubsection{Behaviour of binary liquids}
\label{sec:binary_liquids}

\begin{figure*}
	\centering	
	\includegraphics[width=17cm]{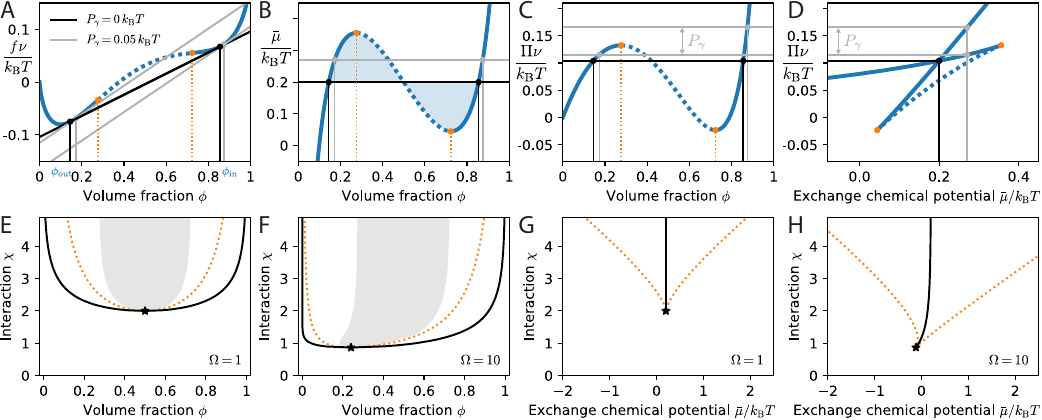}
	\caption{\textbf{Phase behaviour of a binary liquid.}
	(A) Free energy density $f$ given by \Eqref{eqn:free_energy_density_binary} as a function of the fraction~$\phi$ (blue line) for $\size=1$, $\chi=2.5$, and $w=0.2$.
	The common-tangent construction (black line) solves \Eqsref{eqn:coexistence} and thus determines the equilibrium fractions~$\phiOut$ and $\phiIn$ (black disks)  without Laplace pressure~$\Plaplace$.
	Finite $\Plaplace$ requires two parallel tangents offset by $\Plaplace$ (gray lines).
	The inflection points (orange disks) enclose the spinodal region where $f''(\phi)<0$.
	(B) Exchange chemical potential $\bar\mu$ given by \Eqref{eqn:chemical_potential_binary} as a function of $\phi$ corresponding to panel A. 
	The blue regions of equal area indicating a Maxwell construction, equivalent to the common-tangent construction.
	(C) Osmotic pressure $\Pi=\nu^{-1}\phi\bar\mu - f$ as a function of $\phi$ corresponding to panel A.
	(D) $\Pi$ as a function of $\bar\mu$ corresponding to panel A.
	(E, F) Phase diagrams corresponding to panel A highlighting the equilibrium volume fractions (black binodal line) and the spinodal region (within the orange spinodal line) for  a symmetric mixture ($\size=1$, panel E) and for large solutes ($\size=10$, panel~F).
	The critical point is marked by a star and grey areas denotes the region where bi-continuous structures have lower energy than droplets in two dimensions (obtained by comparing interfacial energies of disks and stripes for equal length scales). %
	(G, H) Grand-canonical phase diagram corresponding to panels E and F.
	The binodals (black lines) mark the first-order transition between dilute and dense phases.
	Both phases are linearly stable in the spinodal region (between orange dotted lines).
	}
	\label{fig:phase_diagrams}
\end{figure*}

In many cases, we do not need to describe all individual components of a mixture and can instead focus on droplet material separating from a solvent.
In this case, the mixture is treated as a binary fluid, and only the volume fraction~$\phi$ of one component needs to be specified~(${\Nc=1}$). The local free energy density then reduces to
\begin{equation}
	\label{eqn:free_energy_density_binary}
	f(\phi) = \frac{\kBT}{\nu}\biggl[
		\frac{\phi}{\size}\ln(\phi)
		+ (1-\phi)\ln(1-\phi)
		+ w \phi
		+ \chi \phi(1-\phi)
	\biggr]	, 
\end{equation}
which follows from \Eqref{eqn:free_energy_many} with $\nu=\nu_0$, $\size=\size_1$, $w=w_1 + \frac12 \chi_{11}$, and $\chi = -\frac12\chi_{11}$.
The associated exchange chemical potential reads
\begin{multline}
	\label{eqn:chemical_potential_binary}
	\bar\mu = \kBT\Bigl[
		1 + \ln (\phi ) - \size - \size \ln (1-\phi )
\\
		+\size w 
		+ \size\chi (1-2  \phi )
	\Bigr], 
\end{multline}
and determines the phase behaviour of the liquid; see \figref{fig:phase_diagrams}.
Note that the internal energy~$w$ only shifts the chemical potentials and as such does not influence phase behavior; it will become important when discussing chemical reactions in \secref{sec:chemical_reactions}.

A homogeneous system with volume fraction~$\phi$ is stable with respect to small perturbations when the Jacobian given in \Eqref{eqn:jacobian_multicomp} is negative, i.e., if ${f''(\phi) \propto \bar\mu'(\phi) > 0}$.
This occurs when the interaction parameter~$\chi$ is smaller than
\begin{equation}
	\label{eqn:chi_spinodal}
	\chi_\mathrm{spin}(\phi) 
	= \frac12\left(\frac{1}{\size\phi}  + \frac{1}{1- \phi}\right)
	\;.
\end{equation}
The resulting spinodal line $\chi_\mathrm{spin}(\phi)$ encloses a region in phase space (\figref{fig:phase_diagrams}).
Within this region the homogeneous state is unstable and develops patterns at a dominant length scale $l_\mathrm{max} =\pi[-8 \kappa/f''(\phi )]^{1/2}$ that eventually lead to phase separation, which is known as \textit{spinodal decomposition}~\cite{Weber2019}.
The minimum of the curve $\chi_\mathrm{spin}(\phi)$ corresponds to the critical point,
\begin{align}
	\chi_* &= \frac{\bigl(1 + \sqrt{\size}\bigr)^2}{2 \size}
&
	\phi_* &= \frac{1}{1 + \sqrt{\size}}
	\;,
\end{align}
which reduces to $\chi_* = 2$, $\phi_* = \frac12$, and $\bar\mu_*=w \kBT$ for symmetric mixtures ($\size=1$). %
For $\chi < \chi_*$, only the homogeneous state is stable. %

Homogeneous systems are stable against small fluctuations outside the spinodal region, but they are not necessarily the only stable states or even the state with the lowest free energy.
To see whether heterogeneous states (e.g., droplets) are able to form, we next consider the equilibrium conditions~\eqref{eqn:coexistence_many}.
These conditions can be solved graphically using the \emph{common--tangent construction} (\figref{fig:phase_diagrams}A), or equivalently the equal--area Maxwell construction (\figref{fig:phase_diagrams}B).
Alternatively, the equilibrium point can be read off the $\Pi(\bar\mu)$ plot shown in \figref{fig:phase_diagrams}D, and the corresponding volume fractions then identified via \figsref{fig:phase_diagrams}B--C.
For the chemical potential of the binary system given by \Eqref{eqn:chemical_potential_binary}, these graphical constructions are possible, and the solutions unique, if $\chi > \chi_*$.
The two fractions $\phiBaseOut$ and $\phiBaseIn$ correspond to the equilibrium fractions of the dilute and dense phase, respectively outside and inside droplets.
Note that $\phiBaseOut$ is also known as the \emph{saturation concentration} since droplets can only form when it is exceeded.
The common--tangent construction also illustrates that the phase-separated state generally has lower free energy than the corresponding homogeneous state with equal average composition~$\bar\phi$.
The two states only have equal free energy in the limit where one phase of the phase-separated state vanishes ($\bar\phi = \phiBaseOut$ or $\bar\phi = \phiBaseIn$).
Taken together, the fractions $\phiBaseOut$ and $\phiBaseIn$ as a function of $\chi$ form the \emph{binodal line}.
Since the phase-separated state does not exist outside the binodal line, this line also bounds the region where the homogeneous phase is globally stable; see \figref{fig:phase_diagrams}.
The binodal line can be obtained analytically for the symmetric case,
\begin{equation}
	\label{eqn:chi_binodal_n1}
	\chi_\mathrm{bin}(\phi) = \frac{\ln (1-\phi )-\ln (\phi )}{1-2 \phi }
\qquad \text{for} \quad \size=1
	\;,
\end{equation}
which reiterates that phase separation is independent of the internal energy~$w$.
The coexisting volume fractions cannot be determined analytically, but a fixed-point iteration yields the useful approximation~\cite{Qian2022a}
\begin{align}
	\label{eqn:phi_binodal_approx}
	\phiBase_\mathrm{in/out} &\approx \frac{1}{1 + \exp\left[-\chi \tanh\left(\pm \chi \sqrt{\frac{3\chi - 6}{8}}\right)
	\right]}
\end{align}
for $\size=1$.
In the limit of strong interactions ($\chi \gg 1$), this predicts $\phiBaseOut \approx e^{-\chi}$, which reveals that even a minute increase of the interaction energy by $2 \kBT$ (so $\chi$ is increased by $2$) leads to an almost $10$-fold decrease in the saturation concentration~$\phiBaseOut$.
In contrast, the spinodal branch scales as $\phi^\mathrm{spin}_\mathrm{out} \approx 1/(2\size\chi)$, indicating that a linear stability analysis of the homogeneous state is inadequate to study phase separation in detail.

The entire phase diagram can be separated into three distinct regions.
For low $\chi$ or dilute solutions ($\phi < \phiBaseOut$ or $\phi > \phiBaseIn$) only homogeneous states persist.
Conversely, only the phase-separated state exists in the spinodal region when interactions are sufficiently strong ($\chi > \chi_\mathrm{spin}$).
Between these two regions, homogeneous and phase-separated states are both locally stable, but the phase-separated state has lower energy; nucleation events thus become possible when thermal fluctuations are included.

To see how droplets form and evolve, we next derive the dynamics of heterogeneous systems, akin to the multi-component case discussed in \secref{sec:kinetics_multi-component}.
The associated free energy functional reduces to
\begin{align}
	\label{eqn:free_energy_binary}
	F[\phi] &= \int \left[
		f(\phi) + \frac{\kappa}{2} (\nabla \phi)^2
	\right]\diff V
	\;,
\end{align}
which implies an extra term $-\kappa\nabla^2 \phi$ in the exchange chemical potential $\bar\mu$ given in \Eqref{eqn:chemical_potential_binary}.
The kinetics of the binary liquid follow from \Eqref{eqn:kinetics_many},
\begin{equation}
	\label{eqn:kinetics_binary}
	\partial_t \phi =
		\nabla \cdot \bigl( \Lambda(\phi) \nabla \bar\mu \bigr)
		+ s(\phi, \bar\mu)
	\;,
\end{equation}
where we for simplicity neglect hydrodynamic flows~$\vect v$, and the mobility $\Lambda(\phi) = \Lambda_0 \phi (1-\phi)$ follows from \Eqref{eqn:kramers_mobility}.
Without the source term $s$, which accounts for reactions (see \secref{sec:chemical_reactions}), \Eqref{eqn:kinetics_binary} is know as the \emph{Cahn--Hilliard equation}~\cite{Cahn1958}.
This equation describes many phase separation phenomena and has been studied extensively~\cite{CahnHilliardBook2019}.

\begin{figure*}
	\centering	
	\includegraphics[width=17cm]{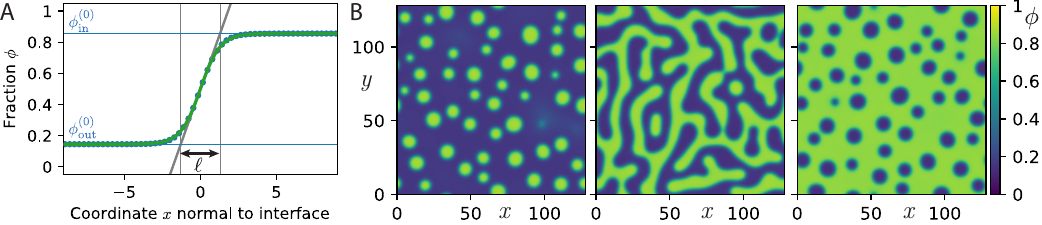}
	\caption{\textbf{Interfaces and morphologies in a binary liquid.}
	(A) Interfacial profile~$\phi(x)$ that minimises $F$ given by \Eqref{eqn:free_energy_binary} in a one-dimensional system (blue dots) compared to a fit of a hyperbolic tangent profile (green line). %
	The slope at the midpoint (grey line) defines the interfacial width~$\width$~\cite{Cahn1958}.
	(B) Numerical simulations of two-dimensional systems from random initial conditions with $\bar\phi=0.3, 0.5, 0.7$ (from left to right), showing droplets, bicontinuous structures, and bubbles, respectively.
	(A, B) Model parameters are $\size=1$ and $\chi=2.5$.
	}
	\label{fig:interface_morphology}
\end{figure*}

\subsubsection{Interfaces and phase morphologies}
In the absence of chemical reactions ($s=0$), the dynamics described by \Eqref{eqn:kinetics_binary} result in equilibrium states that minimise the free energy~$F$ given by \Eqref{eqn:free_energy_binary}.
\Figref{fig:interface_morphology}A shows that the corresponding equilibrium profile~$\phiInt(x)$ smoothly interpolates between the fractions $\phiBaseOut$ and $\phiBaseIn$ of the two co-existing phases.
Denoting by $x$ a coordinate perpendicular to the interface positioned at $x=0$, we approximate the profile as
\begin{align}
	\label{eqn:interfacial_profile}
	\phiInt(x) &\approx \frac{\phiBaseIn + \phiBaseOut}{2} + \frac{\phiBaseIn - \phiBaseOut}2 \tanh\left(\frac{2x}{\width}\right)
\end{align}
and determine the interfacial width $\width$ below. %
This expression is exact for simple polynomial free energies~\cite{Krapivsky2010, Weber2019}, and it captures the sigmoidal profile of more general cases.
For the free energy density given by \Eqref{eqn:free_energy_density_binary} with symmetric molecular sizes ($\size=1$), the interfacial width~$\width$ is approximately
\begin{align}
	\width &\approx
	 \frac{1}{\phiBaseIn - \phiBaseOut}\,\sqrt{\frac{6\nu\kappa}{\kBT}}
	\;,
\end{align}
which is valid for weak segregation ($\chi \gtrsim  2$); see \appref{sec:appendix_interface}.
This expression indicates that stronger segregation, i.e., larger $\phiBaseIn - \phiBaseOut$, implies narrower interfaces.

In the typical case where the region of the interface is thin compared to the linear size of the adjacent phases, one can use a \emph{thin-interface approximation} and replace the sigmoidal profile by a step function.
The excess free energy, which is not accounted for by the bulk phases, defines the surface energy~\cite{Cahn1958}
\begin{align}
	\gamma &= \int_{-\infty}^{\infty} \biggl[
                f\bigl(\phiInt\bigr)
                + \frac{\kappa}{2} \bigl(\phiInt'\bigr)^2
                - \frac{f(\phiBaseIn) + f(\phiBaseOut)}{2}
            \biggr]  \diff x
\notag\\
	&= \sqrt{2\kappa} \int_{\phiBaseOut}^{\phiBaseIn} \left[
		f(\phi)
		- \frac{\muEq}{\nu} \phi
		+ \PEq
	\right]^{\frac12} \diff \phi
	\;,
\end{align}
where $\muEq$ is the exchange chemical potential and $\PEq =\frac{\muEq}{\nu} \phiBaseOut  - f(\phiBaseOut) =\frac{\muEq}{\nu} \phiBaseIn - f(\phiBaseIn)$ denotes the associated osmotic pressure; see \appref{sec:appendix_interface}.
The second line assumes a monotonic profile $\phiInt(x)$, which may not be valid in complex mixtures~\cite{Ji2023}.
However, $\phiInt(x)$ is always monotonic in binary mixtures, and in the symmetric case $\size=1$, we find that (\appref{sec:appendix_interface})
\begin{align}
	\gamma &\approx
		 \sqrt{\frac{\kBT \kappa}{2 \nu}} \cdot \frac{\phiBaseOut - \phiBaseIn}{3}
		 \ln\left(4\phiBaseIn\phiBaseOut\right)
	\;,
\end{align}
which is valid for weak segregation ($\chi \gtrsim 2$).
The surface energy~$\gamma$ increases with larger gradient parameter~$\kappa$, smaller molecular volume~$\nu$, and stronger segregation (larger $\phiBaseIn-\phiBaseOut$).
It can often be measured experimentally~\cite{Michieletto2022} or determined from molecular simulations~\cite{Yamaguchi2023} and thus provide a useful parametrisation of interfacial physics.

The thin-interface approximation simplifies the analysis of equilibrium states in binary liquids:
material conservation fixes the total volume of the dilute and dense phases, so their shape is the only degree of freedom.
The total free energy thus comprises a constant term covering the bulk contributions and an interfacial term, $\gamma A$, where $A$ is the total interfacial area.
Equilibrium states thus minimise $A$, which implies constant mean curvature of interfaces~\cite{Dierkes1992}.
Consequently, surface energies can be interpreted as a mechanical surface tension~\cite{Ip1994}, explaining why the minority phase favours spherical shapes.
Conversely, if both phases occupy similar volumes, bicontinuous structures, such as cylinders and flat lamella, can emerge; see \figref{fig:interface_morphology}B.
These structures predominate in a region of phase space that overlaps significantly with the spinodal region; see \mbox{\figref{fig:phase_diagrams}E,F}.
However, these two phenomena are distinct, so droplets can for instance emerge from spinodal decomposition.
Bicontinuous structures are rarely observed in biological cells, and so we exclusively focus on droplets in this review.

\subsection{Effective description of droplet dynamics} \label{sec:eff_model}
To focus on  droplets, we next consider situations in which the dense phase forms spatially compact regions separated from the dilute phase by a thin interface.
For simplicity, we focus on three-dimensional droplets (key results for two dimensions appear in \appref{sec:appendix_droplets_2d}) without reactions (which are added in \secref{sec:chemical_reactions}).

\subsubsection{Local equilibrium at the interface}
Using the thin-interface approximation, the free energy~$F$ of an isolated droplet reads
\begin{align}
	\label{eqn:free_energy_droplet}
	F \approx (\Vsys - V) f(\phiOut) + V f(\phiIn) + \gamma A
	\;,
\end{align}
where $\Vsys$ is the system volume.
For simplicity, we consider a spherical droplet of radius~$R$ (implying volume $V=\frac{4\pi}{3} R^3$ and surface area $A=4\pi R^2$) with a surface tension~$\gamma$ that is independent of the interface curvature, although this approximation can in principle be lifted~\cite{Tolman1949}.
Material conservation implies
\begin{align}
	\label{eqn:material_conservation_droplet}
	\bar\phi \Vsys = (\Vsys - V) \phiOut + V \phiIn
	\;,
\end{align}
where $\bar\phi$ is the average component fraction in the system.
Minimising $F$ using this constraint and fixed $\Vsys$ results in the coexistence conditions~\cite{Weber2019}
\begin{subequations}
\label{eqn:coexistence}
\begin{align}
	\label{eqn:coexistence_mu}
	\bar\mu(\phiIn) &= \bar\mu(\phiOut)
	\;,
\\
	\label{eqn:coexistence_P}
	\Pi(\phiIn) &= \Pi(\phiOut) + \frac{(d-1)\gamma}{R}
	\;,
\end{align}
\end{subequations}
where $d$ is the dimension of space.
These equations reduce to \Eqsref{eqn:coexistence_many} if surface tension is negligible, $\gamma=0$.
However, mechanical equilibrium as expressed in \Eqref{eqn:coexistence_P} reveals that $\gamma$ generates an extra pressure $\Plaplace = (d-1)\gamma/R$, which is known as \emph{Laplace pressure}.
\EQsref{eqn:coexistence} are solved by the coexisting equilibrium fractions~$\phiEqIn$ and $\phiEqOut$, which depend on the droplet radius~$R$ and only coincide with $\phiBaseIn$ and $\phiBaseOut$ for flat interfaces ($R\rightarrow\infty$).
A pressure increase~$\Delta P$ inside droplets generally implies larger chemical potentials, and thus larger $\phiEqIn$ and $\phiEqOut$.
In the simple case of an incompressible droplet originating from strong phase separation, the fraction inside will hardly change, $\phiEqIn \approx \phiBaseIn$, whereas  the volume fraction outside approximately reads~\cite{Vidal2020}
\begin{equation}
	\label{eqn:ceqout_from_pressure}
	\phiEqOut \approx \phiBaseOut \exp\left(\frac{\Delta P}{\cIn \kBT}\right)
	\;,
\end{equation}
where $\cIn = \phiBaseIn/\nu\size$ is the number concentration inside the droplet and the dilute phase is approximated as an ideal solution.
When only surface tension acts on the droplet, $\Delta P = \Plaplace$, we find the \emph{Gibbs-Thomson relation}~\cite{Thomson1872}
\begin{align}
	\label{eqn:ceqout_laplace}
	\phiEqOut &\approx \phiBaseOut \left(1 + \frac{\capLen}{R}\right)
	\;,
\end{align}
assuming the radius~$R$ is large compared to the capillary length  $\capLen \approx (d-1) \gamma\nu\size/(\phiBaseIn \kBT)$.
This illustrates that smaller droplets require larger concentrations of droplet material in their vicinity to be in equilibrium with their surroundings. %

\subsubsection{Classical nucleation theory}
\label{section:Classical-nucleation-theory}
The free energy of an isolated droplet, given by \Eqref{eqn:free_energy_droplet}, captures the trade-off between bulk terms that promote phase separation and a surface term that suppresses it.
This trade-off becomes particularly important at the initial phase of droplet formation, where a small droplet is embedded in a dilute phase characterised by a fraction~$\phiSup$ away from the droplet.
The free energy difference between such a small droplet and the corresponding homogeneous phase without a droplet can be expressed as~\cite{Xu2014,Kalikmanov2013}
\begin{align}
	\label{eqn:energy_nucleation}
	\DF = \gamma A - \Df V
	\;,
\end{align}
where $\Df$ captures the bulk free energy change when droplet material is moved from the dilute phase to the droplet.
Assuming a small droplet ($V \ll \Vsys$), we find $\Df \approx \Pi(\phiIn) - \Pi(\phiSup)- [\bar\mu(\phiIn) - \bar\mu(\phiSup)]  \cIn$, where $\cIn =\phiIn / (\nu \size)$~\cite{Vidal2021}.
Since pressure differences typically equilibrate quickly, $\Df$ is essentially proportional to the chemical potential difference between the dilute phase and the droplet.
In the simple case where the dilute phase can be described as an ideal solution, $\muSup \approx \kBT \ln(\phiSup)$, we use \Eqref{eqn:coexistence_mu} to find $\Df \approx  \kBT \cIn \ln (\phiSup/\phiBaseOut)$.
Supersaturated solutions ($\phiSup>\phiBaseOut$) thus imply positive $\Df$, which lowers $\Delta F$ and facilitates droplet formation.

\begin{figure}
	\centering	
	\includegraphics[width=\columnwidth]{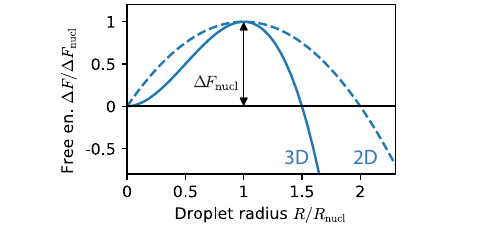}
	\caption{Free energy~$\DF$ of a small droplet of radius~$R$ relative to the homogeneous phase in $2$ and $3$ dimensions; see \Eqref{eqn:energy_nucleation}.
	The maximum marks the critical size~$\Rnuc$ and the nucleation barrier~$\Fnuc$.
	}
	\label{fig:nucleation}
\end{figure}

\Figref{fig:nucleation} shows that $\DF$ is typically non-monotonic since the surface term dominates for small~$R$, while the driving force~$\Df$ eventually leads to spontaneous growth of large droplets.
The transition between these two regimes is characterised by the maximum at
\begin{align}
	\label{eqn:nucl_radius}
	\Rnuc &= \frac{2\gamma}{\Df}
& \text{and} &&
	\Fnuc &= \frac{16 \pi \gamma^3}{3 \Df^2}
	\;,
\end{align} 
for $d=3$ dimensions; \appref{sec:appendix_droplets_2d} gives similar results for two dimensions.
This thermodynamic analysis suggests that small droplets ($R < \Rnuc$) dissolve spontaneously, while larger droplets grow.
However, thermal fluctuations, which become significant for small droplets, can spontaneously generate droplets larger than the critical size, which will then grow macroscopically.
The precise rate at which such \emph{nucleation} happens depends on many microscopic details, but a central result of \emph{classical nucleation theory} is that the number of nucleated droplets per unit volume and time reads~\cite{Kalikmanov2013, Xu2014}
\begin{align}
	\label{eqn:nucleation_rate}
	J_\mathrm{nucl} &= J_0 \exp\left(-\frac{\Fnuc}{\kBT}\right) \;, 
\end{align}
where the pre-factor~$J_0$ contains details about the kinetics of individual molecules~\cite{Kalikmanov2013}.
Equation~\eqref{eqn:nucleation_rate} shows that nucleation is exponentially suppressed if creating a critical nucleus is costly, e.g., because of large surface tension~$\gamma$ or weak driving $\Df$.
Since $\Df$ often depends on temperature, nucleation typically shows a marked temperature dependence, e.g., in biomolecular phase separation~\cite{Wilken2024}.

\subsubsection{Growth of a single droplet}\label{s:Growth of a single droplet}
Once a droplet has nucleated, it can grow by taking up material from its surroundings or by creating material inside itself via chemical reactions.
For simplicity, we postpone the discussion of reactions to \secref{sec:chemical_reactions} and first focus on the case without reactions ($s=0$).
Growth of droplets of any shape can be described by the normal velocity~$\vNorm$ of the droplet interface, which is given by~\cite{Lifshitz1961,Wagner1961, Zwicker2017, Weber2019}
\begin{equation}
	\vNorm = \frac{\jIn - \jOut}{\phiEqIn - \phiEqOut} \cdot \vect{n}	
	\label{eqn:drop_interface_speed}
	\;,
\end{equation}
where $\vect{n}$ is the normal vector to the interface.
Here, $\jIn$ and $\jOut$ denote the fluxes directly inside and outside the interface, and so their difference is related to the net accumulation of material.
The fluxes follow from the Cahn--Hilliard \Eqref{eqn:kinetics_binary}, but the thin-interface limit also provides a useful simplification since $\phi$ typically does not vary much either within or outside droplets.
Consequently, we linearise \Eqref{eqn:kinetics_binary} around the compositions~$\phiBaseIn$ and $\phiBaseOut$ to obtain a simple diffusion equation
\begin{align}
	\label{eqn:effective_model_pde}
	\partial_t \phi_n \approx D_n\nabla^2\phi_n %
	\;,
\end{align}
where we have dropped the fourth-order derivative of $\phi$ and introduced the diffusivity
\begin{equation}
	\label{eqn:effective_model_diffusivity}
	D_n \approx \nu \Lambda (\phiBase_n) f''(\phiBase_n)%
\end{equation}
for $n\in\{\mathrm{in}, \mathrm{out}\}$.
The diffusion \Eqref{eqn:effective_model_pde} is easier to solve than the full Cahn--Hilliard \Eqref{eqn:kinetics_binary}, often permits analytical approximations~\cite{Kulkarni2023}, and thus provides useful approximations for the diffusive fluxes $\vect j_n \approx -D_n \nabla \phi_n$.

As a simple example, we consider a spherical droplet embedded in a large system.
Without chemical reactions, the droplet is essentially homogeneous, implying $\jIn=0$.
Conversely, the fraction~$\phiSup$ far away from the droplet might deviate from its equilibrium fraction~$\phiEqOut$, resulting in diffusive fluxes around the droplet.
Solving \Eqref{eqn:effective_model_pde} at stationary state in three dimensions, we find ${\jOut=(\phiEqOut - \phiSup)\DOut R^{-1}}\vect{e}_r$.
The droplet growth rate following from \Eqref{eqn:drop_interface_speed} then reads~\cite{Kulkarni2023}
\begin{align}
	\difffrac{R}{t} = \frac{\DOut}{R} \cdot \frac{\phiSup - \phiEqOut(R)}{\phiBaseIn - \phiBaseOut}
	\label{eqn:single_drop_growth}
	\;,
\end{align}
where we have neglected the small effects of surface tension in the denominator.
Consequently, droplets grow spontaneously when the surrounding is supersaturated, $\phiSup > \phiEqOut$.
Moreover, droplets exhibit a drift when the surrounding concentration field is anisotropic~\cite{Weber2019},
\begin{equation}
	\difffrac{\vect{x}}{t} = \dfrac{d \, \DOut}{\phiBaseIn - \phiBaseOut}
		\Bigl[\nabla\phiSup - \nabla\phiEqOut\Bigr]_{\vect{x}}
	\label{eqn:single_drop_drift}
	\; , 
\end{equation}
where the gradients are evaluated at the droplet position~$\vect x$, and $d$ is the dimensionality of the system.
Here, the anisotropy can originate from either concentration variations in the surrounding field $\phiSup$ or the equilibrium concentration itself, e.g., due to gradients in additional, regulating components~\cite{Weber2017, Brangwynne2009} or mechanical properties~\cite{Rosowski2019,Rosowski2020}.

Drift caused by anisotropic fields is often dominated by other processes that affect droplets' positions.
For instance, droplets in cells can be actively moved by molecular motors~\cite{Cochard2023} and hydrodynamic flows~\cite{Brangwynne2009}.
Moreover, all droplets exhibit \emph{Brownian motion} characterised by a diffusion constant~\cite{Einstein1905}
\begin{equation}
	\Ddrop \approx \frac{\kBT}{6\pi\eta R}
	\label{eqn:droplet_diffusivity}
	\;,
\end{equation}
where $\eta$ denotes the viscosity of the surrounding liquid, which can depend on the droplet size in complex liquids such as the cytosol~\cite{Garner2022}.
\EQref{eqn:single_drop_drift} illustrates that droplets drift up supersaturation gradients, but Brownian motion caused by thermal fluctuations~\cite{Einstein1905}, hydrodynamic effects (discussed in \secref{sec:internal_complex_interfaces}) and active transport may dominate their trajectory.

Growth stops when the droplet is equilibrated with the dilute phase, $\phiSup = \phiEqOut$.
In the simple case of a finite system of volume~$\Vsys$ and fixed average fraction~$\bar\phi$, the final droplet size is
\begin{align}
	\label{eqn:droplet_size}
	\Veq &= \Vsys \, \frac{\bar\phi - \phiEqOut}{\phiBaseIn - \phiBaseOut}
	\;,
\end{align}
which gives an explicit equation for $V$ in the limit $\phiEqOut\approx \phiBaseOut$.
This equation illustrates that droplet size is controlled by material conservation (equation~\ref{eqn:material_conservation_droplet}), and that no droplets exist for volume fractions outside the binodal region ($\bar\phi < \phiBaseOut$ or $\bar\phi > \phiBaseIn$). %

\subsubsection{Dynamics of many droplets}
\label{s:Dynamics of many droplets}

The dynamical equations derived above for individual droplets also allow us to discuss the interaction of many droplets as long as they are well-separated.
We thus consider a collection of $\Nd$ droplets described  by their positions~$\vect x_n$ and radii $R_n$ for $n=1,\ldots,\Nd$.
These droplets are embedded in a common dilute phase, whose composition is described by a field~$\phi(\vect x)$.
In the typical situation where droplets occupy only a small fraction of the entire system, it is convenient to assume that~$\phi(\vect x)$ is defined across the entire system, and that droplets provide localised perturbations to this field~\cite{Kulkarni2023}.
In this case, we describe the dynamics of the dilute phase by
\begin{equation}
	\label{eqn:many_droplets_dilute_field}
	\partial_t \phi = \DOut \nabla^2 \phi - (\phiBaseIn - \phi) \sum_{n=1}^\Nd \difffrac{V_n}{t} \delta(\vect{x}_n - \vect{x})
	\;,
\end{equation}
where the first term accounts for diffusive transport, while the second term accounts for the material exchange with droplets of volume $V_n = \frac{4\pi}{3} R_n^3$ localised at $\vect{x}_n$.
The growth and drift dynamics of the droplets follow from \Eqsref{eqn:single_drop_growth} and \eqref{eqn:single_drop_drift},
\begin{subequations}
\label{eqn:many_droplets_dynamics}
\begin{align}
	\difffrac{V_n}{t} &= \frac{4\pi R_n \DOut}{\phiBaseIn - \phiBaseOut}  \Bigl(\phi(\vect x_n) - \phiEqOut(R_n)\Bigr) \;, 
\\[4pt]
	\difffrac{\vect{x}_n}{t} &= \dfrac{d \, \DOut}{\phiBaseIn - \phiBaseOut}  \nabla\phi\bigr|_{\vect{x}_n}
	\;,
\end{align}
\end{subequations}
where we assume that the equilibrium concentration~$\phiEqOut$ is independent of position.

\EQsref{eqn:many_droplets_dilute_field}--\eqref{eqn:many_droplets_dynamics} describe how droplets exchange material via diffusion through the dilute phase.
In particular, droplets grow by taking up material from the dilute phase when the fraction~$\phi$ in their surrounding is larger than the saturation concentration~$\phiEqOut$.
\EQref{eqn:ceqout_laplace} shows that $\phiEqOut$ decreases with the droplet radius~$R$, implying that larger droplets grow at the expense of smaller droplets.
The resulting coarsening process, where the droplet count decreases while the mean size increases, is known as \emph{Ostwald ripening}~\cite{Ostwald1897,Voorhees1992}.
In the limit where the diffusion of material between droplets is faster than their growth ($\DOut \gg L \, \diff R/\diff t$, where $L$ is the typical droplet separation), heterogeneities in the dilute phase equilibrate quickly, so that all droplets experience a similar fraction $\phi(\vect x)\approx\phiSup$.
In this case, \Eqref{eqn:ceqout_laplace} implies that droplets grow when their radius is larger than the critical radius 
\begin{align}
	\label{eqn:critical_radius}
	\Rcrit &= \frac{\capLen\phiBaseOut}{\phiSup - \phiBaseOut}
	\;,
\end{align}
while smaller droplets shrink and disappear. 
Lifshitz and Slyozov~\cite{Lifshitz1961}, as well as Wagner~\cite{Wagner1961} showed that the normalised size distribution $P(R/\Rcrit)$ assumes a universal form and that the critical radius evolves as $\partial_t \Rcrit \propto \Rcrit^{-2}$.
Consequently, the mean droplet radius obeys the \emph{Lifshitz--Slyozov--Wagner scaling law},
\begin{equation}
	\mean{R} \propto t^{\frac13}
	\;,
\end{equation}
and so $\mean{V} \propto t$ and $\mean{\Nd} \propto t^{-1}$ in $d=3$ dimensions.

Droplets %
also diffuse and coalesce, leading to fewer, larger droplets over time.
Interestingly, this coarsening process exhibits the same scaling law as Ostwald ripening~\cite{Smoluchowski1918}.
To see this, assume that coalescence roughly doubles the droplet volume, ${\Delta\!V \propto R^3}$, and the time between collisions scales as $\Delta t \propto L^2/\Ddrop$ where $L\propto R$ is the average droplet separation and \Eqref{eqn:droplet_diffusivity} implies $\Ddrop\propto R^{-1}$.
Taken together, $\Delta\!V/\Delta t \propto 1$, so the average volume grows linearly in time, similarly to Ostwald ripening.
Although both coarsening processes obey the same scaling laws, the pre-factors can be very different and depend on the particular details of the system~\cite{Weber2019}.

Droplet coarsening is generally driven by the thermodynamic tendency to reduce interfacial area.
However, the coarsening dynamics depend on details of kinetic processes~\cite{Bray2003}, so observed exponents vary with composition and the strength of phase separation~\cite{Konig2021,vonHofe2025}.
For instance, coarsening can be slowed down by trapped species~\cite{Webster1998}, charges~\cite{Chen2023b}, sub-diffusion in the dilute phase~\cite{Lee2021}, and elastic inhomogeneities~\cite{Zhu2001}.
In contrast, coarsening can be accelerated when it is conversion-limited~\cite{Wagner1961,Lee2021}, when mechanical relaxation is relevant~\cite{Tateno2021}, and when hydrodynamic effects are significant~\cite{Siggia1979, Shimizu2015}, e.g., via coalescence-induced coalescence~\cite{Wagner1998,Wagner2001}. %
In biological cells, coarsening is further complicated by active production and degradation of material, which we consider in section~\ref{sec:chemical_reactions}.
It is currently unclear what governs the size distribution of droplets in cells, although there is evidence that a competition between coalescence and nucleation is crucial~\cite{Lee2023} and that the material properties of the surrounding are important~\cite{Banerjee2024}.
Taken together, this shows that cells can use many physical processes to control droplet size.

\subsection{Passive control of droplet life cycle}

The basic physics of phase separation discussed so far already provides ample opportunity for control of droplet life cycles.
Droplets form by nucleation when the mixture is supersaturated, and then grow by taking up material from the dilute phase.
Multiple droplets compete against each other for this material, resulting in a coarsening process that favours larger droplets, so that a single droplet remains in the final equilibrium state.
The size of this droplet is a function of global parameters, which can also vary in space and time. %

\subsubsection{Global parameters}  %
\label{sec:global_parameters}

Droplet size is directly influenced by the amount of droplet material in the system, $\bar\phi\Vsys$; see \Eqref{eqn:droplet_size}.
Cells can thus influence whether droplets form and how large they get by producing and degrading the relevant droplet material.
Droplet size also depends on the coexisting volume fractions $\phiBaseOut$ and $\phiBaseIn$, which in turn depend on the interaction~$\chi$ and the molecular size~$\size$; see \Eqref{eqn:phi_binodal_approx}.
These aspects can be influenced by modifying the involved molecules, either by post-translational modifications~\cite{Schisa2021,Snead2019} or by multimerisation~\cite{Rossetto2024}, which can induce phase separation \emph{in vitro}~\cite{Rana2024} and \emph{in vivo}~\cite{Shin2017a, Bland2023}.
The interactions between molecules are also sensitive to environmental influences~\cite{Villegas2022}, including 
temperature~\cite{Fritsch2021,Stormo2024,Meyer2025},
pH~\cite{adame2020liquid},
osmotic conditions~\cite{Jalihal2020},
crowding conditions~\cite{Holt2023}, 
CO\textsubscript{2} concentration~\cite{Zhang2022a},
and electric fields~\cite{Agrawal2022}.
Cells can exploit the phase transition associated with changes in these environmental parameters to deploy appropriate responses, e.g., to control flowering in plants~\cite{Emenecker2021}.
There is also evidence that evolution adapts genetic sequences to different temperature niches to control this behaviour~\cite{Keyport-Kik2024}.

\subsubsection{Spatial gradients}
\label{sec:spatial_gradients}
The sensitivity of phase separation to environmental parameters provides exquisite control over droplet positions when external gradients are involved.
For instance, compositional gradients can bias droplets toward one side of a system~\cite{Brangwynne2009,Cejkova2014,Saha2016,Doan2024,Romano2025}, for both passive~\cite{Weber2017, Kruger2018} and active droplets~\cite{Hafner2023a}, in a process akin to diffusiophoresis~\cite{Shim2022}.
In sufficiently large cells, gravity can also become important, so that droplets sediment due to density mismatch~\cite{Feric2013}.
Finally, artificially induced temperature gradients can position droplets~\cite{Fritsch2021}, although such gradients are likely negligible in most cells~\cite{Mabillard2023}.

\subsubsection{Temporal variations}
\label{sec:temporal_variations}
Global parameters can also vary in time.
For instance, the external environment can exhibit wet--dry cycles~\cite{Haugerud2023} or temperature oscillations, which need to be compensated in circadian rhythms~\cite{Tariq2024}.
Alternatively, cells can generate oscillations internally to affect condensates~\cite{Heltberg2022}.
These oscillatory drives can in particular influence chemical reactions controlled by droplets~\cite{Bartolucci2023a}.
Besides structured variations of global parameters, cells are also subject to fluctuations, e.g., from gene expression.
In the simple case of binary mixtures, these fluctuations can be buffered by phase separation since the equilibrium concentrations are independent of the total amount of material~\cite{Klosin2020}.
Buffering effects persist in multi-component mixtures, although concentration buffering may be weaker~\cite{Zechner2024, Devirie2021}. On the other hand, volume buffering is also then possible~\cite{Michaels2025}.

\sectionseparation

\section{Beyond simple liquid droplets}
\label{sec:internal_complexity}

Cells contain an endless variety of biomolecules.
For example, there are an estimated $10^5$ different proteins~\cite{Cooper2000} present in the cell.
It is thus not surprising that condensates typically consist of many different types of molecules~\cite{Updike2009, Jain2016, Hubstenberger2017, Youn2019}, many of which have complex individual properties themselves. Molecules can also form partially ordered structures within droplets, such as filaments~\cite{Scholl2024} and porous networks~\cite{Tollervey2024, Dollinger2025}.
The complex compositional features resulting from this mixture of biomolecules
affects both the bulk and interfacial properties of droplets, and also enables the coexistence of multiple different types of droplet within a cell.

\subsection{Complex molecular interactions}
\label{sec:complex_molecules}

Biomolecular condensates contain a huge variety of components, including water, ions, small peptides, proteins and flexible polymers such as RNA and DNA~\cite{Zhou2024,Holehouse2023}.
Since it is challenging to directly study interactions between these molecules in their native environment, much experimental work has thus far relied on \emph{in vitro} reconstitution~\cite{Currie2023,Arter2022, Qian2022,Currie2021}. %
Here, it is crucial to exploit label-free techniques~\cite{Ibrahim2024,McCall2020} since fluorescent labels typically strongly affect droplet condensation~\cite{McCall2020,Dorner2024}.
One key insight from these approaches is that often a few \emph{scaffold} components control the formation of condensates, while \emph{client} molecules partition into them~\cite{Ning2020, GuillenBoixet2020, Joseph2020, Wheeler2018, Banani2016}.
However, this dichotomy is likely an oversimplification and it will be interesting to more rigorously quantify the contributions of individual components, using theoretical analysis~\cite{Kabalnov1987a,Thewes2023,Desouza2024} or the recently developed dominance metric~\cite{Qian2023}, which can uncover basic interaction motifs between components~\cite{Qian2023a}.

It is clear that biomolecular phase separation involves many different kinds of interactions between molecules~\cite{Nordenskiold2024}.
To distill the essential features of these interactions, it is often helpful to coarse-grain particular microscopic models~\cite{Li2022b, vonBulow2024}. %
For instance, many biomolecules can be conceptualised as \emph{associative polymers}~\cite{Semenov1998, Rubinstein2001, Chang2017, Danielsen2023, Pappu2023,Nordenskiold2024} and described by sticker-and-spacer models~\cite{Semenov1998, Harmon2017, Choi2020a, Holehouse2023,GrandPre2023}, where flexible spacers link stickers with specific interactions.
However, real biomolecules are complex and thus require flexible modelling approaches, which are reviewed in refs.~\cite{Holehouse2023, Jacobs2023, Zhou2024, Adachi2024}.
Such models allow elucidating how particular molecular features, like the amino acid sequence of proteins, affect condensate properties~\cite{Maristany2025,Sundaravadivelu-Devarajan2024,Rekhi2024,Biswas2025}, and can explain why dilute phases often exhibit surprisingly large clusters~\cite{Kar2022,Lan2023}.

Interactions between biomolecules can be roughly divided into weak, transient interactions and strong, persistent bonds.
While weak multivalent interactions are generally responsible for phase separation into liquid-like phases, strong interactions tend to form long-lived assemblies, which can persist as nano-domains inside condensates~\cite{Gao2024}.
Moreover, chemical modifications of biomolecules can affect their interactions.
For instance, proteins can be changed by post-translational modifications~\cite{Cooper2000}, which are implicated in condensate regulation~\cite{Hofweber2019, Snead2019, Soeding2019, Schisa2021}.
Such changes can induce multimerisation, which can in turn enable phase separation.
To understand this phenomenon, we imagine a molecule whose weak interactions are insufficient for phase separation.
A multimer of such molecules, however, can interact with many other multimers, which effectively increases the molecular volume~$\size$ while maintaining the interaction strength~$\chi$ per unit volume~\cite{Rossetto2024}.
This process essentially reduces the entropic cost of concentrating molecules, promoting phase separation.
This feature has been used to induce artificial condensates in cells~\cite{Shin2017a, Shin2018}.
More generally, it is likely that combinations of weak and strong interactions enable fine-tuning of condensate properties~\cite{Schmit2021,Schmit2020}.

\subsection{Complex material properties}

The complex nature of the constituent biomolecules also results in condensates exhibiting complex rheological properties that cannot be captured by simple (Newtonian) droplet models~\cite{Michieletto2022}, due to long, flexible polymers deforming, entangling, and interacting with each other.
Cells exploit these complex material properties to carry out function~\cite{SanfeliuCerdan2025}: for example, centrosomes are able to grow rapidly in a fluid-like manner, but also exhibit solid-like properties to support large stresses during cell division~\cite{Woodruff2021, Paulin2025}. 

\subsubsection{Internal stresses}
\label{sec:ch3_internal_stresses}
The formation, and subsequent growth, of droplets induces stresses both in the droplets themselves and in the surrounding solvent. These stresses may then give rise to hydrodynamic flows (equation~\ref{eqn:navier_stokes}) that in turn affect the composition of the droplet--solvent mixture (equation~\ref{eqn:continuity_many}).  
As such, there is a strong two-way coupling between hydrodynamics and droplet regulation, via the stress field: droplets can induce flows, which then advect and/or deform the droplets themselves. Recent numerical works have highlighted how this interplay can accelerate droplet coarsening~\cite{Hester2023,Gsell2022}, as well as affect the morphology of intermediate droplet states~\cite{Gsell2022}.

For a single-component Newtonian fluid, the total stress can be decomposed into contributions arising from hydrostatic pressure and viscous dissipation~\cite{Landau1959_6}.
The hydrostatic pressure $p$ enforces incompressibility of the fluid, and is determined by momentum conservation (Navier--Stokes; equation~\ref{eqn:navier_stokes}) and the incompressibility condition $\nabla\cdot\vect{v}=0$.
Multi-component fluids additionally induce an \emph{equilibrium stress}, $\tensor\stress_{\textrm{eq}}$, that captures osmotic and interfacial effects. Further, many biological fluids exhibit a complex rheology that results in an elastic contribution to the stress, denoted by $\tensor\stress_{\textrm{el}}$. The overall stress tensor $\tensor\stress$ then comprises the sum of each of these different contributions, 
\begin{equation} \label{eqn:total_stress}
\tensor\stress=-p\identity + \frac{\eta}{2}\left[ \nabla\vect v + \left(\nabla \vect v \right)^\intercal\right]+\tensor\stress_{\textrm{eq}} +\tensor \stress_{\textrm{el}}\;,
\end{equation}
where $\eta$ is the shear viscosity of the mixture.
Note that measurements of the viscosity of both condensates and the cytoplasm vary widely, and in general depend strongly on composition~\cite{Guo2015,Michieletto2022}. Cells may also modulate the viscosity of the cytoplasm to regulate diffusion-controlled processes~\cite{Persson2020}.
We also note the possibility of additional contributions to the total stress, such as active and nematic stresses~\cite{Julicher2018}, which are beyond the scope of this review.
An explicit expression for the equilibrium stress $\tensor\stress_{\textrm{eq}}$ can be derived by considering the variation of the total free energy with respect to changes in volume (see \appref{sec:appendix_equilibrium_stress}), and is given by
\begin{equation}\label{eqn:equilibrium_stress}
	\tensor\stress_{\textrm{eq}} = \left(f - \sum_{i=1}^\Nc \frac{\phi_i \bar\mu_i}{\nu_i} \right) \identity - \sum_{i=1}^\Nc\pfrac{f}{(\nabla \phi_i)} \nabla \phi_i
	\;,
\end{equation}
where we recall that $\identity$ is the identity matrix.
Here, the first term arises from the osmotic pressure, $\Pi=-f + \sum_i\phi_i \bar\mu_i \nu_i^{-1} $, and the second term from the interfacial stress (often referred to as either the Ericksen~\cite{Julicher2018} or Korteweg stress~\cite{Anderson1998}).
Conversely, the elastic stress $\tensor\stress_{\textrm{el}}$ evolves according to a corresponding constitutive equation, as discussed in \secref{sec:viscoelasticity} below.

\subsubsection{Viscoelasticity} \label{sec:viscoelasticity}

Viscoelastic materials can exhibit either liquid-like or solid-like responses to deformation depending on the specific loading conditions. On long timescales (slow deformations), they flow like a liquid, but on shorter timescales (fast deformations), they resist deformation, like an elastic solid. Recent micro-rheological experiments have shown viscoelastic behaviour in a variety of different condensates~\cite{Michieletto2022, Jawerth2018, jawerth2020protein, Cheng2025}, in which stress relaxation occurs over biologically relevant timescales. 

General models of viscoelastic materials can be very complex. Here, we focus on models of \emph{linear viscoelasticity}, in which the viscous and elastic responses of the material are separable.
One viscoelastic model commonly used to describe condensates is the Maxwell model~\cite{Michieletto2022}.
Maxwell materials can be conceptualised as a spring (elastic component) and a damper (viscous component) connected in series. 
The stress--strain relation for a Maxwell material with viscosity $\eta$ is
\begin{equation}
\tau\stackrel{\triangledown}{ \tensor{\sigma}}_{\textrm{el}} + \tensor{\sigma}_{\textrm{el}}= \eta \left[ \nabla\vect v + \left(\nabla \vect v \right)^\intercal\right] \;,
\label{CH3:Eq_Maxwell_fluid}
\end{equation}
where the relaxation time $\tau$ characterises the timescale over which stresses dissipate~\cite{Doi2013}. 
In a kinematically rigorous Maxwell model, the rate of change of  $\tensor{\sigma}_{\textrm{el}}$, accounting for stretching and rotation of the fluid, is given by the upper-convected derivative $\stackrel{\triangledown}{ \tensor{\sigma}}_{\textrm{el}}=\partial_t\tensor{\sigma}_{\textrm{el}}+\vect{v}\cdot\nabla\tensor{\sigma}_{\textrm{el}}-(\nabla\vect{v})^\intercal\cdot\tensor{\sigma}_{\textrm{el}}-\tensor{\sigma}_{\textrm{el}}\cdot(\nabla\vect{v})$.
In one-dimensional flows where deformations are small, this expression can be approximated by the partial derivative $\partial_t \tensor{\sigma}_{\textrm{el}}$.

The response of a viscoelastic material over different timescales can be characterised by considering sinusoidal deformations of different frequency $\omega$. The storage modulus, $G'$, describes in-phase (solid-like) responses to deformation, and the loss modulus, $G''$, describes out-of-phase (liquid-like) responses. 
For low frequency (slow) deformations, the loss modulus is larger than the storage modulus, and viscous effects dominate. For high frequency (fast) deformations, the storage modulus is larger than the loss modulus, and elastic effects dominate. The two curves intersect at the characteristic frequency~$\omega_\mathrm{c}~=~\tau^{-1}$. The viscoelastic material parameters of condensates can be either measured experimentally~\cite{Michieletto2022}, or predicted via molecular dynamics simulations~\cite{Tejedor2023,Cohen2024}.

Tanaka and coworkers~\cite{tanaka2000viscoelastic,tanaka2006viscoelastic,Tanaka2022} provide a comprehensive overview of the coupling between phase separation and viscoelasticity. 
Viscoelastic properties have been shown to stabilise droplets against dissolution~\cite{Meng2023}, provide resistance to external stress, and allow the formation of complex, domain-spanning droplet morphologies~\cite{Tanaka2022}.
Viscoelasticity also has a strong effect on droplet fusion time. For small $\tau$, fusion time increases relative to the Newtonian case, whereas for large $\tau$, fusion time decreases~\cite{Ghosh2021}.

The viscoelastic properties of condensates may change over time, in a process known as ageing~\cite{jawerth2020protein,Ceballos2022,patel2015liquid,lin2015formation,Jawerth2018}. 
During ageing, condensates progressively become less able to deform, with an increase in relaxation time leading to a more solid-like rheology. The molecular mechanisms that drive condensate ageing are currently an active topic of research~\cite{Alberti2021,Takaki2023, Roy2024}, although many theoretical works model ageing condensates as soft glassy systems.
Aged condensates take longer to recover their shape after deformation and show suppressed ability to coalesce~\cite{Garaizar2022}. It is also possible for aged condensates to form complex droplet architectures, in which different parts of the droplet display distinct mechanical properties~\cite{Garaizar2022a}.
Note that the storage and loss moduli of aged condensates at different time points often collapse onto a single curve when scaled by their (time-dependent) characteristic frequency~$\omega_\mathrm{c}$~\cite{jawerth2020protein,Lin2022, Takaki2023}.

\subsubsection{Gelation}

Liquid droplets are characterised by their dynamic, amorphous structure and ability to dissolve in a reversible manner. It is possible, however, for condensates to undergo a phase transition to a gel-like state~\cite{Shin2017a,Woodruff2018,Schmit2020,Mittag2022}. This transition results in kinetic arrest, in which condensates maintain their amorphous structure but lose their dynamic properties~\cite{Zhang2024a,Zhang2024b, Kang2025}. 
Surprisingly, gelation can lead to enhanced ripening relative to liquid droplets~\cite{Chen2024b}.
A further transition can result in the formation of irreversible fibrous aggregates, which are linked to the development of various neurodegenerative conditions~\cite{Visser2024, Alberti2021}. 
Recent work has shown that gelation typically initiates at the interface between the droplet and its surroundings, propagating inwards, such that droplets transiently form a core--shell geometry with a liquid phase surrounded by a gel phase~\cite{Shen2023}.
Similarly, fast changes in composition can result in the formation of kinetically arrested double emulsion structures within droplets\cite{Erkamp2023a}.
Gelation and aggregation are enhanced in droplets compared to dilute proteins as a result of the increased protein concentration~\cite{Michaels2022,Bartolucci2023}, however the formation of metastable non-fibrillar condensates can also suppress aggregation~\cite{Das2025}. 
Gelation can also be driven by shear flows which help to align protein molecules into a kinematically-arrested orientation~\cite{Shen2020}.

\subsection{Complex interfaces}
\label{sec:internal_complex_interfaces}

Interfaces between droplets and the surrounding fluid are characterised by a sharp change in composition over a finite spatial width. Typically, a rapid and continuous exchange of components between droplets and the surrounding fluid occurs across the interface, even at equilibrium. 
Since droplets in biological cells generally comprise long, complex molecules, however, this exchange may be limited by an interface resistance. Such a resistance could result from potential barriers, a reduction in diffusivity~\cite{Hubatsch2024}, saturation of binding sites~\cite{Zhang2024}, or electrostatic effects~\cite{Majee2024}.

Many proteins molecules are amphiphilic, and can thus act as surfactants, preferentially associating to the interface~\cite{Folkmann2021,Boeddeker2022,Oh2025,Favetta2025}. In general, surfactants stabilise droplet interfaces, resisting both dissolution and coalescence. The formation of complex interfaces via preferential protein association can therefore improve structural integrity of condensates and assist in size regulation~\cite{Kelley2021}, as well as leading to dynamical anisotropies at the interface~\cite{Erkamp2025}. Note that although surfactants usually result in kinetic droplet stabilisation, some complex particles may also result in thermodynamic stabilisation to form Pickering emulsions~\cite{Webster1998,Binks2002,Aveyard2012}.

The amphiphilic properties of protein molecules depend on their particular conformational state~\cite{Golani2025}. Correspondingly, the chemical switching of a particular protein from a surfactant to liquid-like form has been implicated in chromosome clustering during mitotic exit~\cite{HernandezArmendariz2023}. Further, interfaces themselves can alter protein conformations, typically leading to extended molecular configurations~\cite{Farag2022}. Variations in either surfactant configuration or compositional state along an interface can give rise to surface tension gradients. These gradients drive Marangoni flows at the surface of the droplet, which may then lead to droplet swimming~\cite{Maass2016,Michelin2023,JambonPuillet2023}, thus highlighting how complex interfacial properties are able to affect the behaviour of the whole droplet.

\subsection{Electrostatic effects} \label{sec:electrostatics}\label{sec:charged_droplets}

Charge is another important biomolecular property that is involved in the regulation of condensates~\cite{Dai2024a}.
On the one hand, charge patterns on protein surfaces can mediate specific, short-ranged interactions~\cite{Zhou2018}.
On the other hand, a net charge induces electrostatic interactions that can potentially range over long distances.
Indeed, many condensates contain charged components~\cite{Dai2024,Pak2016}, which affects interfacial properties~\cite{Wang2024b,Yu2025,Posey2024,Majee2024,Zhang2021d}, slows down coarsening~\cite{Agrawal2022, Chen2022b}, limits droplet size~\cite{Chen2023b, Li2022a, Luo2024a}, influences droplet--environment interactions~\cite{Smokers2024b,Gurunian2024}, and can potentially lead to droplet division~\cite{Deserno2001,Rayleigh1882}.
The dynamics of charged species can be described by the classical electrostatic framework~\cite{Fredrickson2006,Dobrynin2005,Muthukumar2017}.
The central quantity is the charge density, $\varrho(\vect r) = \sum_i q_i \phi_i$, which depends on the volume fractions $\Phi=\bigl(\phi_1, \ldots, \phi_{\Nc}\bigr)$ of all species $i=1,\ldots,\Nc$, each with charge $q_i$ per unit volume.
The charge density~$\varrho$ induces an electrostatic potential~$\pot(\vect r)$, which satisfies the Poisson equation
\begin{align}
	\label{eqn:poisson_equation}
	\nabla^2 \pot = - \frac{\varrho}{\varepsilon}
	\;,
\end{align}
where $\varepsilon$ is the electric permittivity.
The potential~$\pot$ then gives rise to an electric field $\vect E = -\nabla\pot$ and modifies the overall free energy,
\begin{align}
	\label{eqn:free_energy_electrostatics}
	F_\mathrm{e}[\Phi(\vect r)] = F_\mathrm{bulk}[\Phi(\vect r)] + \frac12\int \pot(\vect r) \varrho(\vect r) \diff V
	\;,
\end{align}
where the first term is given by \Eqref{eqn:free_energy_functional}, while the second term captures electrostatic effects.
The associated \emph{electrochemical potential}, $\mu_i^\mathrm{e} = \nu\delta F_\mathrm{e}/\delta\phi_i$, gains a term proportional to $\pot$, $\mu_i^\mathrm{e} = \mu_i + q_i \pot$, relative to the uncharged case.
This electrochemical potential shows that positively charged molecules move down gradients in the electrostatic potential~$\pot$.
In particular, small mobile ions will tend to distribute so that the system becomes charge neutral, inhibiting long-ranged electrostatic effects.
In the simplest case, these ions can be described by \emph{Debye-Hückel theory}, which accounts for their translational entropies and electrostatic interactions~\cite{Debye1923,Wright2007}.
The main result is that electrostatic effects do not extend significantly beyond the Debye length $\lambda_\mathrm{D} = [\varepsilon\kBT/(\sum_i c_i q_i^2)]^{1/2}$, where $c_i$ is the mean number concentration of the ion species $i$ with charge density $q_i$. %
Note that ion concentrations can be phase-dependent~\cite{Dai2024a}, meaning that electrostatic interactions may be longer ranged than would be expected from the average salt concentration~\cite{Luo2024a}.

Over short distances, electrostatic effects mediate attraction between oppositely charged molecules, which can lead to their phase separation from a neutral solvent~\cite{Ahn2024,Sing2020,Sing2017, Obermeyer2016}.
This \emph{complex coacervation} has  been studied for about a century~\cite{Bungenberg1929} and has been implicated in the origin of life~\cite{Oparin1952,Brangwynne2012}.
In the simplest case, the attraction of oppositely charged molecules can be captured by an additional term in the local free energy~$f$ given by \Eqref{eqn:free_energy_many}.
For instance, in $d=3$ dimensions, \emph{Voorn--Overbeek theory} leads to the free energy density~\cite{Overbeek1957, Kumari2022}
\begin{equation}
	f_\mathrm{VO} = f - \frac{\kBT \alpha}{\nu_0} \left[\sum_{i=1}^\Nc \frac{|q_i| \phi_i}{e}\right]^{\frac32}
	\label{eqn:voorn_overbeek}
	\;,
\end{equation}
where the non-dimensional factor $\alpha=\frac13(4\pi)^2 \ell_\mathrm{B}^{3/2}\nu_0^{-1/2}$ governs the strength of electrostatic interactions.
Here, the Bjerrum length $\ell_\mathrm{B} = e^2 / (4 \pi\varepsilon \kBT)$ describes the distance beyond which thermal energy dominates  electrostatic repulsion of two elementary charges~$e$~\cite{Adar2017}.
A typical value is $\ell_\mathrm{B} \approx \unit[0.7]{nm}$ since the permittivity~$\varepsilon$ of cytosol is comparable to water~\cite{Theillet2014}.
We thus have $\alpha \sim 1\ldots100$ for typical molecular volumes of $\nu_0\sim \unit[10^{-1}\ldots10^3]{nm^3}$, and the electrostatic contribution becomes larger than enthalpic contributions in $f$ if $|q_i|\sim e$, indicating that electrostatic effects can lower the free energy significantly. %
The  theory given by \Eqref{eqn:voorn_overbeek} qualitatively explains complex coacervation, but it cannot predict realistic coacervates quantitatively~\cite{Li2018b}, particularly when the oppositely charged molecules are asymmetric~\cite{Dobrynin1995,Chen2022b, Chen2023b, Li2022a, Luo2024a,Rumyantsev2025}.

\subsection{Multiphase coexistence}

\begin{figure}
	\centering	
	\includegraphics[width=\columnwidth]{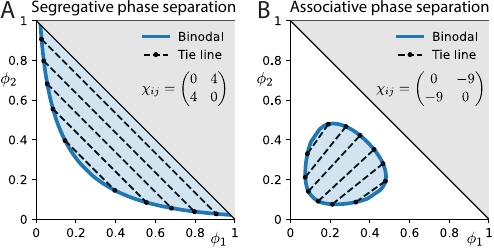}
	\caption{
	\textbf{Multi-component phase diagrams.}
	Phase diagrams of $\Nc=2$ solutes and an inert solvent as a function of the solute fractions $\phi_1$ and $\phi_2$ 
	for two scenarios with different interaction matrices~$\chi_{ij}$.
	A system with an average composition within the blue region can split into two coexisting phases whose compositions follow from the dashed tie lines.
	}
	\label{fig:3_multicomp_phasediagram}
\end{figure}

Cells often comprise many different types of condensates, consistent with Gibbs phase rule, which limits the number of different phases to be smaller than or equal to the number of different components; see \secref{sec:thermodynamics_multi}. %
The number of coexisting phases and their compositions can generally be determined by minimising the free energy of the entire system, which leads to the coexistence conditions given by \Eqsref{eqn:coexistence_many}.
The resulting phase diagrams, showing equilibrium phases as a function of the system's composition, can be complex~\cite{Mao2018}.
Even in the simplest case of two interacting solutes, the topology of the phase diagram depends on the interaction strength; see \Figref{fig:3_multicomp_phasediagram}.
This multi-component, multi-phase problem has been tackled by linear stability analysis of the homogeneous state~\cite{Sear2003, Sear2005, Sear2008, Shrinivas2021,Thewes2023}, grand-canonical simulations~\cite{Jacobs2013, Jacobs2017}, continuation methods~\cite{Binous2021}, convexification of the free energy density~\cite{Lee1992,Wolff2011,Mao2018}, and machine learning methods~\cite{Dhamankar2024}.

\subsubsection{Instability of homogeneous states}

\begin{figure}
	\centering	
	\includegraphics[width=\columnwidth]{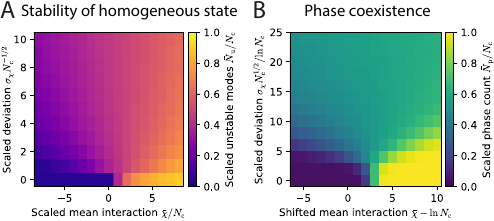}
	\caption{
        \textbf{Scaling behavior of multi-component mixtures with random interactions.}
        (A) Scaled average number of unstable modes, $\Nu/\Nc$, as a function of scaled mean interaction~$\chibar$ and standard deviation $\sigma_\chi$; see \Eqref{eqn:multicomp_unstable_modes_scaling}.
        (B) Scaled average number of coexisting phases, $\NbarP/\Nc$, as a function of scaled $\chibar$ and $\sigma_\chi$; see \Eqref{eqn:multicomp_phasecount_scaling}
		(A, B) Functions were obtained by averaging $300$ samples of random interaction matrices and compositions for each value of $\chibar$ and $\sigma_\chi$ for $\Nc=16,20,24,28,32$; see \refcite{Qiang2024} for details.
	}
	\label{fig:3_multicomp_scaling}
\end{figure}

Phase separation of a multi-component mixture must occur when the homogeneous state is unstable, which is the case when the Hessian $\hessian_{ij} = \partial^2 f/\partial\phi_i\partial\phi_k$  exhibits at least one negative eigenvalue.
If there are multiple negative eigenvalues, the eigenvectors generally point in different directions, suggesting that the system segregates into various phases.
The number of negative eigenvalues, $\Nu$, can be predicted for the Flory--Huggins model given by \Eqref{eqn:free_energy_many}, for which %
\begin{align}
	\hessian_{ij} &= \frac{\kBT} {\nu_0}\left(
		\frac{\delta_{ij}}{\size_i \phi_i} + \frac{1}{\size_0 \phi_0} + \chi_{ij}
	\right)
	\;.
\end{align}
The eigenvalues of $\hessian_{ij}$, and thus $\Nu$, can generally be anlayzed using random matrix theory~\cite{Livan2018,Sear2003,Shrinivas2021}.
In particular, for normally distributed interactions, $\chi_{ij} \sim \normal(\chibar, \sigma_\chi)$, and compositions that are chosen uniformly among all permitted values, the average value of $\Nu$ is approximately~\cite{Qiang2024}
\begin{align}
	\NbarU &\approx \Nc f_\mathrm{u}\!\left(\chibar \Nc^{-1},\, \sigma_\chi \Nc^{-\frac12}\right)
	\;.
	\label{eqn:multicomp_unstable_modes_scaling}
\end{align}
The scaling function $f_\mathrm{u}$ shown in \figref{fig:3_multicomp_scaling}A reveals that homogeneous states are stable if  interactions  tend to be attractive ($\chibar <0$) and vary little.
In contrast, strong repulsive interactions (large $\chibar$) with little variation imply \mbox{$\NbarU \sim \Nc$}, suggesting that all components segregate from each other into $\Nc$ phases, whereas $\NbarU \sim \frac12\Nc$ for very diverse interactions (\mbox{$\sigma_\chi \gg 1$} and $\sigma_\chi\gg \chibar$).
This scaling relation implies that the mean interactions $\chibar$ and the variance $\sigma_\chi^2$ need to scale with the component count $\Nc$ to keep the same fraction of unstable modes, $\Nu/\Nc$.
For biological cells, where $\Nc$ easily exceeds $10^4$, this suggests many unstable modes can only occur when interaction energies are extremely large, on the order of $10^3\,\kBT$.
Linear stability analysis thus suggests that homogeneous mixtures of many components, like in the cytosol, are generally stable~\cite{Sear2005}.

\subsubsection{Coexisting equilibrium phases}
Phase separation can occur even when the homogeneous state is stable, e.g., in the nucleation and growth regime in binary liquids discussed in \secref{sec:binary_liquids}.
This behaviour generalises to multi-component systems.
For example, for a Flory--Huggins model with normally distributed interactions, the number $\Np$ of coexisting phases averaged over all permissible compositions scales as~\cite{Qiang2024}
\begin{align}
	\NbarP &\approx \Nc f_\mathrm{p}\!\left(\chibar  - \ln\Nc, \, \frac{\sigma_\chi \Nc^{\frac12}}{\ln{\Nc}}\right)
	\;,
	\label{eqn:multicomp_phasecount_scaling}
\end{align}
for sufficiently large $\Nc$.
The associated scaling function $f_\mathrm{p}$ shown in \figref{fig:3_multicomp_scaling}B has similar features to the scaling function~$f_\mathrm{u}$ for the unstable modes:
for vanishing variation ($\sigma_\chi=0$) it increases monotonically from $0$ to $1$ with increasing $\chibar$, whereas it approaches~$0.5$ for strong variations (large $\sigma_\chi$).
However, the actual scaling relations, given by the arguments of $f_\mathrm{p}$ in \Eqref{eqn:multicomp_phasecount_scaling}, are fundamentally different between the two cases:
whereas $\chibar$ and $\sigma_\chi^2$ need to increase linearly with component count~$\Nc$ to maintain a fixed fraction of unstable modes, interactions only need to scale logarithmically with $\Nc$ to maintain coexisting phases.
Consequently, for fixed $\chibar$ and $\sigma_\chi$, the fraction of unstable states decreases with larger $\Nc$, whereas $\NbarP/\Nc$ increases.
This result suggests that biological systems, where millions of components interact with energies on the order of a few $\kBT$, exhibit a low phase count compared to the component count in equilibrium (lower left corner of \figref{fig:3_multicomp_scaling}B), consistent with the observed $10^1\ldots10^3$ different condensates~\cite{Rostam2023} that form from $10^5 \ldots 10^7$ different biomolecules~\cite{Cooper2000}.
These systems are thus generically characterised by stable homogeneous states, from which multiple phases can nucleate and stably coexist.
Cells can exploit this fact to create and remove different droplets independently of each other, allowing them to fulfil different functions robustly.

The scaling predicted for the number of unstable modes and the phase count is consistent with the results for binary phase separation discussed in \secref{sec:binary_liquids}:
\EQref{eqn:chi_spinodal} implies that the fraction above which the homogeneous state becomes unstable scales as $\chi^{-1}$, while \Eqref{eqn:phi_binodal_approx} shows that the minimal fraction for forming a droplet  scales as $e^{-\chi}$.
The same scalings are captured by \Eqsref{eqn:multicomp_unstable_modes_scaling} and \eqref{eqn:multicomp_phasecount_scaling}, suggesting that this is a universal feature of phase-separating systems~\cite{Qiang2024}.
This scaling also implies that a stability analysis of the homogeneous state does not carry reliable information about the phase behaviour of multi-component liquids.

\subsubsection{Towards realistic systems}

The scaling result given in \Eqref{eqn:multicomp_phasecount_scaling} predicts some features of condensate coexistence observed in cells, but is likely not quantitatively accurate.
This discrepancy is because the cell's composition is clearly not random, but instead homeostatically controlled, and interactions between biomolecules are also not random, but structured~\cite{Chaderjian2025, DeLaCruz2024, Qian2023a, Chen2023a, Graf2022, Carugno2022} and also under cellular control~\cite{Kilgore2025,Zwicker2022,Chen2023,Teixeira2023,Quinodoz2024}.
Moreover, interactions are typically complex~\cite{Riback2020}, likely involving higher-order interactions~\cite{Luo2024} and the effect of polymer dynamics~\cite{Galvanetto2024, Danielsen2023, Rubinstein2001}.
Note also that the number~$\Np$ of coexisting phases is only the most basic descriptor of multi-component phase separation. The compositions of different phases are also crucial for cellular function since they give rise to the droplets' identities.
Along these lines, it will also be interesting to understand what determines the interfacial parameters $\kappa_{ij}$ in \Eqref{eqn:free_energy_functional}, since these are crucial for predicting surface tensions and thus multiphase morphologies~\cite{Snead2025, Yan2025a, Li2024, Yo2021, Mao2018, Mao2020}.
In the simplest case, surface tension and other related quantities might simply scale as a power law of the distance to a critical point~\cite{Pyo2023}, but whether this situation holds generally is unclear.
Besides surface tension, droplet morphology is also shaped by other objects in the surrounding environment, as discussed in \secref{sec:complex_environment} below.

\sectionseparation

\section{Droplets in complex environments}
\label{sec:complex_environment}

Thus far, we have focussed exclusively on isolated droplets, i.e., dense droplet material fully surrounded by a dilute phase. In reality, biological cells are full of complex structures that interact with droplets, such as the plasma membrane, other organelles, and the cytoskeleton. 
In this section, we explore the effect of interactions with such structures on droplet regulation, detailing in turn interactions with 1D filaments, 2D membranes, and 3D meshes.
Although the specific interactions between different condensates and cellular structures take many forms, we focus here on the key physical processes that affect droplet regulation. Several recent reviews~\cite{Wiegand2020,Gouveia2022,Lee2022} provide comprehensive overviews of the different structures that interact with droplets in cells.

\subsection{Theory of wetting} \label{sec:wett_theory}

The interaction between a multi-component mixture and external structures can be captured by including a surface energy term in the free energy, which acts at the boundary of the mixture. In the continuous description presented in \secref{sec:continuous_desc}, we thus have a total energy given by~\cite{Cahn1977, Bonn2009}
\begin{equation}
	\label{eqn:free_energy_surface_interaction}
	\Ftot[\Phi] = \Fbulk[\Phi] + \oint \fWall(\Phi) \, \diff A \;,
\end{equation}
where $\fWall$ is the free energy density of surface interactions, also known as a \emph{contact potential}. Minimising this extended free energy results in unchanged bulk chemical potentials (\appref{sec:appendix_variation}), in addition to boundary conditions
\begin{equation}
	\sum_{j=1}^{\Nc} \kappa_{ij} \, \partial_n \phi_j
		+  \pfrac{\fWall}{\phi_i} = 0 \;,
\end{equation}
where $\partial_n$ denotes the derivative in the outward normal direction to the boundary. Note that if the external surface interacts identically with all components of the mixture, homogeneous Neumann conditions ($\partial_n \phi_i=0$) are recovered.

\subsubsection{Partial, full and pre- wetting} \label{sec:wett_trans}

\begin{figure}
	\centering	
	\includegraphics[width=\columnwidth]{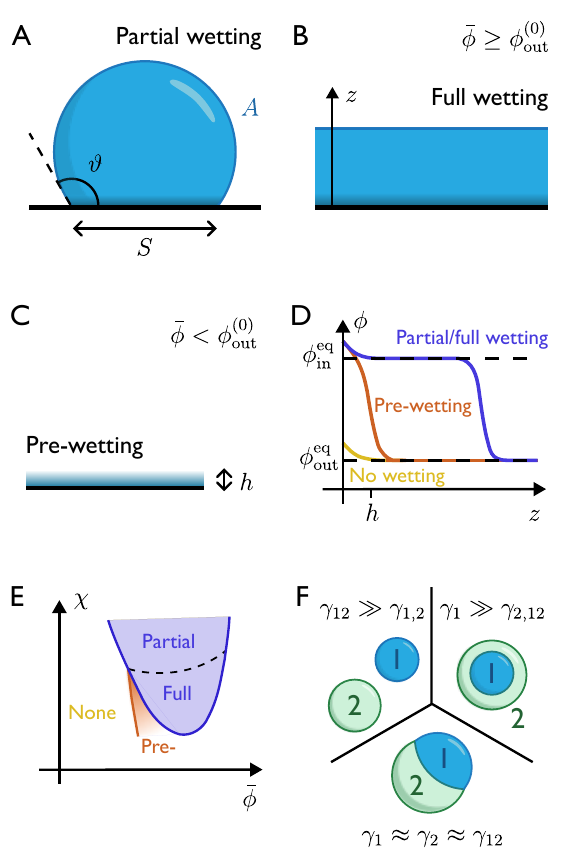}
	\caption{
        \textbf{Different phenomena resulting from droplet--structure interactions.}
        (A) Dense phase partially wets the external boundary, forming a compact droplet whose interface meets the boundary at a contact angle $\vartheta$. The areas of the interfaces between the droplet and the dilute phase, and the droplet and the boundary are denoted by $A$ and $S$, respectively.
        (B) Dense phase completely wets the boundary.
        (C) In an undersaturated system, droplet material pre-wets the boundary, forming a layer of thickness $h$.
        (D) Droplet volume fraction $\phi$ as a function of distance~$z$ from the boundary. Note that the gradient of $\phi$ at $z=0$ is the same for all three cases.
        (E) Phase diagram showing the pre-, full and partial wetting regimes. The typical size of the pre-wetting regime has been exaggerated for clarity.
        (F) Droplet--droplet interactions between two immiscible phases result in either droplet separation, engulfment, or partial wetting (clockwise from top left).
Panels (A) to (E) inspired by figures in~\cite{Zhao2021} and panel (F) inspired by figure in~\cite{Shin2017a}.
	}
	\label{fig:4_wetting_basics}
\end{figure}

To simplify the discussion, we focus on a binary fluid mixture in contact with a single external boundary (be this a rigid solid, soft solid, or liquid).
We then extend the effective droplet model introduced in \secref{sec:eff_model}, in which a dense (droplet) phase of volume $V$ and composition~$\phiEqIn$, coexists with a dilute phase of composition~$\phiEqOut$. 
The energetic interactions of the binary mixture with the boundary are described by surface energies $\gammaIn = \fWall(\phiEqIn)$ and $\gammaOut = \fWall(\phiEqOut)$, for the dense and dilute phases, respectively. 
Assuming that the boundary has a fixed surface area $S_0$, the total free energy of the mixture is then~\cite{deGennes1985, deGennes2003}
\begin{equation}\label{eqn:energy_surfaces}
F=F_0 (\phiEqIn, V) + \gamma A + \gammaIn S + \gammaOut (S_0-S)\;,
\end{equation}
where $V$ is the total droplet volume, $F_0$ is the free energy contribution due to the bulk dense and dilute phases, $A$ is the area of the interface between these phases, $\gamma$ is the corresponding surface tension, and $S$ is the area of the droplet--boundary interface (\figref{fig:4_wetting_basics}A).
In general, droplets can form if the free energy difference, $\Delta F$, between a phase-separated state and the corresponding homogeneous system is negative. When an external boundary is present,
\begin{equation} \label{eq:deltaF_surface}
\Delta F = \gamma A - \Df V + (\gammaIn-\gammaOut) S\;,
\end{equation}
where we recall that $\Df$ is the driving force for macroscopic phase separation, as defined in \Eqref{eqn:energy_nucleation}.

In undersaturated systems ($\bar{\phi}<\phiBaseOut$), ${\Df<0}$, and so isolated droplets cannot form.
However, if $\gammaIn<\gammaOut$, and $\gammaOut - \gammaIn$ is sufficiently large, a layer of droplet material can be stabilised along the external surface (\figref{fig:4_wetting_basics}C). This phenomenon is known as pre-wetting. As the concentration of the homogenous phase approaches the saturation concentration, the thickness of the pre-wetted layer diverges as ${h\sim-\log(\phiBaseOut - \bar{\phi})}$ ~\cite{cahn1977critical}.
Typically, pre-wetting is only possible in a very narrow portion of the parameter space, but recent work has shown that binding interactions between droplet molecules and the surface can significantly enhance the size of the pre-wetted regime~\cite{Zhao2021,Zhao2024}

In a supersaturated system ($\bar{\phi}>\phiBaseOut$), $\Df>0$, and so macroscopic droplets are able to form regardless of the presence of external boundaries. Nevertheless, surface interactions still have a significant impact on droplet regulation, both in terms of droplet nucleation (see \secref{sec:het_nucl} below) and droplet morphology.
In particular, if $\gamma < \gammaOut - \gammaIn$ the dense phase will completely `wet' the surface, forming a macroscopic layer between the surface and the dilute phase.
This configuration is known as perfect, or full, wetting. For systems in which $\gamma > |\gammaOut - \gammaIn |$, the dense phase will instead partially wet the surface to form spatially compact droplets. The crossover between these two cases is known as the wetting transition~\cite{cahn1977critical}. The different phenomenologies of pre-, partial, and full wetting are highlighted in \figsref{fig:4_wetting_basics}{A-D}. \Figref{fig:4_wetting_basics}E shows a phase diagram demarcating the transitions between the different wetting regimes.
 
\subsubsection{Solid substrates} \label{sec:solid_substrates}

For droplets that partially wet a rigid external boundary, droplet shape is determined by minimising the total free energy $F$ given by \Eqref{eqn:energy_surfaces}.
For macroscopic droplets, $V$ is not affected by interactions with the surface. As such, the equilibrium droplet shape is that which minimises interfacial energy,
\begin{align}
	\Fint = \gamma A + (\gammaIn - \gammaOut) S \; .
\end{align}
This energy is minimised by droplets with constant mean curvature, thus resulting in spherical cap shapes.
For planar substrates, $\Fint$ is then minimal when the contact angle at the droplet interface, $\vartheta$, obeys Young's law
\begin{align}
	\cos(\vartheta) &= \frac{\gammaOut - \gammaIn}{\gamma}
	\;.
	\label{eqn:partial_wetting_angle}
\end{align}
Note that when $\gammaOut>\gammaIn$, $0<\vartheta<\pi/2$ and the surface is preferentially wetted by the droplet phase. When $\gammaOut<\gammaIn$, $\pi/2<\vartheta<\pi$ and the surface is preferentially wetted by the dilute phase. 

Capillary forces resulting from differential wetting may deform sufficiently soft solid substrates. The elasto-capillary length $\Lelasto=\gamma/E$ characterises the relative strength of capillary interactions and elastic deformation, for a solid with stretching modulus $E$.
For very small $\Lelasto$, resistance to stretching dominates capillary forces, and the substrate is effectively rigid.
If $\Lelasto$ is comparable to the interfacial width $\width$ of the droplet, the surface will deform at the contact line, forming small ridges which point towards the droplet~\cite{style2013universal, Andreotti2020}. 
For very soft materials, $\Lelasto$ is comparable to droplet radius $R$, and hence the rigidity of the solid is negligible compared to surface tension. In this case, the substrate deforms on the scale of the entire droplet, which thus acquires a lens-like shape~\cite{Andreotti2020}, with angles between each interface determined by Neumann's law ${\gamma\vect{t} + \gammaIn \vect{t}_{\textrm{in}} + \gammaOut \vect{t}_{\textrm{in}} = 0}$, where $\vect{t}$, $\vect{t}_{\textrm{in}}$, and $\vect{t}_{\textrm{out}}$ are the tangential unit vectors along the droplet--dilute, droplet--substrate, and dilute--substrate interfaces, respectively.

\subsubsection{Interactions with other droplets}
\label{sec:ch4_droplet_droplet_interactions}

If multiple immiscible droplets are present in a given system, they may come into contact, and thus interact. The morphology of the droplet--droplet interaction is governed by the relative strengths of surface tension between the different phases (\figref{fig:4_wetting_basics}F), and may be regulated by surfactant proteins that associate to interfaces and modify these surface tensions~\cite{Lu2025}. To describe the interaction between two immiscible droplets, we denote the surface tension between the droplets by~$\gamma_{12}$, and the respective surface tensions between each droplet and the dilute phase as $\gamma_1$ and $\gamma_2$. If $\gamma_{12}\gg\gamma_1\approx\gamma_2$, then droplet--droplet interfaces are relatively energetically costly, and so the droplets will remain separated. If $\gamma_{1}\approx\gamma_2\approx\gamma_{12}$, the droplets will partially wet each other. The contact angle of the droplet--droplet interface is then determined by Neumann's law, analogously to droplets wetting a very soft surface as described above.
Finally, if either $\gamma_1$ or $\gamma_2$ is sufficiently large compared to the other two surface tensions, perfect wetting conditions may occur, resulting in the engulfment of the droplet with large surface tension by the other.
Such `double emulsion' structures have been observed in several sub-cellular structures, including nucleoli and stress granules~\cite{Feric2016}.

\subsubsection{Heterogeneous nucleation} \label{sec:het_nucl}

Interactions with external surfaces promote droplet nucleation by lowering the energy barrier associated with droplet formation. 
This phenomenon is known as heterogeneous nucleation~\cite{volmer1939kinetik,turnbull1950kinetics}. 
For a droplet to nucleate at the surface of a planar wall, the energy barrier is given by~\cite{turnbull1950kinetics}
\begin{equation}
	\label{eqn:nucl_energy_heterogeneous}
	\Fnuc = \dfrac{4 \pi \gamma^3}{3\Df^2}(2+\cos\vartheta)(1-\cos\vartheta)^2 \;.
\end{equation}
This expression is obtained analogously to the energy barrier for homogeneous nucleation (equation \ref{eqn:nucl_radius}) by finding the maximum value of $\Delta F$ prescribed by \Eqref{eq:deltaF_surface}.
For perfectly non-wetting conditions ($\vartheta =\pi$) the classic homogeneous nucleation result is recovered.
Any other contact angle results in preferential nucleation at the surface.
Analogous expressions can also be obtained for more complex surface shapes~\cite{fletcher1958size}. By tuning the strength of interactions between droplet material and other cellular structures, cells are thus able to regulate the location and rate at which condensates are nucleated~\cite{Shevtsov2011,Shimobayashi2021,Schede2023, Visser2024a}.

\subsection{Wetting of filaments}

\begin{figure*}
	\centering	
	\includegraphics[width=\textwidth]{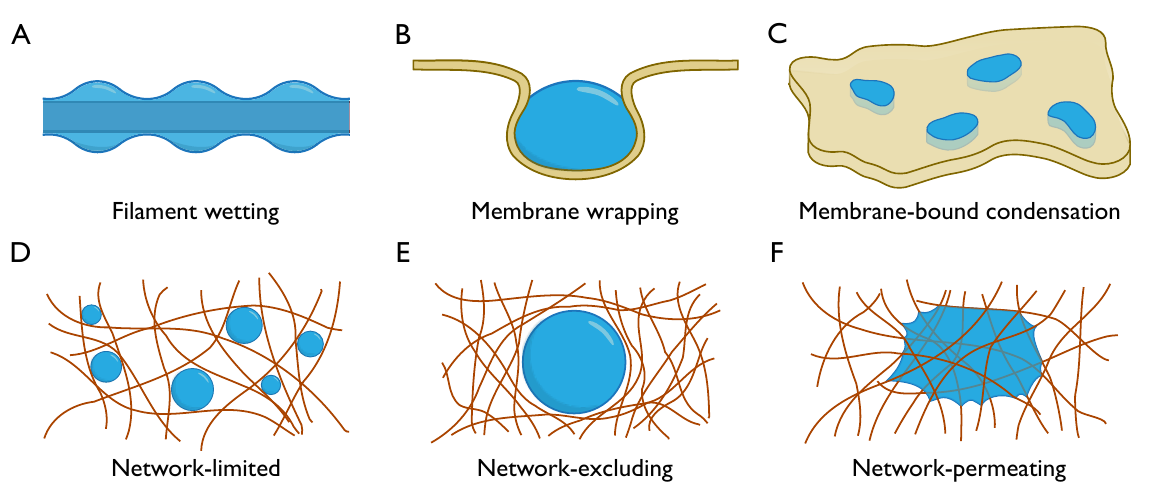}
	\caption{
        \textbf{Interactions between droplets and various external structures.}
	(A) 1D structures: droplet beading on a filament. (B-C) 2D structures: membrane deformation due to droplet capillary forces; formation of lipid rafts within a membrane. (D-F) 3D structures: droplets limited by the mesh size; droplets deforming the mesh; droplets permeating the mesh.
	}
	\label{fig:4_droplet_structure_interaction}
\end{figure*}

Throughout the cell, droplets interact with various semi-flexible filaments and fibres, including chromatin in the nucleus and cytoskeletal filaments such as microtubules~\cite{Wiegand2020,Broedersz2014,Mohapatra2023}. In this section, we discuss filaments which can be considered to be one-dimensional structures with regard to their interactions with droplets ({i.e.}, when droplets interact with single, or few, individual filaments). We consider the interaction of droplets with a mesh-like network of many filaments in \secref{sec:ElasticMeshes} below.

Under full wetting conditions, proteins may condense around a filament to form a cylindrical layer of droplet material with uniform thickness. Over time, however, surface tension may cause this layer to break up into a series of smaller beaded droplets with finite spacing, via the Plateau--Rayleigh instability (see \figref{fig:4_droplet_structure_interaction}A). %
Recent works have demonstrated this effect for certain microtubule-binding proteins~\cite{setru2021hydrodynamic,Jijumon2022}. 
The spacing of beaded droplets increases with the original thickness of the wetted cylinder~\cite{setru2021hydrodynamic}, and in turn controls the spacing of microtubules that subsequently nucleate from these droplets.

If multiple filaments are simultaneously wetted by the same droplet, the capillary bridges that form between the filaments give rise to adhesive forces. These adhesive forces then lead to bundling~\cite{Hernandez-Vega2017,Feigeles2025} and repositioning~\cite{Strom2024} of adjacent filaments. Adhesive forces between droplets and filaments can also control the positioning of droplets within the cell, concentrating them within filament-rich regions~\cite{Boeddeker2022,Boddeker2023}. 
Droplet position and shape may also be regulated by active filament polymerisation within the droplet, resulting in droplet translocation and deformation~\cite{Wiegand2020}.
Finally, droplet formation on individual strands of semi-flexible filaments can result in a co-condensation process, in which the filament is pulled into the condensate by capillary forces. This effect can be resisted if the filament is held at a sufficiently high tension which opposes the capillary pull~\cite{Quail2021}, and provides an additional method of force generation within the cell. Droplet--filament co-condensation has also been implicated in the spatial organisation of chromatin~\cite{Takaki2025}, with protein--DNA co-condensates preferentially nucleating at DNA loops.

\subsection{Interaction with membranes}

Cellular membranes take the form of phospholipid bilayers, and allow the cell to partition different regions of space. Membranes are typically very thin and flexible, and contain an array of proteins and other biomolecules that can reorganise laterally in a fluid-like manner.
Droplets come into contact with many different membranes within the cell, including the plasma membrane, membrane-bound organelles, vesicles, and the endoplasmic reticulum. 
Droplet--membrane interactions are mediated by both physical ({e.g.}, wetting) and biomolecular ({e.g.}, binding) effects~\cite{botterbusch2021interactions, Lipowsky2023,Mondal2023,Mangiarotti2024, Mangiarotti2025, Lipowsky2025}. In this section, we first discuss interactions between macroscopic droplets and membranes, before considering the possibility of phase separation phenomena within the membrane itself.

\subsubsection{Condensation at membranes}
\label{sec:condensates_at_membranes}

Macroscopic three-dimensional droplets readily come into contact with membrane structures within the cell. The condensation of droplets at membrane surfaces can be influenced by binding between the membrane and membrane-associated proteins, leading to enhanced droplet nucleation~\cite{botterbusch2021interactions}, modified contact angles~\cite{GrandPre2024a}, and spatiotemporal regulation of droplet assembly~\cite{Snead2022,Liu2023c}. In turn, droplet condensation can drive the generation of membrane curvature~\cite{Agudo2021}, which may result in arrested droplet coarsening~\cite{Winter2024}, and even initiate endocytosis via adhesive forces~\cite{Bergeron2021endocytic,Kozak2022}. 

One striking example of membrane--droplet interactions is the complex deformation and mutual remodelling that results from wetting effects, as highlighted by recent experiments with synthetic analogue systems~\cite{Mondal2023,Mangiarotti2024, Mangiarotti2025, Nair2025} and detailed numerical simulations~\cite{Mokbel2023}. %
This two-way coupling causes membrane budding, interfacial ruffling, and nanotube formation. As discussed in \secref{sec:solid_substrates} above, capillary forces are able to deform soft surfaces over the scale of the elasto-capillary length $\Lelasto$. For deformable solids which are extremely thin in one dimension, such as membranes, the relevant elastic energy which resists capillary forces is the bending energy, characterised by bending rigidity $B$.
Redefining the elasto-capillary length as $\Lelasto=\gamma/B$ then gives the length scale over which capillary forces give rise to membrane--droplet deformations.

When the influence of wetting is sufficiently strong, capillary forces result in the membrane partially wrapping around a droplet, to increase the droplet--membrane contact area (\figref{fig:4_droplet_structure_interaction}B). In extreme cases, this effect can lead to complete engulfment of the droplet by the membrane and subsequent fission of the droplet-containing membrane section~\cite{Wang2024a}. Similarly, when a droplet wets the outside of a membrane, engulfment may allow the droplet to be transported to the interior side of the membrane. An alternative droplet regulation mechanism occurs when excess membrane area results in the protrusion of membrane nanotubes along the droplet's contact line~\cite{Agudo2021,kusumaatmajae2021wetting}. These protrusions may then wrap around a small portion of droplet material and remove it from the cell.

\subsubsection{Condensation within membranes}

Spatial heterogeneities in the surface organisation of cellular membranes have an important role in compartmentalising specific membrane processes, similarly to how protein droplets compartmentalise the interior of the cell. In particular, the lipid raft hypothesis proposes that cholesterol-enriched regions of the plasma membrane preferentially associate with protein receptors to form dynamic domains that float within the membrane~\cite{Pike2009}. Studies based on extracted and synthetic membranes have implicated lipid raft formation with macroscopic phase separation of the membrane into lipid-ordered and lipid-disordered domains, with protein complexes then partitioning into the ordered phase~\cite{Veatch2005}. Phase separation in a two-dimensional system (\figref{fig:4_droplet_structure_interaction}C) follows the same principles as those discussed in \secref{sec:continuous_desc} for three-dimensional systems. Appendix \ref{sec:appendix_droplets_2d} briefly discusses alterations to the theory of phase separation resulting from the reduction in dimensionality to 2D.

Despite extensive work investigating the membrane-bound phase separation of lipid domains in model membranes, it remains unclear whether such phenomena occur in intact biological cells.
Recent work, however, suggests that membrane heterogeneities result from correlations induced by the proximity of the membrane to a critical point~\cite{Shaw2021}. As such, the ability of a lipid bilayer to phase separate can affect membrane organisation, regardless of the fact that macroscopic phase separation may not occur. A detailed discussion of the role of critical phenomena in cellular membranes is provided by~\cite{Shaw2021}.

In addition to compartmentalising different membrane functionalities, lipid domains couple to various bulk processes within the cytoplasm. Binding interactions between membrane domains and cytoplasmic proteins can induce the formation of quasi-2D condensates adjacent to the membrane, which resemble a pre-wetting layer~\cite{Litschel2023,Rouches2021}. Similarly, protein condensates may influence the spatial localisation of membrane domains themselves~\cite{Wang2022a}, in turn leading to a transmembrane coupling of phase-separated protein condensates~\cite{Lee2023a}. Finally, exchange of components between membranes and the surrounding bulk environment allows enhanced regulation of phase-separated domains~\cite{Rossetto2024,Caballero2023}.

\subsection{Droplets in elastic meshes} \label{sec:ElasticMeshes}

The interior of the cell is criss-crossed by a series of elastic meshes, such as the cytoskeleton in the cytoplasm, and chromatin in the nucleus.
In order to grow larger than the characteristic pore size of these meshes, droplets must necessarily interact with the mesh, which may deform as a result.
Droplet--mesh interactions can lead to restricted droplet growth~\cite{Zhang2021b}, suppressed droplet coalescence~\cite{Zhang2021b, Qi2021}, sub-diffusive droplet motion~\cite{Lee2021}, and spatial localisation of droplets~\cite{Shin2018}. The presence of meshes may also prevent condensates from settling in very large cells in which gravitational effects are relevant~\cite{Feric2013}. In turn, the formation of droplets may affect the mechanical properties of the mesh itself~\cite{Williams2020, Style2015a}.

The formation of droplets within macroscopic meshes broadly falls into one of two categories: network-excluding droplets (\figref{fig:4_droplet_structure_interaction}E) and network-permeating droplets (\figref{fig:4_droplet_structure_interaction}F). Which of these possibilities might be expected for a given example depends on both the mesh size and the wetting properties of the droplet~\cite{Ronceray2022}. 
Whilst direct imaging of droplet--mesh interactions within the cell is currently at the forefront of experimental capabilities, recent experiments investigating droplet formation in synthetic polymer gels have provided a useful analog for elucidating the key physics involved~\cite{Style2018,Fernandez-Rico2021}.

\subsubsection{Network-excluding droplets}
\label{sec:mesh_excluding}

In many situations, droplets continue to exclude, and thus deform, the mesh network as they grow past the characteristic pore size of the mesh~\cite{Style2018}. In this case, we can develop a relatively simple theory to demonstrate the effect of elastic confinement on droplet growth by extending the effective droplet model introduced in \secref{sec:eff_model}.

The energy required to deform the network introduces an additional contribution to the free energy of the system, denoted by $F_{\textrm{el}}$. This energy corresponds to the elastic energy stored in the network due to the stretching induced by the growing droplet.
The energy difference between a droplet of volume $V$  and surface area $A$ and a corresponding homogeneous state is then approximately given by~\cite{Vidal2020,Vidal2021}
\begin{equation} \label{eqn:energy_elastic}
	\Delta F = \gamma A -\Df V+F_\mathrm{el}\;, 
\end{equation}
where $\gamma$ is surface tension, and $\Df$ is the driving force behind droplet formation; see \Eqref{eqn:energy_nucleation}.
The elastic pressure in the surrounding mesh is given by $P_{\mathrm{el}}=-\partial F_{\textrm{el}}/\partial V$, and is a function of droplet size.
The specific form of $P_{\mathrm{el}}(R)$ depends on the particular rheological response of the mesh. 
For a droplet growing in a neo-Hookean mesh with elastic modulus $E$, $P_{\mathrm{el}}$ is a monotonically increasing function of $R$ that converges to $P_{\mathrm{el}}=5E/6$ at large $R$~\cite{Vidal2020}.
Minimising \Eqref{eqn:energy_elastic}, we see that elastic stresses can stabilise droplets, with a stable droplet radius $R_*$ given by the implicit equation
\begin{equation} \label{eqn:eq_radius_elastic}
	R_* = \frac{2\gamma}{\Df + P_{\mathrm{el}}(R_*)}\;.
\end{equation}
In general, stiffer materials lead to smaller droplets than softer materials~\cite{Wei2020,Kothari2020}. 

For droplets to grow significantly larger than the mesh size, it is necessary to break individual strands of the mesh network as they reach the limit of their stretchability~\cite{Kim2020}. Introducing a cavitation pressure $P_{\mathrm{cav}}$ that quantifies the maximum stress before material breakage results in an additional energy barrier in the free energy landscape~\cite{Vidal2021}. If the driving force of phase separation is not sufficient to overcome this cavitation barrier, then the network will not break, and droplet sizes are well predicted by \Eqref{eqn:eq_radius_elastic}. On the other hand, if this barrier is overcome, droplets are able to rapidly expand beyond the mesh size, depleting the surrounding fluid. The final droplet size is then set by the amount of droplet material locally available~\cite{Vidal2021}.

To capture the dynamics of droplets embedded in elastic meshes, we describe the dilute and dense phases separately, as in \secref{sec:eff_model}. If multiple droplets exist, then the evolution of droplet sizes is given by \Eqref{eqn:many_droplets_dynamics}, but with the equilibrium volume fraction of the dilute phase, $\phi_{\textrm{out}}^{\textrm{eq}}$, now a function of $P_{\textrm{el}}$, which in turn depends on the size of the droplet. The dependence of $\phi_{\textrm{out}}^{\textrm{eq}}$ on $P_{\textrm{el}}$ can be found by considering a pressure balance between the dilute and dense phases~\cite{Vidal2021}.
In general, $P_{\textrm{el}}$ can also depend on space, as discussed in \secref{sec:complex_mesh} below. Note that the presence of the mesh restricts droplet motion, such that droplets can be assumed to be fixed in place once they have nucleated.
The evolution of the dilute phase follows \Eqref{eqn:many_droplets_dilute_field}. 

\subsubsection{Network permeation}

If the elastic network is sufficiently stiff and/or the mesh size is sufficiently large, it may be energetically favourable for droplets to permeate, rather than exclude, the network.  As droplets permeate the network, a further contribution to the free energy is induced, resulting from the differential surface energies of network--droplet and network--solvent interfaces. Following the work of Ronceray \textit{et al.}~\cite{Ronceray2022}, we capture this contribution by introducing a `wetting energy' of the form
\begin{equation}\label{eqn:energy_permeation}
F_{\textrm{wet}}\propto\frac{\gamma_\mathrm{in}-\gamma_\mathrm{out}}{\ell_\mathrm{p}}V\;,
\end{equation} 
where $\ell_\mathrm{p}$ is the characteristic pore size of the network, and we recall that $\gamma_{\mathrm{in}/\mathrm{out}}$ are the surface energies corresponding to network--droplet and network--solvent interfaces, respectively.
This wetting energy represents an averaging of capillary interactions between the droplet and individual filaments over a representative volume element of the mesh.
A simple scaling analysis shows that the transition between network permeation and network exclusion is controlled by the `permeo-elastic number' $\mathcal{P}=(\gamma_\mathrm{in}-\gamma_\mathrm{out})/(\ell_\mathrm{p}E)$, for elastic modulus~$E$.

Recent experiments in fibrillar gels have shown how network permeation can affect both the dynamics of droplet growth and the final shape of droplets~\cite{Liu2023b}. Further, dynamic transitions between network permeation and network exclusion have been observed~\cite{Liu2023b}. Such a process can be modelled by coupling the Cahn--Hilliard equation describing droplet growth with elastic deformation of the mesh via a poromechanical framework~\cite{Paulin2022,Paulin2023}. Note that network permeation still results in a slight deformation of the elastic mesh, owing to differential capillary stresses. If the network is too stiff (or phase separation is too weak) to accommodate such deformations, then droplets will be limited to the pore size of the mesh (\figref{fig:4_droplet_structure_interaction}D), and droplet growth will be suppressed~\cite{Ronceray2022,Paulin2023}.

\subsubsection{Complex properties of the mesh} \label{sec:complex_mesh}

So far, we have focussed on homogeneous meshes; in reality, meshes within cells display both small-scale and large-scale heterogeneity~\cite{Furthauer2022}.
When droplets are the same size as characteristic mesh heterogeneities, bulk elasticity is not a good representation of the mesh's mechanical response. Instead, non-local effects from neighbouring mesh filaments must be taken into account. Non-local elasticity can be included in the current framework by including an elastic contribution to the free energy given by a convolution of the elastic pressure in the mesh with a representative non-locality kernel~\cite{Qiang2024a}.
Non-local elasticity has been shown to facilitate the formation of equilibrium patterned phases during phase separation~\cite{Qiang2024a, Mannattil2024, Oudich2025, Paulin2025a}, as observed in recent experiments with synthetic polymer meshes~\cite{Fernandez-Rico2023}.

\EQref{eqn:eq_radius_elastic} highlights the dependence of the equilibrium droplet size on the elastic properties of the mesh, via the elastic pressure. Typically, droplets will be larger in softer materials. If the elastic mesh has long-range variations in its stiffness, we would thus expect larger droplets to form in softer regions of the mesh~\cite{Vidal2020,Kothari2022}. Over long timescales, it is then energetically favourable for droplets in stiffer regions to dissolve in favour of droplet growth in softer regions, in order to equilibrate chemical potential across the system. This process has been termed \emph{elastic ripening}~\cite{Rosowski2020}, and is distinct from the classical Ostwald ripening effect. In particular, it has been shown that larger droplets can dissolve at the expense of smaller droplets in appropriate stiffness gradients, in contrast to Ostwald ripening~\cite{Rosowski2020}.

Many biological meshes, such as the cytoskeleton, also display viscoelastic behaviour over sufficiently long timescales. As a result, elastic stresses can relax over time, facilitating further droplet growth, which may in turn lead to accelerated ripening~\cite{Curk2023}.
Finally, although we have focussed on passive meshes here, molecular motors may also drive active contraction of the cytoskeleton~\cite{Prost2015}, in turn deforming droplets that bind to the mesh~\cite{Tayar2023}, and potentially leading to droplet size selection~\cite{Bodinilefranc2025}. The interaction between biomolecular condensates and such active meshes remains relatively unexplored.

\sectionseparation

\section{Chemically active droplets}
\label{sec:chemical_reactions}

Cells can control the properties of biomolecules by chemical modifications, such as post-translational modifications in the case of proteins.
Such modifications can affect physical interactions with other biomolecules and thus also phase separation.
Many condensates in cells are regulated by chemical modifications~\cite{rai2018kinase, Soeding2019,Snead2019,Hofweber2019, Hondele2020}, including centrosomes~\cite{Zwicker2014} and stress granules~\cite{Hondele2019}.
To develop a physical understanding of these regulatory mechanisms, we conceptualise biomolecular modifications as chemical transitions between two forms of the same molecule, differing in their interaction parameters encoded by $w_i$ and~$\chi_{ij}$.
In this section, we discuss the complex interplay between chemical reactions and phase separation.
After introducing a general framework for describing (active) chemical reactions, we couple reactions to phase separation, identifying two distinct classes of chemically active droplets in the simplest case of a binary mixture.

\subsection{Chemical reactions of many components}

We start by considering a homogeneous mixture of $\Nc$ components $X_i$ that participate in $\Nr$ reactions~\cite{Atkin2010},
\begin{equation}
	\sum_{i = 1}^\Nc \sigmaA_{\rightarrow , i} X_i \rightleftharpoons \sum_{j = 1}^\Nc \sigmaA_{\leftarrow, j} X_j 
	\;, 
\end{equation}
where $\sigmaA_{\rightarrow , i}$ and $\sigmaA_{\leftarrow , i}$ are the respective stoichiometry coefficients of reactants and products for ${\alpha=1,\ldots,\Nr}$.
If the system is at thermodynamic equilibrium, the respective free energy is minimized, implying that chemical potentials~$\mu_i$ are balanced for each reaction,
\begin{equation}
	\label{eqn:reaction_equilibrium}
	\sum_{i=1}^\Nc \sigmaA_i \mu_i = 0
	\qquad \forall \; \alpha=1,\ldots, \Nr
	\;,
\end{equation}
where $\sigmaA_{ i} = \sigmaA_{\leftarrow , i} - \sigmaA_{\rightarrow , i}$ is the stoichiometry matrix.
Note that chemical equilibrium can always be reached in closed systems, whereas open systems, allowing particle exchange across boundaries, may be continuously driven away from equilibrium.

\begin{figure*}%
	\centering
	\includegraphics[width=\linewidth]{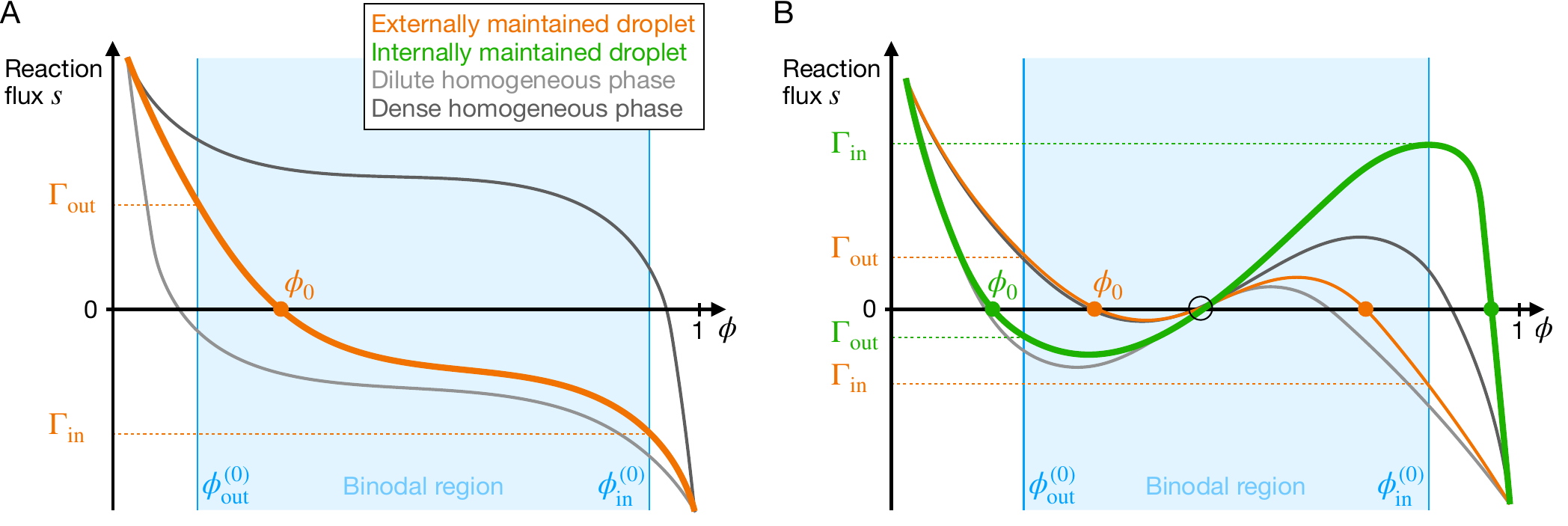}
	\caption{\textbf{Qualitatively different classes of chemically active droplets.}
	Reaction flux $s$ as a function of the volume fraction~$\phi$ of droplet material for various qualitatively different functions $s(\phi)$.
	The flux diverges for extreme fractions (equation~\ref{eqn:chemical_flux_linearized}) and vanishes for either one fraction (panel A) or multiple fractions (panel B).
	Homogeneous states with fraction~$\phi$ are stable with respect to reactions if $s'(\phi)<0$, implying that only the outer two roots are stable in panel B.
	The relative positions of these roots in comparison to the equilibrium fractions $\phiBase_\mathrm{in/out}$ delimiting the blue binodal region determines the basal reaction rates $\sBase_\mathrm{in/out}$, and thus the two distinct classes of externally maintained droplets (orange, $\sBaseIn < 0$, $\sBaseOut > 0$) and internally maintained droplets (green, $\sBaseIn > 0$, $\sBaseOut < 0$); all other cases lead to homogeneous systems (grey lines).
	}
	\label{fig:reaction_rates}
\end{figure*}

Reaction kinetics describe how the mixture composition $\Phi = (\phi_1, \ldots,  \phi_{N_c})$ changes over time.
For a homogeneous mixture, there are no diffusive fluxes, implying $\partial_t \phi_i = s_i$, where $s_i = \sum_{\alpha=1}^\Nr \sigmaA_i \sA$ is the net production of component~$i$.
The net flux $\sA$ of reaction $\alpha$ can be split into a forward and a backward direction, $\sA = \sA_\rightarrow - \sA_\leftarrow$, which must obey \emph{detailed balance of the rates}~\cite{Weber2019,JulicherAjdari1997},
\begin{equation}
	\frac{\sA_\rightarrow}{\sA_\leftarrow} = \exp \biggl[ - \frac{\sum_i \sigmaA_i \mu_i }{\kBT} \biggl] 
	\;.
\end{equation}
Detailed balance of the rates ensures that the reaction ceases (\mbox{$\sA=0$}) when chemical equilibrium as described by \Eqref{eqn:reaction_equilibrium} is reached, and also that the reaction proceeds in the forward direction when products are energetically favoured over reactants.
A simple theory that obeys these constraints is \emph{transition state theory} (\appref{sec:transition_state_theory} and \refcite{Atkin2010,Hanggi1990,Pagonabarraga1997,Bazant2013,Kirschbaum2021,Zwicker2022a}), %
\begin{equation}
	\sA  =  k_\alpha(\Phi) \sinh\biggl[
		 -\frac{\sum_i \sigmaA_i\, \mu_i }{2\kBT}
	 \biggr]
	\;,
	\label{eqn:transition_state_theory}
\end{equation}
where the second term describes thermodynamic constraints, whereas kinetic effects are captured by the positive rate coefficient~$k_\alpha$, which can depend on the composition~$\Phi$.
The functional form of $k_\alpha$ can be constrained by requiring that $\sA$ reduces to mass action kinetics for ideal solutions, implying that (\appref{sec:transition_state_theory})
\begin{equation}
	\label{eqn:mass_action_kinetics_rate}
	k_\alpha(\Phi) = \hat k_\alpha \prod_{i=1}^\Nc\phi_i^{\frac12\bigl(\sigmaA_{\rightarrow,i} + \sigmaA_{\leftarrow,i}\bigr)}
\end{equation}
with constant coefficient $\hat k_\alpha$. %
This form of $k_{\alpha}$ can also describe reaction pathways that include an enzyme~$E$. %
Equation~\eqref{eqn:mass_action_kinetics_rate} then predicts that the reaction rate is scaled by the corresponding fraction $\phi_E$ if a single enzyme participates in the reaction (${\sigmaA_{\rightarrow,E} = \sigmaA_{\leftarrow,E}=1}$).
Note, however, that biologically realistic reactions might exhibit more complex dependencies on the volume fractions.

The reaction kinetics described by \Eqref{eqn:transition_state_theory} generally drive the system towards equilibrium.
However, biological systems are active and driven away from equilibrium by their metabolism~\cite{qian2007phosphorylation}.
A theoretical representation of metabolic processes involves the conversion of high-energy fuel molecules into low-energy waste molecules via chemical reactions.
These reactions allow ordered structures within the cell to be maintained while increasing the total entropy of the cell and its environment. Note that although the energy difference between fuel and waste molecules is dissipated as heat, resulting temperature changes are typically negligible in cellular systems~\cite{Mabillard2023}.
If fuel molecules are plentiful, they essentially provide an infinite particle reservoir with fixed chemical potential, also known as a \emph{chemostat}.
If the production of waste molecules does not significantly change their concentration in the environment, they can also be considered to be chemostatted. The energy difference between chemostatted species drives chemical reactions out of equilibrium~\cite{Avanzini2021,Aslyamov2023,Avanzini2024}.
This framework can also be used to describe replenishing systems, where waste molecules (e.g., ADP) are turned back into fuel molecules (ATP) by some mechanism (ATP synthase) whose energy input is not described.
In both of these situations, the supplied energy is used locally to drive processes against their thermodynamic tendency, implying that the resulting system is an example of \emph{active matter}~\cite{Vrugt2024}.

Chemical reactions also take place in heterogeneous systems, where the composition~$\Phi$ varies in space and time.
The general dynamics of such systems are discussed in \secref{sec:basic_LLPS}, where we already introduced the source term~$s_i$ in \Eqref{eqn:kinetics_many} to account for chemical reactions.
The combination of \Eqref{eqn:kinetics_many} with \Eqref{eqn:transition_state_theory} for $s^{(\alpha)}$ implies that chemical reactions affect composition only locally, whereas spatial transport takes place via diffusion and advection.

\subsection{Chemically active binary mixtures}
\label{sec:continuous-theory}
To discuss the influence of chemical reactions on droplets, we now focus on the simplest case of a binary mixture comprising droplet material~$D$ and precursor material~$P$.
The composition of the system is then quantified by the volume fraction~$\phi(\vect r, t)$ of droplet material, while precursor material occupies the fraction~${1-\phi}$.
Thermodynamic properties of the mixture are captured by the exchange chemical potential $\bar\mu$; see \secref{sec:binary_liquids}.
If the two species can freely convert into each other, ${P \rightleftharpoons D}$, the system can in principle attain the minimum free energy (with ${\bar\mu=0}$; see \figref{fig:phase_diagrams}A) at every point in space, implying a homogeneous mixture as the only possible equilibrium state.

More interesting dynamics arise when the system is driven out of equilibrium by introducing an additional chemical reaction, $P + F \rightleftharpoons D + W$, where the fuel $F$ and waste $W$ are chemostatted.
For simplicity, we assume that fuel and waste distribute fast and are dilute, and so do not affect phase separation~\cite{Bauermann2022}.
These components therefore just provide a chemical potential difference $\Dmu = \mu_F-\mu_W$, which drives an active reaction.
The combined reaction flux $s$ then reads %
\begin{equation}
	\label{eqn:chemical_flux_linearized}
	s(\phi, \bar\mu) = \kp\sinh\!\left[\frac{-\bar\mu}{2\kBT}\right] + \ka \sinh\!\left[\frac{\Dmu - \bar\mu}{2\kBT}\right]
	\;,
\end{equation}
where $\kp$ and $\ka$ are the respective reaction coefficients for the passive and active reactions. Both $\kp$ and $\ka$ are positive and may depend on the composition~$\phi$ and the chemical potential $\bar\mu$. %

The chemical reaction flux given by \Eqref{eqn:chemical_flux_linearized} results in the creation of droplet material when the droplet fraction~$\phi$ is very small; i.e., \mbox{$s\rightarrow\infty$} in the limit \mbox{$\phi\rightarrow0$}.
This result emerges because the exchange chemical potential diverges (\mbox{$\bar\mu \rightarrow -\infty$}) as \mbox{$\phi\rightarrow0$} (see \Eqref{eqn:chemical_potential_binary} and \figref{fig:phase_diagrams}B).
Conversely, \mbox{$s\rightarrow -\infty$} as \mbox{$\phi\rightarrow1$}, implying that droplet material is converted to precursor if droplet concentration is too high.
As a result, the reaction flux~$s$ can be approximated as an explicit function of the composition $\phi$.
Recall from \Eqref{eqn:kinetics_binary} that the dynamics of the binary mixture are then governed by
\begin{equation}
	\label{eqn:kinetics_binary-sec5}
	\partial_t \phi =
		\nabla \cdot \bigl( \Lambda(\phi) \nabla \bar\mu \bigr)
		+ s(\phi)
	\;,
\end{equation}
where $\Lambda$ denotes the diffusive mobility. %
The limiting behavior of $s(\phi)$ for large and small volume fraction are constrained by the thermodynamic arguments discussed above, but the behaviour at intermediate volume fractions depends on the precise choice of parameters and in particular the composition dependence of $\kp$ and $\ka$.
\Figref{fig:reaction_rates} shows various different functional forms for $s(\phi)$, which we categorise below based on the qualitative behaviour that they induce.

\subsubsection{Stability of homogeneous states}
\label{sec:active_droplets_homogeneous_state}

\begin{figure}%
	\centering
	\includegraphics[width=.8\linewidth]{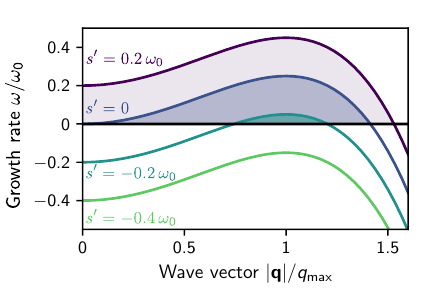}
	\caption{\textbf{Stability of homogeneous states.}
	Growth rate~$\omega$ given by \Eqref{eqn:perturbation_homogeneous_state} as a function of wavenumber~$|\vect q|$ for various reaction sensitivities $s'(\phi_0)$ with ${\omega_0=\kappa^{-1}\Lambda(\phi_0)[f''(\phi_0)]^2}$ and $q_\mathrm{max} = (-f''(\phi_0)/2\kappa)^{1/2}$.
	}
	\label{fig:stability_analysis}
\end{figure}

To build intuition, we first analyse the stability of homogeneous stationary states, which obey $\phi(\vect r) = \phi_0$ with $s(\phi_0)=0$.
Considering harmonic perturbations with wavevector~$\vect q$, the associated growth rate~$\omega$ is equal to the single entry of the Jacobian given by \Eqref{eqn:jacobian_multicomp}, 
\begin{align}
	\label{eqn:perturbation_homogeneous_state}
	\omega = - \Lambda(\phi_0)\vect{q}^2 \Bigl[ f''(\phi_0 ) + \kappa \vect{q}^2\Bigr] + s'(\phi_0)
	\;, 
\end{align}
where we have used the exchange chemical potential ${\bar\mu=\nu\delta F/\delta \phi}$ corresponding to the free energy $F$ given in \Eqref{eqn:free_energy_binary}.
The stationary state is unstable if perturbations grow, i.e., if $\omega(\vect q) > 0$ for any value of
~$\vect{q}$.
Without chemical reactions, $s'(\phi_0)=0$, this is the case if $f''(\phi_0) < 0$, corresponding to the spinodal instability described in \secref{sec:binary_liquids}.

The influence of chemical reactions on the stability of the homogeneous system is more subtle (\Figref{fig:stability_analysis}).
For autocatalytic reactions where $s'(\phi_0) > 0$, the homogeneous state is always unstable since increasing $\phi$ further accelerates its production, destabilizing the overall composition, $\omega(\vect q =0) > 0$.
In contrast, the overall composition is stable $\omega(\vect q=0)<0$ for auto-inhibitory reactions where $s'(\phi_0)< 0$.
In this case, the fastest growing mode corresponds to a length scale $\ell_\mathrm{max} = 2\pi/q_\mathrm{max} = \pi[-8 \kappa/f''(\phi_0 )]^{1/2}$ and grows at rate $\omega_\mathrm{max} =s'(\phi_0) +  \Lambda [f''(\phi_0 )]^2/4\kappa$.
Consequently, auto-inhibitory reactions only exhibit instabilities in the spinodal region where $f''(\phi_0)<0$.
However, even in the spinodal region, the homogeneous state can be completely stabilised if the reactions are sufficiently strong, $s'(\phi_0) <  k_\mathrm{stab}$, where $k_\mathrm{stab}=-\Lambda [f''(\phi_0)]^2 / 4\kappa$.
For weaker reactions, $k_\mathrm{stab} < s'(\phi_0) < 0$, only a compact band of wavelengths is unstable (\Figref{fig:stability_analysis}).
If $f''(\phi_0) < 0$, the  length scale $\ell_\mathrm{max}$ of the most unstable mode is independent of $s'(\phi_0)$, so we expect standard spinodal decomposition to initially develop in the intermediate regime.
In particular, droplets will form and Ostwald ripening (\secref{s:Dynamics of many droplets}) may drive coarsening.
However, the fact that $\omega$ is negative for long wavelengths (small $\vect q$) in this intermediate regime suggests that late stage dynamics may deviate from the classical ripening behaviour.
To investigate this possibility, we next analyse the dynamics of multiple droplets.

\subsubsection{Effective droplet model}
\label{sec:effective-droplet-model}

To understand the dynamics of chemically active droplets in more detail, we next discuss the thin-interface model introduced in \secref{sec:eff_model}.
The equilibration of the interfacial region is not affected significantly by reactions since the interfacial volume is typically small.
Instead, reactions affect the dynamical equation in the bulk phases, which after linearisation led to the diffusion \Eqref{eqn:effective_model_pde}.
Repeating this linearisation for \Eqref{eqn:kinetics_binary-sec5}, we find
\begin{equation}
	\label{eqn:reaction_diffusion}
	\partial_t \phi_n  \approx  D_n \nabla^2\phi_n + \sBase_n - k_n(\phi_n - \phiBase_n)
	\;,   
\end{equation}
both inside droplets ($n=\mathrm{in}$) and in the surrounding dilute phase ($n=\mathrm{out}$), with the familiar diffusivities ${D_n\approx \nu \Lambda (\phiBase_n) f''(\phiBase_n)}$.
Here, we also define the basal reaction flux~$\sBase_n=s(\phiBase_n)$, which quantifies the production of droplet material at the equilibrium fraction $\phiBase_n$.
We also introduce reaction rates~${k_n=-s'(\phiBase_n)}$, which specify how the production of droplet material changes when concentrations deviate from equilibrium.

The reaction--diffusion \Eqref{eqn:reaction_diffusion} is linear in $\phi$ and can be solved analytically for spherical droplets of radius~$R$~\cite{Kulkarni2023}.
The resulting concentration profiles display spatial variations over characteristic length scales given by the reaction--diffusion lengths $\LRD_n = \sqrt{D_n/k_n}$, which are defined for $k_n>0$ (stabilising reactions).
In general, these variations drive diffusive fluxes $\vect j_n$ on both sides of the interface, which affect the droplet growth speed given by \Eqref{eqn:drop_interface_speed}.
In the limit of small droplets, $R \ll \LRDin$, the diffusive flux across the interface is balanced by the reaction in the entire droplet volume, resulting in the approximate growth rate~\cite{Zwicker2024}
\begin{equation}
	\difffrac{R}{t}  \approx \frac{\DOut }{R(\phiBaseIn - \phiBaseOut)} \biggl( \phi_\infty - \phiEqOut(R) + \frac{R^2\sBaseIn}{3 \DOut}\biggl)
	\;,
	\label{eqn:single_active_drop_growth}
\end{equation}
where $\phi_\infty=\phiBaseOut + \sBaseOut/\kOut$ is the volume fraction far away from the droplet and the equilibrium fraction~$\phiEqOut$ is given by \Eqref{eqn:ceqout_laplace}. %
Equation~\eqref{eqn:single_active_drop_growth} differs in two crucial aspects from \Eqref{eqn:single_drop_growth} describing passive droplets.
First, the fraction $\phi_\infty$ is now a constant value controlled by the reactions in the dilute phase, in contrast to the passive case where it was determined by material conservation and depended on the size of all droplets.
Indeed, the dynamics of multiple droplets are independently described by \Eqref{eqn:single_active_drop_growth}, as long as the droplet separation is large enough compared to $\LRDout$ such that they do not disturb their corresponding concentration profiles.
Second, \Eqref{eqn:single_active_drop_growth} contains a term related to the reactions inside the droplet, which can dominate growth over the surface tension effects captured by the other terms in this equation.

\subsubsection{Two classes of active droplets}
Equation \eqref{eqn:single_active_drop_growth} suggests that the basal fluxes $\sBase_n$ can affect the dynamics of chemically active droplets significantly.
This is not surprising since these fluxes determine whether droplet material is produced ($\sBase_n > 0$) or degraded ($\sBase_n < 0$) inside the droplet ($n=\mathrm{in}$) and in its surrounding ($n=\mathrm{out}$).
If droplet material is degraded (produced) everywhere, the system will end up in a dilute (dense) homogeneous state (see grey lines in \figref{fig:reaction_rates}).
To have stable droplets, material thus needs to be produced in one phase while it is degraded in the other, leaving two distinct classes of active droplets: externally maintained droplets and internally maintained droplets.
\emph{Externally maintained droplets} degrade droplet material inside the droplet ($\sBaseIn < 0$), while it is produced outside the droplet ($\sBaseOut > 0$).
A simple example of a possible reaction rate for this case is shown in \figref{fig:reaction_rates}A.
On the other hand, \emph{internally maintained droplets} produce droplet material inside the droplet ($\sBaseIn > 0$), while it is degraded outside the droplet ($\sBaseOut<0$).
The green line in \figref{fig:reaction_rates}B shows that in this case there are stable states outside the blue binodal region, implying that a dilute and a dense homogeneous state can be stable.
If the two stable points are instead inside the binodal region, we obtain a more complex version of externally maintained droplets (orange line), whereas just one homogeneous state survives if exactly one stable point is inside the binodal region (grey lines).
Taken together, this qualitative analysis suggests that there are two classes of chemically active droplets, which have very different behaviours as summarised in Table~\ref{tab:active_droplets}, and which we discuss separately in the following sections.

\newcommand{\specialcell}[2][c]{%
  \renewcommand{\arraystretch}{1}
  \begin{tabular}[#1]{@{}l@{}}#2\end{tabular}}
  \renewcommand{\arraystretch}{1.4}
\begin{table*}

\caption{Summary of the two classes of chemically active droplets. The supersaturation is defined as $\supSat = \sBaseOut/(\kOut\phiBaseOut)$.
\\}
{\renewcommand{\arraystretch}{1.4} %
\begin{tabularx}{\textwidth}{lXX}   %
\toprule
Property & Externally maintained droplets &Internally maintained droplets \\
\midrule
Basal fluxes & Production outside: $\sBaseOut > 0 , \sBaseIn < 0$ & Production inside: $\sBaseOut < 0 , \sBaseIn >  0$ \\
Dilute phase & Supersaturated: $\phi_\infty > \phiBaseOut$, $\supSat > 0$   & Undersaturated: $\phi_\infty < \phiBaseOut$, $\supSat < 0$ \\[3pt]
Critical radius  &
	$\Rcrit \approx \dfrac{\capLen}{\supSat}$  &
	$\Rcrit \approx \biggl|\dfrac{3 \DOut \supSat \phiBaseOut}{\sBaseIn} \biggr|^\frac12 - \dfrac{\capLen}{2\supSat}$  \\[8pt]
Nucleation     &Suppressed by reactions~\cite{Ziethen2023}  & Deterministic by active cores~\cite{Zwicker2015} \\
Collective growth & Suppressed Ostwald ripening~\cite{Zwicker2015} & Accelerated Ostwald ripening~\cite{Tena-solsona2019}   \\[3pt]
Final size &
	$R_* \approx \biggl|\dfrac{3 \DOut \supSat \phiBaseOut}{ \sBaseIn}\biggr|^\frac12 -  \dfrac{\capLen}{2\supSat}$  &
	Limited by available precursor \\[8pt]
Additional features &
	\specialcell[t]{%
		Droplet division~\cite{Zwicker2017} \\
		Active shells~\cite{Bauermann2023} \\
		Self-organised drift~\cite{Demarchi2023}
	}  & \specialcell[t]{
	Stabilisation by active cores~\cite{Zwicker2014} \\
	Centring of active cores~\cite{Zwicker2018b}
	} \\
\bottomrule
\end{tabularx}}
\label{tab:active_droplets}
\end{table*}

\subsection{Externally maintained droplets}
\label{sec:externally-Maintained}

\begin{figure}%
	\centering
	\includegraphics[width=\linewidth]{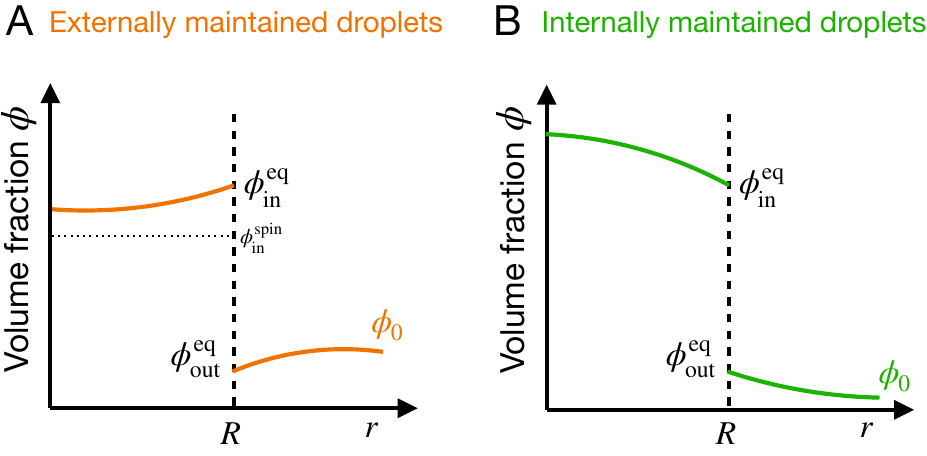}
	\caption{ Volume fraction $\phi$ of droplet material as a function of the radial distance $r$ from the droplet centre. The volume fractions are controlled by phase equilibrium at the droplet interface $r = R$ which causes gradients inside and outside the droplet. For the two cases of externally and internally maintained droplets, these gradients are opposite in sign.} 
	\label{fig:concentration_profiles}
\end{figure}

Externally maintained droplets form when the volume fraction of the dilute phase away from droplets is maintained at a volume fraction $\phi_\infty>\phiBaseOut$, where we recall that $\phiBaseOut$ is the volume fraction imposed at the interface by phase equilibrium.
The associated supersaturation $\supSat=\sBaseOut/(\kOut\phiBaseOut)$ results from a net production of droplet material in the dilute phase ($\sBaseOut>0$), which is balanced by degradation of droplet material within the dense phase (${\sBaseIn < 0}$).
\Figref{fig:concentration_profiles}A demonstrates that these opposing reactions cause diffusive fluxes across the interface, transporting droplet material into the droplet in exchange for precursor material.
This transfer of material governs the dynamic behaviour described by \Eqref{eqn:single_active_drop_growth}. For externally maintained droplets, this equation has two stationary solutions given by,
\begin{subequations}
\begin{align}
	\Rcrit  &\approx  \frac{\capLen}{\supSat} - \frac{\sBaseIn\capLen^3}{3\DOut\phiBaseOut \supSat^4}
	\;,
	\label{eqn:ext_droplets_radius1}
\\[5pt]
	R_* &\approx - \frac{\capLen}{2\supSat} +  \biggl( \frac{3 \DOut \supSat \phiBaseOut}{- \sBaseIn} \biggl)^\frac12 
	\;.   
	\label{eqn:ext_droplets_radius2}
\end{align}
\end{subequations}
Linear stability analysis of \Eqref{eqn:single_active_drop_growth} reveals that the first of these solutions (equation~\ref{eqn:ext_droplets_radius1}) is unstable and thus corresponds to the critical radius that must be exceeded for droplet nucleation to occur.
The second term in \Eqref{eqn:ext_droplets_radius1} indicates that the critical radius increases for stronger reactions (more negative $\sBaseIn$) even when the supersaturation~$\supSat$ is kept fixed.
We analyse in detail the suppression of droplet nucleation by reactions in section~\ref{s:SuppresedNucleation} below.
The second stationary solution (equation~\ref{eqn:ext_droplets_radius2}) is stable, implying that externally maintained droplets prefer to attain a fixed finite size.

\subsubsection{Droplet size control}
\label{sec:active_droplets_size_control}

The expression for the stable stationary radius~$R_*$ derived above relies on the assumption that a single active droplet is surrounded by a large dilute phase at an average fraction~$\phi_0$.
In this case, droplets that are larger than $R_*$ will shrink, since they lose more droplet material via internal degradation reactions than they gain via the diffusive influx from the dilute phase.
Equation~\eqref{eqn:ext_droplets_radius2} shows that this size control ceases in passive systems since $R_* \rightarrow \infty$ for $\sBaseIn\rightarrow0$.
The fact that isolated droplets attain a fixed size suggests that droplet coarsening via Ostwald ripening is suppressed.
A more detailed calculation shows that multiple externally maintained droplets ($\sBaseIn<0$) can stably coexist if they are larger than the threshold radius~\cite{Zwicker2015}
\begin{align}
	R_{\rm stab} = \biggl( \frac{3 \DOut \capLen \phiBaseOut}{-2 \sBaseIn} \biggl)^\frac13
	\;.
\end{align}
This result, together with numerical simulations~\cite{Zwicker2015}, analytical arguments \cite{Wurtz2018}, and recent experiments~\cite{Sastre2024} show that externally maintained droplets can suppress Ostwald ripening and tend to form regularly patterned states with a well-defined length scale.

Whether multiple externally maintained droplets coexist or a single droplet dominates the system not only depends on physical parameters such as reaction rates, but also on initial conditions~\cite{Zwicker2015, bauermann2024}. %
Since a range of droplet sizes can be stable, the precise initialisation protocol determines the final state.
In particular, starting with a single small droplet, there exists two qualitatively different regimes~\cite{bauermann2024}:
either the droplet grows until it reaches a fixed radius comparable to $R_*$, or it grows extensively, occupying a fixed fraction of the system. %
In the latter case, the droplet radius eventually exceeds the reaction-diffusion length $\LRDin$, so that the approximation leading to \Eqref{eqn:single_active_drop_growth} breaks down and the solution given in \Eqref{eqn:ext_droplets_radius2} no longer applies.
Essentially, these large active droplets have chemical fluxes restricted to a region of size $\LRD_n$ around their interfaces, permitting extensive growth overall. 
Taken together, these arguments imply that externally maintained droplets can attain multiple stable states, although the details of this situation are currently not well understood. %

\subsubsection{Electrostatic analogy} 
\label{sec:electro_static_analogy}
Some behaviours of externally maintained droplets can be discussed intuitively in the special case of linearised reactions, ${s(\phi) = k(\phi_0-\phi)}$, where $k$ determines the reaction rate, and $\phi_0$ denotes the fraction in the dilute phase. %
In this case, the dynamics of the entire system obey the Cahn--Hilliard--Onoo equation~\cite{Oono1988}
\begin{equation}
	\label{eqn:CH_Onoo}
	\partial_t \phi = \Lambda \nabla^2 \bar\mu + k(\phi_0 - \phi)
	\;,
\end{equation}
where $\Lambda$ is a constant mobility.
Phase-separated structures emerging from these dynamics generically show size selection~\cite{Glotzer1994,Glotzer1994a,Christensen1996,Muratov2002}.

\begin{figure}
	\centering
	\includegraphics[width=1.0\linewidth]{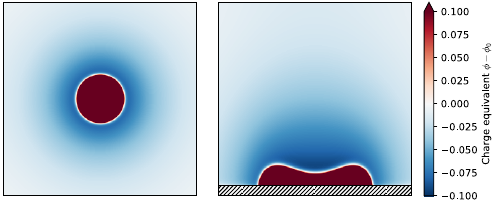}
	\caption{\textbf{Electrostatic analogy of externally maintained droplets}.
	Density plot of the field $\phi(\vect r)-\phi_0$, which plays the role of charge density in the electrostatic analogy discussed in \secref{sec:electro_static_analogy}.
	The left panel displays a single droplet in a large system, whereas the system on the right includes a neutral wall at the bottom.
	}
	\label{fig-Ch5electrostatics}
\end{figure}

Although \Eqref{eqn:CH_Onoo} describes an active system, it can also be written in the form of gradient dynamics, $\partial_t \phi = \Lambda\nabla^2 \delta \tilde{F}/\delta \phi$, with a modified free energy
\begin{equation}
	\tilde F[\Phi] = F[\Phi] + \frac12 \int (\phi- \phi_0) \pot \, \diff V
	\;, 
\end{equation}
where the potential $\pot(\vect r, t)$ obeys the Poisson equation
\begin{align}
	\nabla^2 \pot &= -\frac{k}{\Lambda}(\phi - \phi_0)
\end{align}
with appropriate boundary conditions~\cite{Liese2023}.
In analogy to the electrostatic interactions discussed in \secref{sec:complex_molecules}, $\pot$ can thus be interpreted as the electrostatic potential, $\phi(\vect r)-\phi_0$ plays the role of the charge density~$\varrho(\vect r)$, and $\Lambda/k$ is equivalent to the electric permittivity~$\varepsilon$.
\Figref{fig-Ch5electrostatics} shows how the effect of the chemical reactions can thus be interpreted as a charge cloud surrounding oppositely charged droplets.
In general, large droplets will accrue a large amount of charge, increasing the energy corresponding to the equivalent electrostatic interaction, such that further growth is halted.

A major strength of this electrostatic analogy is that states that minimise $\tilde F$ are necessarily stationary states of \Eqref{eqn:CH_Onoo}, and so we can use equilibrium thermodynamics to analyse their behaviour.
This method has previously been successful in analysing length scale selection \cite{Glotzer1994a,Christensen1996,Muratov2002,Liese2023,Ziethen2024}, in analogy with the seminal Ohta--Kawasaki model \cite{Ohta1986,Liu1989} and other pattern-forming models based on long-range repulsion~\cite{Seul1995}.
In general, patterns tend to emerge in systems that combine short-range attraction by phase separation with long-range repulsion due to chemical reactions~\cite{Huberman1976,Puri1994, Glotzer1995, Carati1997, Tran-Cong1996,Motoyama1996,Motoyama1997,Toxvaerd1996, TranCongMiyata2017,Tran-Cong-Miyata2011}.

\subsubsection{Suppressed nucleation}
\label{s:SuppresedNucleation}

The electrostatic analogy introduced above also explains why droplet nucleation is suppressed by reactions~\cite{Ziethen2023}.
Nucleation kinetics in this system are well described by the classical nucleation theory described in \secref{section:Classical-nucleation-theory}~\cite{Cho2023a}.
Following the electrostatic analogy, the energy penalty~$\DF$ associated with forming a small nucleus of radius~$R$ approximately reads~\cite{Ziethen2024}
\begin{align}
	\DF &\approx  \gamma A - \Df V +
		\frac{4\pi k(\phiBaseIn - \phi_0)^2}{15\Lambda} R^5
	\;,
\end{align}
where the first two terms are equivalent to the passive case given by \Eqref{eqn:energy_nucleation} and the last term captures the influence of reactions in the limit $\LRDin,\LRDout \gg R$.
If droplet formation is favourable ($\Df > 0$) and reactions are weak, $\DF$ exhibits a maximum at $R\approx \Rcrit$ given by \Eqref{eqn:ext_droplets_radius1}, which reads
\begin{align}
	\Fnuc &\approx	\frac{16\pi \gamma^3}{3\Df^2} \left[ 1+ 
		\frac{8\gamma^2k(\phiBaseIn - \phi_0)^2}{5\Lambda \Df^2}
	\right]
\end{align}
to linear order in $k$.
The pre-factor here is identical to the passive energy barrier given by \Eqref{eqn:nucl_radius}.
Reactions thus increase $\Fnuc$, implying a smaller nucleation rate~$J_\mathrm{nucl} \propto \exp(-\Fnuc/\kBT)$; see \Eqref{eqn:nucleation_rate}.
Since chemical reactions can be interpreted as a separation of charge, this suppression of nucleation follows from the fact that forming a charged nucleus is energetically unfavourable compared to forming a neutral nucleus.
This process can also be understood from the fact that chemical reactions want to maintain a fixed concentration locally and thus dampen fluctuations originating from diffusive fluxes, making the formation of a critical nucleus less likely.

Chemical reactions also affect heterogeneous droplet nucleation  at surfaces~\cite{Ziethen2024}. %
In \secref{sec:het_nucl}, we discuss how surfaces promote nucleation, and this behaviour is retained when reactions describing externally maintained droplets are added~\cite{Ziethen2024}.
However, sufficiently large reaction rates will still suppress nucleation, as for the homogeneous case.
Further, reactions may cause deformation of sessile droplets at surfaces (\figref{fig-Ch5electrostatics}B), as discussed in more detail in \secref{s:ChemActiveWetting} below.
The interactions of droplets with both surfaces and active reactions thus provide flexible control over droplet nucleation.

\subsubsection{Spontaneous division}
Externally maintained \\ droplets grow by taking up material from the surrounding supersaturated dilute phase.
If droplets are slightly deformed from a spherical state, e.g., due to thermal fluctuations, bumps on their surface have larger curvature and thus receive a higher material influx. %
This influx then amplifies these bumps, leading to an instability similar to the \emph{Mullins-Sekerka instability} of growing fronts~\cite{Mullins1963}.
Additionally, externally maintained droplets lose material within their bulk, which reduces their total volume and causes a self-constriction in conjunction with the uneven material influx. As a result, externally maintained droplets can divide spontaneously~\cite{Zwicker2017}.
This shape instability can set in for droplets smaller than the stationary radius~$R_*$ of isolated droplets predicted by \Eqref{eqn:ext_droplets_radius2}.
Starting with an otherwise empty, and sufficiently large, system, a single droplet will thus grow and split multiple times until the entire system is filled with droplets.
These duplications cease once droplets are so dense that they compete for material in the dilute phase.
This competition lowers the effective supersaturation $\supSat$ and thus $R_*$, until the system finally settles into a stationary state with many droplets~\cite{Zwicker2017}.
Spontaneous droplet division can also be explained using the electrostatic analogy since charged droplets split in a similar fashion due to the \emph{Rayleigh instability}~\cite{Rayleigh1882, Golestanian2017}.

\subsection{Internally maintained droplets}
\label{sec:internally-Maintained}

Internally maintained droplets essentially have the opposite behaviour to their externally maintained counterparts; see Table~\ref{tab:active_droplets}.
In particular, internally maintained droplets create droplet material inside ($\sBaseIn > 0$) while it is degraded outside ($\sBaseOut < 0$), implying an under-saturated dilute phase ($\supSat<0$).
Droplet growth is still described by \Eqref{eqn:single_active_drop_growth}, but this equation now only has a single stationary solution,
\begin{equation}
	\label{Rcrit-internally}
	\Rcrit \approx \biggl( \frac{3 \DOut \supSat \phiBaseOut}{- \sBaseIn} \biggl)^\frac12 - \frac{\capLen}{2\supSat}
	\;,
\end{equation}
which is unstable.
This critical radius is typically much larger than that of passive droplets (equation \ref{eqn:nucl_radius}), implying that spontaneous nucleation of internally maintained droplets is strongly suppressed.

Internally maintained droplets of radius~$R$ exhibit an efflux of droplet material $J \approx 4 \pi D R \left(\phiEqIn- \phi_\infty\right)$ (\figref{fig:concentration_profiles}B), whose functional form follows from solving \Eqref{eqn:reaction_diffusion} in the stationary state.
Here, $\phiEqIn = \phiBaseIn(1 + \capLen/R)$ is the equilibrium fraction inside the droplet and \mbox{$\phi_\infty = \phiBaseOut + \sBaseOut/\kOut = \phiBaseOut(1+\supSat)$} denotes the volume fraction in the dilute phase.
To nucleate a droplet, the efflux~$J$ needs to be compensated for by local production of droplet material, which could originate from localised enzymes, e.g., at the surface of a catalytically active core of radius~$a$.
Since the integrated reaction flux~$Q$ at such a core needs to exceed $J$, there is a minimal flux~$Q_\mathrm{min}$ beyond which an internally maintained droplet is nucleated.
For weak surface tension effects ($\capLen\ll a$), we find $Q_\mathrm{min} \approx 4\pi D a (\phiBaseIn - \phi_\infty)$, whereas strong surface tension ($\capLen \gg a$) implies $Q_\mathrm{min} \approx 4\pi D \capLen \phiBaseOut$.
In both cases, nucleation of internally maintained droplets around the core is ensured, while spontaneous nucleation anywhere else in the system is virtually impossible.

If internally maintained droplets exceed the radius $\Rcrit$ given by \Eqref{Rcrit-internally}, they grow even in the absence of a catalytic core, and their final size is only limited by the system size.
Since larger droplets produce more droplet material, an emulsion of multiple internally maintained droplets is unstable, and coarsening is faster than that exhibited in passive Ostwald ripening~\cite{Zwicker2015,Tena-solsona2019}.
Interestingly, however, sufficiently strong catalytically active cores can suppress Ostwald ripening, such that final droplet volumes approach values proportional to the respective reaction fluxes~$Q$~\cite{Zwicker2014}.
The diffusive fluxes created by catalytically active cores also automatically centre these cores within droplets~\cite{Zwicker2018b}.
Taken together, these properties allow exquisite control over the positioning of internally maintained droplets~\cite{Soeding2019}.

\subsection{More complex chemically active droplets}

In this section, we have presented a continuous theory of binary chemically active droplets, which neglects many molecular details of the biomolecules involved.
Including such details, even for simple effects such as molecular crowding~\cite{Fries2025a}, can lead to additional complex behaviour~\cite{Zippo2025}.
Similarly, when many interacting species are involved, we expect to find further interesting effects.
In the future, it will thus be important to study thermodynamically consistent models of chemically active multicomponent mixtures, as well as how these mixtures interact with their environment. 
Since this topic represents a relatively new field of research, we highlight only a few recent discoveries in this direction below.

\subsubsection{Chemically active wetting} %
\label{s:ChemActiveWetting}

Chemically active droplets typically do not exist in isolation but interact with their surrounding.
While such interactions naturally lead to the wetting phenomena discussed in \secref{sec:complex_environment} and \secref{s:SuppresedNucleation}, the additional fluxes created by chemical reactions can also distort the droplet%
; see \figref{fig-Ch5electrostatics} for an example. 
The electrostatic analogy introduced in \secref{sec:electro_static_analogy} explains this deformation for externally maintained droplets since the overall energy is lowered by spatially separating parts of positively charged droplets~\cite{Ziethen2024}.
Deformed droplets can also emerge when the wetted structure enriches catalytic enzymes, so that reactions themselves become localised, which further distorts the diffusive fluxes~\cite{Liese2023} and can position the droplet relative to the wetted structure~\cite{Zwicker2018b}.
Complex interactions between wetted structures and reactions thus allow for an intricate control over when and where condensates appear~\cite{Case2019a}.

\subsubsection{Self-organized drift}
\label{sec:active_droplet_drift}

The combination of chemical turnover and diffusion generically creates concentration gradients, which can move particles and droplets by diffusiophoresis~\cite{Goychuk2024,Demarchi2023,Hafner2023a, Shim2022, Golestanian2009}.
Concentration gradients also develop around chemically active droplets (\figref{fig:concentration_profiles}), implying that they can influence the positioning of other droplets.
For instance, externally maintained droplets naturally repel each other, which stabilises their hexagonal stationary pattern.
Their repulsion follows directly from the electrostatic analogy discussed in \secref{sec:electro_static_analogy}.

Chemical gradients also induce surface tension gradients, which can drive movement via Marangoni flows, as discussed in \secref{sec:internal_complex_interfaces}.
While droplets generically react to external gradients, there is also the exciting possibility that they may create their own gradients.
Moreover, they may exploit an instability where movement amplifies this gradient, so that the entire system settles into a moving state, either via diffusiophoresis~\cite{Demarchi2023} or Marangoni flows~\cite{Maass2016, Michelin2023}.
Alternatively, chemical reactions can create an asymmetric state, which then moves due to isotropies in surrounding gradients~\cite{Hafner2023a}, and even induce active capillary waves~\cite{Rasshofer2025}.
In all these cases, activity needs to exceed a threshold, above which isotropy is spontaneously broken and swimming begins.
In contrast, isotropy can also be broken in multiphasic systems, which can exhibit self-propulsion at arbitrarily low activity~\cite{Qiang2025a}.

\subsubsection{Active multi-component mixtures}

Many of the complex behaviours of chemically active droplets rely on multiple interacting components.
The general theory of active multi-component mixtures~\cite{Julicher2024, Zwicker2022a} is obtained by inserting the reaction flux ${s_i=\sum_{\alpha=1}^\Nr \sigmaA_i \sA}$ of species~$i$ into \Eqref{eqn:kinetics_many},
\begin{equation}
	\label{eqn:kinetics_many_reactions}
	\partial_t \phi_i %
	=
		\nabla \cdot \biggl(\! \size_i\!\sum_{j=1}^{\Nc}\Lambda_{ij} \nabla \bar\mu_j \!\biggr)
		+ \sum_{\alpha=1}^\Nr \sigmaA_i \sA
	\;,
\end{equation}
where $\sA$  denotes the flux of reaction~$\alpha$, e.g., given by \Eqref{eqn:transition_state_theory}, and we neglect hydrodynamic effects and thermal noise for simplicity.
These equations capture how chemical reactions affect droplets and how droplets affect chemical reactions~\cite{Laha2024}.

Active multi-component mixtures can exhibit various instabilities and patterns, ranging from simple Turing patterns~\cite{Menou2023, Aslyamov2023, Avanzini2024, Carati1997} to systems with multiple length scales~\cite{Tran-Cong1999a}, to more complex spatiotemporal behaviour~\cite{Luo2023,Coupe2024,Haugerud2025}.
Particularly interesting is the case where phase separation interferes with chemical reactions~\cite{Miangolarra2022,Sang2022}, so that behaviours beyond those described above can emerge.
For instance, droplets may now be stable even in the presence of passive reactions, which is impossible in binary systems~\cite{Kirschbaum2021}.
In this case, internally maintained droplets do not fill the entire system, although their size typically still scales with system size~\cite{Zwicker2015, Zwicker2014}.
Externally maintained droplets, however, can either scale with system size or exhibit the size control discussed in \secref{sec:externally-Maintained}~\cite{bauermann2024,Bauermann2025}.
In both of these cases, the droplet size distribution relaxes to a monodisperse (and universal) distribution at late times~\cite{Bauermann2025}.
Externally maintained droplets can also assume shell-like structures~\cite{Bauermann2023,Bergmann2023}, essentially because chemical reactions deplete the inside of the droplet so that concentration (\figref{fig:concentration_profiles}A) drops below the spinodal concentration and a bubble emerges.
Chemically active droplets may be further controlled by enzyme positioning, either because they are mobile~\cite{Fries2024} or they are fixed in particular locations~\cite{Banani2024}. Recent work has also shown that enzymes themselves may form phase-separated clusters with a characteristic equilibrium size~\cite{MartinezCalvo2024}, similar to the size control exhibited by externally maintained droplets discussed in~\secref{sec:active_droplets_size_control}.
As this field becomes more developed, there will surely be many additional phenomena that we are yet to unveil.

\sectionseparation

\section{Summary}

Physical interactions between biomolecules lead to the spontaneous emergence of phase-separated droplets within cells. The formation, growth, and dissolution of these \emph{biomolecular condensates} depends on a delicate balance between entropic and enthalpic effects.
Condensates are utilised by cells to structure their interior, allowing them to concentrate molecules, mediate forces, and control reactions.
To achieve their desired function, cells must maintain exquisite control over droplet size, location, and timing.
In this review, we have discussed in depth our current knowledge of physical mechanisms that mediate the interplay between droplets and their surrounding environment, and how cells utilise these mechanisms to regulate droplet behaviour.

\subsection{Life cycle of droplets in biological cells}
The different physical processes discussed in this review provide the cell with intricate control over the entire life cycle of droplets, as outlined below.

\paragraph{Birth by nucleation:}
In many situations, cells need to control when and where droplets form.
Such control is difficult to achieve via classical spontaneous nucleation (\secref{section:Classical-nucleation-theory}).
Although heterogeneous nucleation sites can bias droplet position (\secref{sec:het_nucl}), true spatial control might require active catalytic sites (\secref{sec:internally-Maintained}), which drive localised droplet formation in an undersaturated system.
The timing of droplet formation can be controlled by actively driven bulk reactions, which can suppress phase separation (\secref{sec:active_droplets_homogeneous_state}).
This process requires a continuous expenditure of fuel such as ATP, but has the advantage that droplets form immediately when fuel ceases, thus providing a robust stress response~\cite{wurtz2018stress}.

\paragraph{Growth by material uptake:}
Droplets generally grow by taking up material from their surrounding if they are not otherwise restricted.
Growth thus naturally stops once the surrounding environment is no longer supersaturated.
However, growth may be limited if droplets are charged (\secref{sec:charged_droplets}), the surrounding provides an elastic resistance (\secref{sec:mesh_excluding}), or chemical reactions restrict further growth (\secref{sec:active_droplets_size_control}).
Interestingly, all three of these mechanisms can be conceptualised as long-ranged repulsion due to electrostatics, elastic deformation, and reaction-diffusion systems, respectively.
These long-ranged repulsive forces oppose the short-ranged attraction due to phase separation, resulting in a preferentially selected droplet size~\cite{Seul1995}.

\paragraph{Coarsening by competition:}
Droplets compete with each other for material. 
Typically, larger droplets grow at the expense of smaller droplets due to their smaller Laplace pressure (\secref{s:Dynamics of many droplets}).
Such competition allows cells to randomly select a particular droplet from a collection of initial droplets, which is for instance crucial in selecting crossover sites during meiosis~\cite{Morgan2021, Zhang2021c, Durand2022, Girard2023, Ernst2024}.
Despite this, competition between droplets can also be inhibited by the size-selection mechanisms discussed in the previous paragraph, providing cells with a complex machinery for controlling the size distribution of multiple droplets.

\paragraph{Positioning by heterogeneities:}
Droplets need to be in the right place to fulfil their function.
To control position, cells can use wetting (\secref{sec:complex_environment}) to anchor droplets at particular structures, such as the plasma membrane or chromosome loci, so that they remain in place.
If droplets are not anchored, cells can reposition them, e.g., using external gradients (\secref{sec:spatial_gradients}), hydrodynamic fluxes (\secref{sec:ch3_internal_stresses}), or active transport~\cite{Chauhan2025}.
Even more exciting are droplets that swim themselves, either because they self-organise chemical gradients (\secref{sec:active_droplet_drift}) or they exert active stresses~\cite{Cates2024, Marchetti2013, Ramaswamy2010}.

\paragraph{Death by dissolution:}
Droplets need to be dissolved when they are no longer needed~\cite{Buchan2024}. %
Cells can achieve dissolution by breaking down the involved biomolecules into their basic constituents to lower concentrations below saturation, so that droplets spontaneously dissolve.
This process is costly, time-consuming, and difficult to revert, but it removes droplets permanently.
Alternatively, active reactions can be used to modify physical properties of the involved molecules, thus achieving faster, cheaper, and more flexible dissolution, which might only be temporary.

\subsection{Droplet--environment interactions}

Droplets react sensitively to their environment.
Changes in external parameters, such as temperature, pH, and cellular composition, affect whether droplets form and how large they can grow (\secref{sec:global_parameters}).
This ability to sense environmental cues allows droplets to react to changes in the internal states of cells, such as stress~\cite{Franzmann2018a}, which can then trigger appropriate responses.
In addition, the first-order phase transition associated with phase separation implies hysteresis: droplets will not dissolve easily once formed, but also will not re-form easily once dissolved.
This switch-like behaviour limits how much disruption can be caused by noise, allowing cells to make committed decisions~\cite{Dine2018,Badia2021,Zhang2024c}.

Droplets also contribute to the manipulation of the cellular environment.
First and foremost, droplets are dynamic membrane-less compartments providing spatial organisation.
They can sequester material from the rest of the cell while also concentrating components, facilitating controlled chemical reactions.
Moreover, droplets can exert capillary forces on cellular structures (\secref{sec:condensates_at_membranes}), potentially causing significant deformations.
Droplets can even perform mundane tasks such as plugging leaks~\cite{Bussi2023}.
Taken together, droplets can thus directly affect chemical and mechanical processes in cells.

The interplay of droplets with each other and their environment leads to complex interaction networks, which can process information in cells and enable communication between droplets~\cite{Ji2025,Ye2025}.
Such interaction networks typically involve multiple feedback loops, which give rise to additional dynamic behaviours~\cite{Modi2023}.
At an abstract level, droplets might thus play an integral part in information processing and learning in biological cells~\cite{Teixeira2023, Zentner2025, Murugan2025, Hilbert2025}.

\subsection{Outlook}

The study of droplets inside biological cells is a relatively new field of research with many open questions~\cite{Spruijt2023}.
One major challenge is that cellular environments are incredibly more complex than those which can be studied in the lab:
biomolecular droplets display a myriad of complex compositional and material properties (\secref{sec:internal_complexity}), and live in a continuously evolving, heterogeneous environment (\secref{sec:complex_environment}).
Moreover, a hallmark of life is its non-equilibrium nature, which sets it apart from traditional synthetic matter (\secref{sec:chemical_reactions}).

Beyond exploring how droplets enable cellular function, understanding the physical principles that biology exploits so effectively has application beyond cell biology. %
For example, it has been hypothesised that phase separation could be implicated in the origin of life by enabling the formation of basic protocells~\cite{Zwicker2017}.
Indeed, droplets provide well-defined entities that grow, move, and divide by metabolising material from their environment~\cite{Zwicker2023,Wenisch2025}.
On the more theoretical side, chemically active droplets comprise an exciting class of non-equilibrium system~\cite{Cates2024}, with similarities to proliferating active matter~\cite{Hallatschek2023} and general dissipative structures~\cite{Goldbeter2018}.
Finally, understanding droplet regulation in biological cells will allow us to leverage similar principles in designing novel synthetic systems~\cite{Cao2024, Harrington2023}.

\ack
We would like to thank Jan Kirschbaum and Estefania Vidal-Henriquez for early input and drafting initial concepts for the review.
We thank Kristian Blom, Rumiana Dimova, Marcel Ernst, Stefan K\"{o}stler, Guido Kusters, Chengjie Luo, Yicheng Qiang, Riccardo Rossetto, Henri Schmidt, Filipe Thewes, Gerrit Wellecke, Ned Wingreen, Mengmeng Wu, and Noah Ziethen for helpful comments and discussions of the manuscript.
We further thank the entire Zwicker group for invaluable scientific  discussions.
We gratefully acknowledge funding from the Max Planck Society and the European Union (ERC, EmulSim, 101044662).

\sectionseparation

\phantomsection
\section*{Appendices}
\addcontentsline{toc}{section}{Appendices}
\setcounter{subsection}{0}

\renewcommand{\thesubsection}{\Alph{subsection}}
\numberwithin{equation}{subsection}
\renewcommand{\theequation}{\thesubsection.\arabic{equation}}

\subsection{Compressible multi-component fluid}
\label{sec:appendix_compressible}

We consider a compressible fluid consisting of a solvent and $\Nc$ different interacting components, with molecular masses~$m_i$ for $i=0,\ldots, \Nc$.
Assuming local equilibrium, we describe the state of the system using number densities $c_i = N_i/\Vsys$.
The mass density $\rho= \sum_{i=0}^\Nc m_i c_i$ is conserved, ${\partial_t \rho + \nabla\cdot(\vect{v} \rho)=0}$,  where $\vect v = \frac1\rho \sum_{i=0}^\Nc m_i \vect{\hat J}_i$ is the centre-of-mass velocity, and $\vect{\hat J}_i$ are particle fluxes for each component.
Momentum conservation implies $\partial_t(\rho \vect{v}) = \nabla \cdot \tensor\sigma^\mathrm{tot}$, where $\tensor\sigma^\mathrm{tot}$ denotes the total stress. 
Moreover, particle conservation  is described by
\begin{align}
	\label{eqn:continuity_compressible}
	\partial_t c_i + \nabla \cdot \vect{\hat J}_i = s_i
	\;,
\end{align}
where $s_i$ are source terms related to chemical reactions.

We first consider a homogeneous system at constant temperature~$T$ and system volume~$\Vsys$.
The free energy $\hat F=\Vsys \hat f(c_0,\ldots, c_\Nc)$ is then proportional to the free energy density~$\hat f$.
For simplicity, we consider a regular solution model,
\begin{multline}
	\label{eqn:free_energy_many_compressible}
	\hat f(c_0,\ldots,c_\Nc) =  
		\frac{\rho \vect{v}^2}{2}
		+ \kBT\sum_{i=0}^{\Nc} c_i \ln\left(\frac{\nu_i c_i}{\sum_j \nu_j c_j}\right)
\\
		+ h(\{c_i\})
		+ \frac{K}{2}\biggl(1 - \sum_{i=0}^\Nc \nu_i c_i\biggr)^2
	\;,
\end{multline}
where the first term captures the kinetic energy of the centre-of-mass motion,  
the second term describes the mixing entropy, $h(\{c_i\})$ accounts for the enthalpy of all components, and the last term models the compressibility with a bulk modulus~$K$~\cite{Julicher2009}.
Note that we have assumed that all particles have well-defined molecular volumes~$\nu_i$, but we allow for particle overlap to describe the compressibility of the fluid.

Thermodynamic equilibrium implies balanced intensive quantities, including the total chemical potentials~$\hat \mu_i^\mathrm{tot} = (\partial \hat  F/\partial N_i) = \frac12m_i \vect{v}^2 + \hat\mu_i$, where
\begin{align}
	\hat \mu_i &= %
		\kBT\left[
		 \ln\left(\frac{ \nu_i c_i}{\sum_j \nu_j c_j}\right)
		+ \frac{\sum_j (\nu_j - \nu_i) c_j}{\sum_j \nu_j c_j}
	\right]
\notag\\&\quad 
	 + \pfrac{h}{c_i} + \nu_i K \biggl(\sum_{j=0}^\Nc \nu_j c_j - 1\biggr)
	\label{eqn:chemical_potential_compressible}
\end{align}
is the chemical potential in a co-moving frame~\cite{Julicher2018}.
We also define the overall pressure $\hat P = - (\partial\hat  F/\partial \Vsys)$,
\begin{align}
	\hat P &= \frac{K}{2} \Biggl[\biggl(\sum_{i=0}^\Nc \nu_i c_i\biggr)^2 - 1\Biggr]  + \sum_{i=0}^\Nc c_i \pfrac{h}{c_i} - h%
	\label{eqn:pressure_compressible}
	\;.
\end{align}
The particle fluxes~$\vect{\hat J}_i$ can be split into a centre-of-mass motion and relative particle fluxes ${\vect{\hat j}_i = \vect{\hat J}_i - c_i \vect v}$~\cite{Julicher2018}, implying the constraint
\begin{equation}
	\sum_{i=0}^{\Nc} m_i \vect{\hat j}_i = \vect{0}
	\label{eqn:rel_flux_constraint}
	\;.
\end{equation}
The dynamics of the flow field $\vect{v}$ follow from momentum conservation, $\partial_t(\rho\vect{v}) = \nabla \cdot \tensor{\sigma}^\mathrm{tot}$, where the total stress \mbox{$\tensor{\sigma}^\mathrm{tot} = \tensor{\sigma} - \rho \vect{v} \vect{v}$} includes the stress~$\tensor\sigma$ in the co-moving reference frame~\cite{Landau1959_6}.
The relative fluxes~$\vect{\hat j}_i$ can be expanded assuming a linear response to thermodynamic forces~\cite{Julicher2018},
\begin{align}
	\label{eqn:diffusive_fluxes_compress_all}
	\vect{\hat j}_i &= - \sum_{j=0}^{\Nc} L_{ij} \nabla \bigl(\hat{\mu}_j + \zeta \bigr) - \vect{\hat\xi}_i
	&& \text{for}
	& i=0,\ldots,\Nc
	\;,
\end{align}
where $L_{ij}$ denotes the positive definite mobility matrix, $\zeta$ is a Lagrange multiplier ensuring constraint \eqref{eqn:rel_flux_constraint}, and $\vect{\hat\xi}$ is additive noise that we discuss below.

Since only $\Nc$ relative fluxes are independent, we can eliminate the flux of one component, where we arbitrarily choose the solvent ($i=0$).
In particular, we can then eliminate $\zeta$ by inserting \Eqref{eqn:diffusive_fluxes_compress_all} into \Eqref{eqn:rel_flux_constraint} and solving the deterministic part for $\nabla\zeta$.
Inserting the solution into \Eqref{eqn:diffusive_fluxes_compress_all} yields
\begin{align}
	\label{eqn:diffusive_fluxes_compress_many}
	\vect{\hat j}_i &= - \sum_{j=1}^{\Nc} \hat\Lambda_{ij} \nabla \hat{\bar\mu}_j - \vect{\hat\xi}_i
	&& \text{for}
	& i=1,\ldots,\Nc
	\;,
\end{align}
where we have defined exchange chemical potentials $\hat{\bar\mu}_j = \hat\mu_j - m_j m_0^{-1} \hat \mu_0$ relative to solvent and mobilities
\begin{equation}
	\label{eqn:mobility_compressible}
	\hat\Lambda_{ij} = \hat L_{ij} - \frac{\sum_{k,l=0}^\Nc m_k m_l \hat L_{il}\hat L_{kj}}{\sum_{k,l=0}^\Nc m_k m_l \hat L_{kl} }
	\;,
\end{equation}
which obeys Onsager's symmetry condition ($\hat\Lambda_{ij} = \hat\Lambda_{ji}$) and additionally $\sum_{i=0}^\Nc m_i \hat\Lambda_{ij}=0$.
Finally, we use the  fluctuation--dissipation relation~\cite{Julicher2018} to determine the variance of the  noise $\vect{\hat\xi}_i$ for $i=1,\ldots,\Nc$,
\begin{align}
	\mean{\vect{\hat\xi}_i(\vect r, t)\vect{\hat\xi}_j(\vect r', t')}
		= 2 \kBT \hat\Lambda_{ij} \identity \delta(\vect r-\vect r')\delta(t-t')
	\;,
\end{align}
and the solvent flux reads $\vect{\hat j}_0 = -m_0^{-1}\sum_{i=1}^\Nc m_i \vect{\hat j}_i$.

To describe incompressible liquids, we assume a large bulk modulus ($K\rightarrow\infty$), which enforces a constant volume density, $\sum_i \nu_i c_i = 1$.
In the simple case where all components have equal mass density $m_i/\nu_i$, the total mass density~$\rho$ is then also conserved, implying incompressible flows ($\nabla.\vect v=0$) enforced by hydrostatic pressure.
The system state is then conveniently expressed in terms of volume fractions $\phi_i = \nu_i c_i$, with the solvent fraction $\phi_0 = 1-\sum_{i=1}^\Nc \phi_i$.
The kinetics described by \Eqsref{eqn:continuity_many}--\eqref{eqn:fluctuation_dissipation_relation} follow directly by substituting $\Lambda_{ij} = \nu_i\hat\Lambda_{ij}$ and $\vect j_i = \nu_i \vect{\hat j_i}$.
The associated mobility matrix given by \Eqref{eqn:mobility_matrix} follows directly from \Eqref{eqn:mobility_compressible}.
In the simple case of a non-interacting system ($h=0$), diagonal mobilities, $L_{ij} = \lambda_i \phi_i \delta_{ij}$, dilute concentrations ($\phi_i \ll 1$ for $i\ge1$), and a fast solvent ($\lambda_0\gg \lambda_i$), the kinetics reduce to ideal diffusion, $\vect j_i = -D_i \nabla \phi_i$, with diffusivity $D_i = \kBT \lambda_i$.

To obtain explicit expressions for the thermodynamic quantities, we consider the enthalpy
\begin{align}
	h(c_0,\ldots,c_\Nc) &= \sum_{i=0}^\Nc e_i c_i +  \sum_{i,j=0}^\Nc e_{ij} c_i c_j
	\;,
\end{align}
where $e_i$ are internal energies and $e_{ij}$ accounts for pairwise interactions with $e_{ij} = e_{ji}$.
This model can be derived from a lattice model or can be viewed as a simple polynomial expansion up to second order, which is sufficient to describe phase separation, although higher-order effects are also interesting~\cite{Luo2024}.
Taken together, this results in the incompressible free energy density given in \Eqref{eqn:free_energy_many}, where
\begin{subequations}
\label{eqn:enthalpy_coefficients_with_solvent}
\begin{align}
	w_i &= \frac{1}{\kBT} \left[
		\frac{e_i}{\size_i} - \frac{e_0}{\size_0} 
		+ \frac{e_{i0} + e_{0i}}{\nu_0 \size_i}
		- \frac{2e_{00}}{\nu_0} 
	\right]
	\;,
\\
	\chi_{ij} &= \frac{2}{\nu_0 \kBT} \left[
		\frac{e_{ij}}{\size_i\size_j} 
		- \frac{e_{i0}}{\size_i} - \frac{e_{0 j}}{\size_j}
		+ e_{00}
	\right]
	\;,
\end{align}
\end{subequations}
with $\chi_{ij} = \chi_{ji}$, but $\chi_{ii}\neq0$.
The exchange chemical potentials in \Eqref{eqn:chemical_potential_many} follow from \Eqref{eqn:chemical_potential_compressible} using $\bar\mu_i  = \hat \mu_i - \size_i \hat \mu_0$ with $\size_i = \nu_i/\nu_0$.
Similarly, the osmotic pressure~$P=\hat P - \hat \mu_0\nu_0^{-1}$  in \Eqref{eqn:pressure_many} follows from \Eqref{eqn:pressure_compressible}, illustrating  that $P$ involves the chemical potential of the solvent.

\subsection{Free energy variation with surface terms}
\label{sec:appendix_variation}
We consider free energy functionals of the structure
\begin{align}
	F[\Phi] = \int \! \fBulk(\Phi, \nabla\Phi) \diff V + \oint \fWall(\Phi) \diff A
	\;,
\end{align}
where $\fBulk$ and $\fWall$ describe the bulk and surface contributions of $n$ fields, $\Phi=(\phi_1, \phi_2, \ldots, \phi_n)$.
Applying an infinitesimal perturbation $\delta\Phi$,
\begin{align}
	F[\Phi + \delta \Phi] &= \int \! \fBulk(\Phi + \delta\Phi, \nabla\Phi + \nabla\delta\Phi) \diff V 
\notag\\&\quad
	+ \oint \fWall(\Phi + \delta\Phi) \diff A
\notag\\
	 &= \int \left[
		\fBulk
		+ \pfrac{\fBulk}{\phi_i} \delta\phi_i
		+ \pfrac{\fBulk}{(\nabla \phi_i)} \cdot \nabla \delta\phi_i
	\right]\diff V 
\notag\\&\quad
	+ \oint \left[
		\fWall
		+ \pfrac{\fWall}{\phi_i} \delta \phi_i
	\right]\diff A
	\;,
\end{align}
and using Einstein's summation convention, the first variation is %
\begin{align}
	\delta F &= F[\Phi + \delta \Phi] - F[\Phi]
\notag\\ &=
	\int \left[
		\pfrac{\fBulk}{\phi_i} \delta\phi_i
		+ \pfrac{\fBulk}{(\nabla \phi_i)} \cdot \nabla \delta\phi_i
	\right] \! \diff V 
\notag\\&\quad 
	+ \oint \! \pfrac{\fWall}{\phi_i} \delta \phi_i \, \diff A
\notag\\&=
	\int \left[
		\pfrac{\fBulk}{\phi_i} \delta\phi_i
		- \nabla \cdot \left(\pfrac{\fBulk}{(\nabla \phi_i)}  \delta\phi_i\right)
	\right]\diff V
\notag\\&\quad 
	+ \oint \left[
		\pfrac{\fBulk}{(\nabla \phi_i)} \cdot \vect{n} \, \delta\phi_i
		+ \pfrac{\fWall}{\phi_i} \delta \phi_i
	\right] \diff A
	\;,
\end{align}
where the last equality follows from partial integration.
Local equilibrium at the boundary implies that the surface term does not contribute to the variation~\cite{Zhao2024}, which is only possible if
\begin{align}
	\pfrac{\fBulk}{(\nabla \phi_i)}  \cdot \vect{n} 
	+ \pfrac{\fWall}{\phi_i} &=0
	\;.
\end{align}
Since the first term involves normal derivatives of the fields~$\phi_i$, this equation provides boundary conditions for the fields~$\phi_i$.
The remaining term gives the bulk contribution of the functional derivative, 
\begin{align}
	\frac{\delta F}{\delta \phi_i} &= 
		\pfrac{\fBulk}{\phi_i} 
		- \nabla \cdot \pfrac{\fBulk}{(\nabla \phi_i)}
	\;,
	\label{eqn:appendix_func_deriv_mu}
\end{align}
which are related to the bulk chemical potentials~$\mu_i$.

\subsection{Interface properties of binary liquids}
\label{sec:appendix_interface}
We study a flat interface described by the equilibrium profile~$\phiInt(x)$, where the $x$--coordinate is perpendicular to the interface.
The profile $\phiInt(x)$ connects two phases, $\phiInt(x\rightarrow -\infty)=\phiBaseOut$ and $\phiInt(x\rightarrow \infty)=\phiBaseIn$, and exhibits a constant chemical potential~$\muEq$, 
\begin{align}
	\label{eqn:appendix_equilibrium_condition}
	f'\bigl(\phiInt(x)\bigr) - \kappa \phiInt''(x) = \frac{\muEq}{\nu}
	\;.
\end{align}
Material conservation implies 
\begin{align}
	\label{eqn:appendix_interface_location}
	\int_{-\infty}^\infty \phiInt(x) \diff x = \int_{-\infty}^0 \phiBaseOut \diff x + \int_0^{\infty} \phiBaseIn \diff x
	\;,
\end{align}
which implicitly positions the interface at $x=0$.
The interfacial width~$\width$ is then defined as~\cite{Cahn1958, Tang1991}
\begin{align}
	\width = \frac{\phiBaseIn - \phiBaseOut}{\phiInt'(x=0)}
	\;.
\end{align}
To determine $\phiInt'(x=0)$, we multiply \Eqref{eqn:appendix_equilibrium_condition} by $\phiInt'(x)$ and integrate once, yielding
\begin{align}
	\label{eqn:appendix_interfacial_steepness}
	\phiInt'(x) = \sqrt{\frac{2}{\kappa}\left[f\bigl(\phiInt(x)\bigr) - \frac{\muEq}{\nu} \phiInt(x) + \PEq\right]}
	\;,
\end{align}
where the integration constant $\PEq$ is chosen such that $\phiInt'(x\rightarrow \pm \infty)=0$.
Hence, $\PEq =\frac{\muEq}{\nu} \phiBaseOut  - f(\phiBaseOut) =\frac{\muEq}{\nu} \phiBaseIn - f(\phiBaseIn)$ is the bulk osmotic pressure.

The surface tension~$\gamma$ associated with the interface can be defined mechanically via the excess stress~\cite{Kirkwood1949},
\begin{align}
	\label{eqn:appendix_surface_tension_definition_mechanical}
	\gamma &= \int_{-\infty}^\infty \left(\stressEq_\mathrm{yy} - \stressEq_\mathrm{xx}\right) \diff x
	\;,
\end{align}
where the first term describes the uniform equilibrium pressure, whereas the second term captures the stress normal to the interface. %
Using \Eqref{eqn:appendix_equilibrium_stress} and the free energy given by \Eqref{eqn:free_energy_binary}, we thus find
\begin{align}
	\label{eqn:appendix_surface_tension_common_integral}
\gamma	&= 
	\kappa \int_{-\infty}^\infty \bigl(\phi'(x)\bigr)^2 \diff x
	\;.
\end{align}
Alternatively, $\gamma$ is defined as the excess free energy, 
\begin{align}
	\label{eqn:appendix_surface_tension_definition}
	\gamma &= \int_{-\infty}^\infty \left[
		f(\phiInt) + \frac\kappa2\bigl(\phiInt'\bigr)^2
		- \frac{f(\phiBaseOut) + f(\phiBaseIn)}{2}
	\right] \diff x
	\;.
\end{align}
Using \Eqsref{eqn:appendix_interfacial_steepness}, \eqref{eqn:appendix_interface_location}, and  \eqref{eqn:appendix_surface_tension_definition}, we again find \Eqref{eqn:appendix_surface_tension_common_integral}, showing that the mechanical and thermodynamic definitions of $\gamma$ agree~\cite{Ip1994}.

In simple mixtures, where the interfacial profile $\phiInt(x)$ is monotonic, we can use a change of variable,
\begin{align}
		\gamma &= \kappa \int_{\phiBaseOut}^{\phiBaseIn} \phiInt' \diff \phiInt
		\;,
\end{align}
and \Eqref{eqn:appendix_interfacial_steepness} to obtain
\begin{align}
	\label{eqn:appendix_surface_tension_integral}
	\gamma = \sqrt{2\kappa} \int_{\phiBaseOut}^{\phiBaseIn} \left[
		f(\phi)
		- \frac{\muEq}{\nu} \phi
		+ \PEq
	\right]^{\frac12} \diff \phi
	\;,
\end{align}
which can be evaluated without solving for the  profile~$\phiInt(x)$ if the  fractions $\phiBaseOut$ and $\phiBaseIn$ are known.

We obtain concrete expressions for $\width$ and $\gamma$ using the free energy \eqref{eqn:free_energy_density_binary} in the symmetric case ($\size=1$), implying
$\phiBaseIn = 1- \phiBaseOut$, $\phiInt(x=0)=\frac12$, and $\muEq =w \kBT $. 
Hence,
\begin{align}
	\width &= \frac{\sqrt{\frac{2\nu\kappa}{\kBT}} \left(\phiBaseIn - \phiBaseOut\right)}
		{\left[-2 \phiBaseIn \ln (\phiBaseIn)-2 \phiBaseOut \ln (\phiBaseOut) - \ln(16\phiBaseIn\phiBaseOut)\right]^{\frac12}}
\notag\\
	&\approx \frac{1}{\phiBaseIn - \phiBaseOut}\,\sqrt{\frac{6\nu\kappa}{\kBT}}
	\;,
\end{align}
where the approximation is only valid for weak phase separation.
Approximating the integrand in \Eqref{eqn:appendix_surface_tension_integral} as a parabola around $\phi=\frac12$, we find 
\begin{align}
	\gamma &\approx
		 \sqrt{\frac{\kBT \kappa}{2 \nu}} \cdot \frac{\phiBaseOut - \phiBaseIn}{3} 
		 	\ln\left(4\phiBaseIn\phiBaseOut\right)
		\;.
\end{align}
The fractions $\phiBaseOut$ and $\phiBaseIn = 1 - \phiBaseOut$ can be obtained self-consistently~\cite{Qian2022a}.
Similar approximations for $\gamma$ are proposed in \cite{Joanny1978, Tang1991}.
In all cases, the interfacial width~$\width$ diverges and $\gamma$ vanishes at the critical point;
For stronger phase separation, the interface becomes thinner and $\gamma$ increases.

\subsection{Droplets in two dimensions}
\label{sec:appendix_droplets_2d}
While the main text focuses primarily on three-dimensional systems, phase separation can also take place within two-dimensional membranes inside biological cells~\cite{Litschel2023,Lee2023a,Mangiarotti2023,Mangiarotti2024}.
Here, we briefly summarise the key differences from the three-dimensional theory presented above resulting from this reduction in dimensionality.
In two dimensions, surface tension now demands circular (rather than spherical) droplets of radius~$R$; droplet size~${V=4\pi R^2}$ now has dimensions of area; and the size of the interface~${A=2\pi R}$ is now a length.
The Laplace pressure reduces to $\Plaplace=\gamma/R$ (equation \ref{eqn:coexistence}), and classical nucleation theory implies $\Rnuc = \gamma/\Df$ and $\Fnuc = \pi \gamma^2/\Df$, compared to \Eqref{eqn:critical_radius}.

Droplet growth is still described by \Eqref{eqn:drop_interface_speed}, but the diffusive fluxes $\vect j$ are more difficult to determine in two dimensions since steady state concentration profiles diverge logarithmically. As a result, one must introduce a cut-off length in numerical simulations~\cite{Kulkarni2023} or use asymptotic matching~\cite{Bressloff2020, Bressloff2020a}.
The latter approach shows that Ostwald ripening is suppressed~\cite{Bressloff2020} and that droplets drift in gradients~\cite{Bressloff2020a}, similar to their 3D counterparts.
The diffusivity~$\Ddrop$ associated with Brownian motion also exhibits a logarithmic scaling due to a hydrodynamic coupling mediated via the surrounding liquid.
For a droplet of radius $R$ embedded in a membrane of thickness~$h$, the Saffman--Delbr\"uck model predicts~\cite{Saffman1975}
\begin{align}
	\Ddrop &\approx \frac{\kBT}{4\pi \eta_\mathrm{m} h}\left(\ln\frac{\eta_\mathrm{m}h}{\eta_\mathrm{b} R} - \gamma_\mathrm{EM} \right)
	\;,	
\end{align}
where $\gamma_\mathrm{EM} \approx 0.5772$ is the Euler--Mascheroni constant and $\eta_\mathrm{m}$ and $\eta_\mathrm{b}$ are the viscosities of the membrane and the surrounding bulk liquid, respectively.

\subsection{Derivation of equilibrium stress tensor}
\label{sec:appendix_equilibrium_stress}
The equilibrium stress tensor $\stressEq_\ab$ follows from the variation of the energy
\begin{align}
	F = \int \! \left[ 
		\frac{\rho v_\alpha^2}{2} + 
		f(\{\phi_i\}, \{\partial_\alpha \phi_i\})
	\right] \diff V
\end{align}
when the control volume~$V$ is deformed~\cite{Julicher2018}.
Note that in this appendix we write vectorial and tensorial quantities in index notation for clarity, adopting Einstein's summation convention. Greek indices (e.g., $\alpha,\beta$) indicate components of vectors and tensors, whereas Latin indices (e.g., $i$) indicate different components of the mixture, as in the main text.
The variation~$\delta F$ of the energy reads
\newcommand{\partialFpartialGradPhi}{f^{(i)}_\alpha }
\begin{align}
	\delta F &= 
		\!\int_V \! \left[
			\rho v_\alpha\delta v_\alpha
			+ \frac{v_\alpha^2}{2} \delta \rho
			+ \pfrac{f}{\phi_i} \delta \phi_i
			+ \partialFpartialGradPhi \partial_{\alpha}\delta\phi_i
		\right]\!\diff V
\notag\\&\quad
		+ \int_{\delta V} \left[ \frac{\rho v_\alpha^2}{2} + f\right]\diff V
	\;,
\end{align}
where $f^{(i)}_\alpha = \partial f / \partial(\partial_\alpha\phi_i)$.
The first line describes contributions from the perturbed fields, whereas the second line accounts for the deformed volume.
We consider a displacement field $u_\alpha$ conserving volume ($\partial_\alpha u_\alpha=0$), so that reference positions $x_\alpha$ map to $x_\alpha' = x_\alpha + u_\alpha$.
The perturbation of any field $B$ is given by $\delta B = B'(x_\alpha) - B(x_\alpha)  \approx B(x_\alpha - u_\alpha) - B(x_\alpha) \approx -u_\alpha \partial_\alpha B$ to linear order in $u_\alpha$.
Hence,
\begin{align}
	\delta F &= 
		- \int_V  \biggl[
			\rho v_\alpha u_\beta \partial_\beta v_\alpha
			+ \frac{v_\alpha^2}{2} u_\beta \partial_\beta \rho
			+ \pfrac{f}{\phi_i} u_\beta \partial_\beta \phi_i
\notag\\&\qquad\quad
			+ \partialFpartialGradPhi u_\beta \partial_\beta(\partial_{\alpha}\phi_i)
		\biggr]\diff V
\notag\\&\quad
		+ \oint_{\partial V} \left[ \frac{\rho v_\alpha^2}{2} + f\right] u_\beta \diff A_\beta
	\;.
\end{align}
Integrating by parts and using $\partial_\alpha u_\alpha=0$ yields
\begin{align}
	\delta F &= 
		\int_V  \biggl[
			v_\alpha \partial_\beta (\rho v_\alpha)
			+ \rho v_\alpha \partial_\beta v_\alpha
			+  \phi_i\partial_\beta \pfrac{f}{\phi_i} 
\notag\\&\quad
			+ (\partial_\beta\phi_i) \partial_\alpha\partialFpartialGradPhi
		\biggr]u_\beta \diff V
		- \oint_{\partial V}  
			 \partialFpartialGradPhi  (\partial_{\beta}\phi_i)
		u_\beta  \diff A_\alpha
\notag\\&\quad
		+ \oint_{\partial V}  \left[
			f
			- \rho v_\alpha^2 
			- \pfrac{f}{\phi_i} \phi_i
		\right]u_\beta  \diff A_\beta
	\;.
\end{align}
The divergence theorem applied to $\rho v_\alpha^2$ implies terms involving $v_\alpha$ cancel.
Integrating by parts then yields
\begin{align}
	\delta F &= 
		\oint_{\partial V}  \left[
			\left(f - \phi_i \hat\mu_i \right)\delta_{\alpha\beta}
			- \partialFpartialGradPhi  \partial_{\beta}\phi_i
		\right]u_\beta  \diff A_\alpha
\notag\\&\quad
		+ \int_V  
			\phi_i(\partial_\beta \hat\mu_i)
		u_\beta\diff V
	\;,
\end{align}
with $\hat\mu_i = \partial f/\partial\phi_i - \partial_\alpha \partialFpartialGradPhi$ proportional to the exchange chemical potentials given in \Eqref{eqn:chemical_potential_many_kappa}.
We identify the equilibrium stress,
\begin{align}
	\label{eqn:appendix_equilibrium_stress}
		\stressEq_\ab &= 
			\left(f - \phi_i \hat\mu_i \right)\delta_{\alpha\beta}
			- \pfrac{f}{(\partial_{\alpha}\phi_i)}  \partial_{\beta}\phi_i
		\;,
\end{align}
from the surface contributions.
To obey translational invariance, the free energy must be constant ($\delta F=0$) for a uniform displacement ($u_\beta=\text{const}$).
Using Gauss's theorem, this yields the \emph{Gibbs--Duhem relation}
\begin{align}
	\partial_\alpha \stressEq_\ab &= - \phi_i\partial_\beta \hat\mu_i
	\;,
\end{align}
which generalizes \Eqref{eqn:gibbs_duhem}.
In summary, the energy variation reads
\begin{align}
	\delta F &= 
			\oint_{\partial V}  \stressEq_\ab u_\beta  \diff A_\alpha
		- \int_V  \left(
			\partial_\alpha \stressEq_\ab
		\right)u_\beta\diff V
	\;,
\end{align}
where the bulk term vanishes in equilibrium.

\subsection{Transition state theory}
\label{sec:transition_state_theory}
The rate of the chemical conversion between reactants~$R$ and products $P$ depends on their thermodynamic state described by the respective chemical potentials $\mu_R$ and $\mu_P$, as well as the kinetic details of the reaction $R \rightleftharpoons P$.
Typically, reactions follow a reaction path that involves an energetic barrier limiting the rate.
This barrier can be conceptualised by a \emph{transition state}~$T$, leading to the extended reaction~\cite{Atkin2010,Hanggi1990,Pagonabarraga1997,Bazant2013,Kirschbaum2021}
\begin{align}
	R \rightleftharpoons T \rightleftharpoons P
	\;.
\end{align}
The net flux between $R$ and $P$ then follows from two separate assumptions.
First, the individual transition fluxes $s_{R \rightleftharpoons T}$ and $s_{P \rightleftharpoons T}$ obey detailed balance of the rates (see Appendix C in \refcite{Weber2019}),
\begin{align}
	\frac{s_{R \rightarrow T}}{s_{R \leftarrow T}} &=e^{\beta(\mu_R - \mu_T)}
& \text{and} &&
	\frac{s_{P \rightarrow T}}{s_{P \leftarrow T}} &=e^{\beta(\mu_P - \mu_T)}
	\;,
\end{align}
where $\beta=1/\kBT$ and $\mu_T$ is the chemical potential of the transition state.
Second, we assume that the transition state decays fast, implying \mbox{$s_{R \rightarrow P} \approx \frac12 s_{R \rightarrow T}$} and \mbox{$s_{P \rightarrow R} \approx \frac12 s_{P \rightarrow T}$}.
Here, the factor $\frac12$ reflects that the transition state is defined such that it has equal probability to decay towards the reactants and products, \mbox{$s_{R \leftarrow T} = s_{P \leftarrow T} = s_T$}. %
Taken together,
\begin{subequations}
\begin{align}
    s_{R \rightarrow P}  &\approx \frac{s_T}{2} e^{\beta(\mu_R - \mu_T)} \;, 
\\
	s_{P \rightarrow R} &\approx \frac{s_T}{2}  e^{\beta(\mu_P - \mu_T)}
	\;.
\end{align}
\end{subequations}
The reaction fluxes used in refs.~\cite{Kirschbaum2021,Bauermann2022,Zwicker2022a,Julicher2024} then follow when the chemical potential of the transition state is constant ($\mu_T=w_T$).
However, typical transition states physically resemble both the reactants and the product, so a simple model for its chemical potential $\mu_T$ is
\begin{align}
	\mu_T = (1 - \alpha) \mu_R + \alpha \mu_P + w_T
	\;,
\end{align}
where the symmetry factor $\alpha$ is also known as a generalised Brønsted coefficient~\cite{Bazant2013}, and $w_T$ quantifies the barrier height.
Taken together, the net reaction flux, $s=s_{R \rightarrow P}  - s_{P \rightarrow R}$, thus reads
\begin{align}
	s = k \left(
		e^{-\beta\alpha\mu} - e^{\beta(1 - \alpha)\mu}
	\right)
	\;,
\end{align}
where $\mu = \mu_P - \mu_R$ is the thermodynamic driving force of the reaction, and $k=\frac12 s_T e^{\beta w_T}$ can be interpreted as a kinetic pre-factor, which can depend on composition.

A simple model for the rate $k$ follows from considering ideal solutions, where $\mu_i = \beta^{-1}(\ln \phi_i + w_i)$.
Demanding that $s$ reduces to mass action kinetics, 
\begin{align}
	s = k_\rightarrow \prod_{i=1}^\Nc \phi_i^{\sigma_{\rightarrow,i}} -  k_\leftarrow \prod_{i=1}^\Nc \phi_i^{\sigma_{\leftarrow,i}}
	\;,
\end{align}
with constant reaction rates $k_\rightarrow$ and $k_\leftarrow$, implies
\begin{equation}
	s=\underbrace{\hat k\,\prod_{i=1}^\Nc\phi_i^{(1-\alpha)\sigma_{\rightarrow,i} + \alpha\sigma_{\leftarrow,i}}}_{\text{kinetics}} \;
\underbrace{\vphantom{\prod_{i=1}^\Nc}\!\!\left(e^{-\beta\alpha\mu} - e^{\beta(1 - \alpha)\mu}\right)\!}_{\text{thermodynamics}}
	\;,
\end{equation}
where the first term ensures that the kinetics obey mass action kinetics in the ideal limit, whereas the second term captures the thermodynamic constraints of detailed balance of the rates.

The additional requirement of  symmetry between reactants and products, i.e., that swapping reactants and products must turn $s$ into $-s$, implies $\alpha=\frac12$, leading to \Eqsref{eqn:transition_state_theory} and \eqref{eqn:mass_action_kinetics_rate}.

\sectionseparation
\section*{References}
\addcontentsline{toc}{section}{References}

\bibliographystyle{nar}%
\bibliography{references}

\begin{thebibliography}{100}

\bibitem{Banani2017}
Banani, S.~F., Lee, H.~O., Hyman, A.~A., and Rosen, M.~K. (05, 2017)
Biomolecular condensates: organizers of cellular biochemistry.
{\em Nat. Rev. Mol. Cell Biol.,} {\bf 18}(5), 285--298.

\bibitem{Kilgore2023}
Kilgore, H.~R., Mikhael, P.~G., Overholt, K.~J., Boija, A., Hannett, N.~M., Van~Dongen, C., Lee, T.~I., Chang, Y.-T., Barzilay, R., and Young, R.~A. (2024)
Distinct chemical environments in biomolecular condensates.
{\em Nat. Chem. Biol.,} {\bf 20}, 291--301.

\bibitem{Papp2025}
Papp, M., Stoffel, F., Gil-Garcia, M., Faltova, L., and Arosio, P. (2025)
Biomolecular condensates as regulators of enzymatic reactions.
{\em Nat. Chem. Eng.,} {\bf 2}(7), 394--397.

\bibitem{Oflynn2021}
O'Flynn, B.~G. and Mittag, T. (2021)
The role of liquid--liquid phase separation in regulating enzyme activity.
{\em Curr. Opin. Cell Biol.,} {\bf 69}, 70--79.

\bibitem{Glauninger2024}
Glauninger, H., Bard, J.~A., Hickernell, C. J.~W., Airoldi, E.~M., Li, W., Singer, R.~H., Paul, S., Fei, J., Sosnick, T.~R., Wallace, E.~W., and Drummond, D.~A. (2024)
Transcriptome-wide mRNA condensation precedes stress granule formation and excludes stress-induced transcripts.
{\em bioRxiv,}.

\bibitem{Wiegand2020}
Wiegand, T. and Hyman, A.~A. (10, 2020)
{Drops and fibers - how biomolecular condensates and cytoskeletal filaments influence each other}.
{\em Emerg. Top. Life Sci.,} {\bf 4}, 247--261.

\bibitem{Hong2024}
Hong, L., Zhang, Z., Wang, Z., Yu, X., and Zhang, J. (Jun, 2024)
Phase separation provides a mechanism to drive phenotype switching.
{\em Phys. Rev. E,} {\bf 109}, 064414.

\bibitem{So2021}
So, C., Cheng, S., and Schuh, M. (2021)
Phase Separation during Germline Development.
{\em Trends Cell Biol.,} {\bf 31}(4), 254--268.

\bibitem{Pei2024}
Pei, G., Lyons, H., Li, P., and Sabari, B.~R. (2024)
Transcription regulation by biomolecular condensates.
{\em Nat. Rev. Mol. Cell Biol.,} {\bf 26}, 213--236.

\bibitem{Hirose2022}
Hirose, T., Ninomiya, K., Nakagawa, S., and Yamazaki, T. (2022)
A guide to membraneless organelles and their various roles in gene regulation.
{\em Nat. Rev. Mol. Cell Biol.,} {\bf 24}, 288--304.

\bibitem{Rajendran2025}
Rajendran, A. and Casta{\\textasciitilde n}eda, C.~A. (2025)
Protein quality control machinery: regulators of condensate architecture and functionality.
{\em Trends Biochem. Sci.,} {\bf 50}(2), 106--120.

\bibitem{Jaqaman2021}
Jaqaman, K. and Ditlev, J.~A. (2021)
Biomolecular condensates in membrane receptor signaling.
{\em Curr. Opin. Cell Biol.,} {\bf 69}, 48 -- 54.

\bibitem{Alberti2021}
Alberti, S. and Hyman, A.~A. (2021)
Biomolecular condensates at the nexus of cellular stress, protein aggregation disease and ageing.
{\em Nat. Rev. Mol. Cell Biol.,} {\bf 22}, 196--213.

\bibitem{Shin2017a}
Shin, Y. and Brangwynne, C.~P. (09, 2017)
Liquid phase condensation in cell physiology and disease.
{\em Science,} {\bf 357}(6357).

\bibitem{Mathieu2020}
Mathieu, C., Pappu, R.~V., and Taylor, J.~P. (Oct, 2020)
Beyond aggregation: Pathological phase transitions in neurodegenerative disease.
{\em Science,} {\bf 370}(6512), 56--60.

\bibitem{Cai2021}
Cai, D., Liu, Z., and Lippincott-Schwartz, J. (2021)
Biomolecular Condensates and Their Links to Cancer Progression.
{\em Trends Biochem. Sci.,} {\bf 46}(7), 535--549.

\bibitem{Lafontaine2020}
Lafontaine, D. L.~J., Riback, J.~A., Bascetin, R., and Brangwynne, C.~P. (2020)
The nucleolus as a multiphase liquid condensate.
{\em Nat. Rev. Mol. Cell Biol.,} {\bf 22}, 165--182.

\bibitem{Azaldegui2020}
Azaldegui, C.~A., Vecchiarelli, A.~G., and Biteen, J.~S. (2020)
The Emergence of Phase Separation as an Organizing Principle in Bacteria.
{\em Biophys. J.,} {\bf 120}(7), 1123--1138.

\bibitem{Kim2021}
Kim, J., Lee, H., Lee, H.~G., and Seo, P.~J. (2021)
Get closer and make hotspots: liquid--liquid phase separation in plants.
{\em EMBO Rep.,} {\bf 22}(5), e51656.

\bibitem{Gibbs1876}
Gibbs, J.~W. (1876)
On the Equilibrium of Heterogeneous Substances.
{\em Trans. Conn. Acad. Arts Sci.,} {\bf 3}, 1--329.

\bibitem{Wilson1899}
Wilson, E.~B. (1899)
The Structure of Protoplasm.
{\em Science,} {\bf 10}(237), 33--45.

\bibitem{Hyman2014}
Hyman, A.~A., Weber, C.~A., and J{\"u}licher, F. (2014)
Liquid-liquid phase separation in biology.
{\em Annu. Rev. Cell Dev. Biol.,} {\bf 30}, 39--58.

\bibitem{Holehouse2023}
Holehouse, A.~S. and Kragelund, B.~B. (2024)
The molecular basis for cellular function of intrinsically disordered protein regions.
{\em Nat. Rev. Mol. Cell Biol.,} {\bf 25}, 187--211.

\bibitem{Weber2019}
Weber, C.~A., Zwicker, D., J{\"u}licher, F., and Lee, C.~F. (2019)
Physics of Active Emulsions.
{\em Rep. Prog. Phys.,} {\bf 82}, 064601.

\bibitem{Dignon2020}
Dignon, G.~L., Best, R.~B., and Mittal, J. (Apr, 2020)
Biomolecular Phase Separation: From Molecular Driving Forces to Macroscopic Properties.
{\em Annu. Rev. Phys. Chem.,} {\bf 71}, 53--75.

\bibitem{Sing2020}
Sing, C.~E. and Perry, S.~L. (2020)
Recent progress in the science of complex coacervation.
{\em Soft Matter,} {\bf 16}, 2885--2914.

\bibitem{Rippe2025}
Rippe, K. and Papantonis, A. (2025)
RNA polymerase II transcription compartments---from factories to condensates.
{\em Nat. Rev. Genet.,}.

\bibitem{Visser2024}
Visser, B.~S., Lipi{\'n}ski, W.~P., and Spruijt, E. (2024)
The role of biomolecular condensates in protein aggregation.
{\em Nat. Rev. Chem.,} {\bf 8}, 686--700.

\bibitem{Buchan2024}
Buchan, J.~R. (2024)
Stress granule and P-body clearance: Seeking coherence in acts of disappearance.
{\em Semin. Cell Biol.,} {\bf 159-160}, 10--26.

\bibitem{Mohapatra2023}
Mohapatra, S. and Wegmann, S. (2023)
Biomolecular condensation involving the cytoskeleton.
{\em Brain Res. Bull.,} {\bf 194}, 105--117.

\bibitem{Millar2023}
Millar, S.~R., Huang, J.~Q., Schreiber, K.~J., Tsai, Y.-C., Won, J., Zhang, J., Moses, A.~M., and Youn, J.-Y. (2023)
A New Phase of Networking: The Molecular Composition and Regulatory Dynamics of Mammalian Stress Granules.
{\em Chem. Rev.,} {\bf 123}(14), 9036--9064.

\bibitem{Gormal2023}
Gormal, R.~S., Martinez-Marmol, R., Brooks, A.~J., and Meunier, F.~A. (jan, 2024)
Location, location, location: Protein kinase nanoclustering for optimised signalling output.
{\em eLife,} {\bf 13}, e93902.

\bibitem{Lee2022}
Lee, D. S.~W., Strom, A.~R., and Brangwynne, C.~P. (2022)
The mechanobiology of nuclear phase separation.
{\em APL Bioengineering,} {\bf 6}(2), 021503.

\bibitem{Wang2021}
Wang, B., Zhang, L., Dai, T., Qin, Z., Lu, H., Zhang, L., and Zhou, F. (2021)
Liquid--liquid phase separation in human health and diseases.
{\em Signal Transduct. Target. Ther.,} {\bf 6}(1), 290.

\bibitem{Su2021}
Su, Q., Mehta, S., and Zhang, J. (2021)
Liquid-liquid phase separation: Orchestrating cell signaling through time and space.
{\em Mol. Cell,} {\bf 81}(20), 4137--4146.

\bibitem{Schisa2021}
Schisa, J.~A. and Elaswad, M.~T. (2021)
An Emerging Role for Post-translational Modifications in Regulating RNP Condensates in the Germ Line.
{\em Front. Mol. Biosci.,} {\bf 8}, 230.

\bibitem{Ghosh2021a}
Ghosh, B., Bose, R., and Tang, T.-Y.~D. (2021)
Can coacervation unify disparate hypotheses in the origin of cellular life?.
{\em Curr. Opin. Colloid Interface Sci.,} {\bf 52}, 101415.

\bibitem{Currie2021}
Currie, S.~L. and Rosen, M.~K. (2021)
Using quantitative reconstitution to investigate multi-component condensates.
{\em RNA,} {\bf 28}, 27--35.

\bibitem{Bhat2021}
Bhat, P., Honson, D., and Guttman, M. (2021)
Nuclear compartmentalization as a mechanism of quantitative control of gene expression.
{\em Nat. Rev. Mol. Cell Biol.,} {\bf 22}, 653--670.

\bibitem{Swain2020}
Swain, P. and Weber, S.~C. (2020)
{Dissecting the complexity of biomolecular condensates}.
{\em Biochem. Soc. Trans.,} {\bf 48}(6), 2591--2602.

\bibitem{Adekunle2020}
Adekunle, D.~A. and Hubstenberger, A. (12, 2020)
The multiscale and multiphase organization of the transcriptome.
{\em Emerg. Top. Life Sci.,} {\bf 4}(3), 265--280.

\bibitem{Sabari2020a}
Sabari, B.~R. (Oct, 2020)
Biomolecular Condensates and Gene Activation in Development and Disease.
{\em Dev. Cell,} {\bf 55}(1), 84--96.

\bibitem{Ong2020}
Ong, J.~Y. and Torres, J.~Z. (2020)
Phase Separation in Cell Division.
{\em Mol. Cell,} {\bf 80}(1), 9--20.

\bibitem{Hondele2020}
Hondele, M., Heinrich, S., De~Los~Rios, P., and Weis, K. (Aug, 2020)
Membraneless organelles: phasing out of equilibrium.
{\em Emerg. Top. Life Sci.,} {\bf 4}(3), 343--354.

\bibitem{Cohan2020}
Cohan, M.~C. and Pappu, R.~V. (08, 2020)
Making the Case for Disordered Proteins and Biomolecular Condensates in Bacteria.
{\em Trends Biochem. Sci.,} {\bf 45}(8), 668--680.

\bibitem{Zhao2020a}
Zhao, Y.~G. and Zhang, H. (2020)
Phase Separation in Membrane Biology: The Interplay between Membrane-Bound Organelles and Membraneless Condensates.
{\em Dev. Cell,} {\bf 55}(1), 30--44.

\bibitem{Greening2020}
Greening, C. and Lithgow, T. (Jul, 2020)
Formation and function of bacterial organelles.
{\em Nat. Rev. Microbiol.,} {\bf 18}, 677--689.

\bibitem{Wu2020}
Wu, X., Cai, Q., Feng, Z., and Zhang, M. (2020)
Liquid-Liquid Phase Separation in Neuronal Development and Synaptic Signaling.
{\em Dev. Cell,} {\bf 55}(1), 18--29.

\bibitem{Wheeler2020}
Wheeler, R.~J. (2020)
Therapeutics---how to treat phase separation-associated diseases.
{\em Emerg. Top. Life Sci.,} {\bf 4}(3), 307--318.

\bibitem{Levental2020}
Levental, I., Levental, K.~R., and Heberle, F.~A. (May, 2020)
Lipid Rafts: Controversies Resolved, Mysteries Remain.
{\em Trends Cell Biol.,} {\bf 30}(5), 341--353.

\bibitem{Chen2020}
Chen, X., Wu, X., Wu, H., and Zhang, M. (2020)
Phase separation at the synapse.
{\em Nat. Neurosci.,} {\bf 23}, 301--310.

\bibitem{Zhang2020}
Zhang, Y., Narlikar, G.~J., and Kutateladze, T.~G. (2021)
Enzymatic Reactions inside Biological Condensates.
{\em J. Mol. Biol.,} {\bf 433}(12), 166624.

\bibitem{Lyon2020}
Lyon, A.~S., Peeples, W.~B., and Rosen, M.~K. (2020)
A framework for understanding the functions of biomolecular condensates across scales.
{\em Nat. Rev. Mol. Cell Biol.,} {\bf 22}, 215--235.

\bibitem{Snead2019}
Snead, W.~T. and Gladfelter, A.~S. (2019)
The Control Centers of Biomolecular Phase Separation: How Membrane Surfaces, PTMs, and Active Processes Regulate Condensation.
{\em Mol. Cell,} {\bf 76}(2), 295--305.

\bibitem{Youn2019}
Youn, J.-Y., Dyakov, B.~J., Zhang, J., Knight, J.~D., Vernon, R.~M., Forman-Kay, J.~D., and Gingras, A.-C. (2019)
Properties of Stress Granule and P-Body Proteomes.
{\em Mol. Cell,} {\bf 76}(2), 286--294.

\bibitem{Spannl2019}
Spannl, S., Tereshchenko, M., Mastromarco, G.~J., Ihn, S.~J., and Lee, H.~O. (2019)
Biomolecular condensates in neurodegeneration and cancer.
{\em Traffic,} {\bf 20}(12), 890--911.

\bibitem{Alberti2019a}
Alberti, S. and Dormann, D. (2019)
Liquid--Liquid Phase Separation in Disease.
{\em Annu. Rev. Genet.,} {\bf 53}(1), 171--194.

\bibitem{Alberti2017}
Alberti, S. (Sep, 2017)
The wisdom of crowds: regulating cell function through condensed states of living matter.
{\em J. Cell Sci.,} {\bf 130}(17), 2789--2796.

\bibitem{Simons2010}
Simons, K. and Gerl, M.~J. (2010)
Revitalizing membrane rafts: new tools and insights..
{\em Nat. Rev. Mol. Cell Biol.,} {\bf 11}(10), 688--699.

\bibitem{Liang2023}
Liang, J. and Cai, D. (2023)
Membrane-less compartments in the nucleus: Separated or connected phases?.
{\em Curr. Opin. Cell Biol.,} {\bf 84}, 102215.

\bibitem{Alberti2025}
Alberti, S., Arosio, P., Best, R.~B., Boeynaems, S., Cai, D., Collepardo-Guevara, R., Dignon, G.~L., Dimova, R., Elbaum-Garfinkle, S., Fawzi, N.~L., Fuxreiter, M., Gladfelter, A.~S., Honigmann, A., Jain, A., Joseph, J.~A., Knowles, T. P.~J., Lasker, K., Lemke, E.~A., Lindorff-Larsen, K., Lipowsky, R., Mittal, J., Mukhopadhyay, S., Myong, S., Pappu, R.~V., Rippe, K., Shelkovnikova, T.~A., Vecchiarelli, A.~G., Wegmann, S., Zhang, H., Zhang, M., Zubieta, C., Zweckstetter, M., Dormann, D., and Mittag, T. (2025)
Current practices in the study of biomolecular condensates: a community comment.
{\em Nat. Commun.,} {\bf 16}(1), 7730.

\bibitem{Sabari2024}
Sabari, B.~R., Hyman, A.~A., and Hnisz, D. (2025)
Functional specificity in biomolecular condensates revealed by genetic complementation.
{\em Nat. Rev. Genet.,} {\bf 26}, 279--290.

\bibitem{Zhou2024}
Zhou, H.-X., Kota, D., Qin, S., and Prasad, R. (2024)
Fundamental Aspects of Phase-Separated Biomolecular Condensates.
{\em Chem. Rev.,} {\bf 124}(13), 8550--8595.

\bibitem{Mangiarotti2024}
Mangiarotti, A. and Dimova, R. (2024)
Biomolecular Condensates in Contact with Membranes.
{\em Annu. Rev. Biophys.,} {\bf 53}(1).

\bibitem{Cao2024}
Cao, S., Song, S., Ivanov, T., Doan-Nguyen, T.~P., Caire~da Silva, L., Xie, J., and Landfester, K. (2024)
Synthetic Biomolecular Condensates: Phase-Separation Control, Cytomimetic Modelling and Emerging Biomedical Potential.
{\em Angew. Chem. Int. Ed. Engl.,} {\bf 64}(8), e202418431.

\bibitem{Romero-Perez2023}
Romero-Perez, P.~S., Dorone, Y., Flores, E., Sukenik, S., and Boeynaems, S. (2023)
When Phased without Water: Biophysics of Cellular Desiccation, from Biomolecules to Condensates.
{\em Chem. Rev.,} {\bf 123}(14), 9010--9035.

\bibitem{Donau2023}
Donau, C. and Boekhoven, J. (2023)
The chemistry of chemically fueled droplets.
{\em Trends Chem.,} {\bf 5}(1), 45--60.

\bibitem{Michelin2023}
Michelin, S. (2023)
Self-Propulsion of Chemically Active Droplets.
{\em Annu. Rev. Fluid Mech.,} {\bf 55}(1), 77--101.

\bibitem{Holt2023}
Holt, L.~J. and Delarue, M. (2023)
Macromolecular crowding: Sensing without a sensor.
{\em Curr. Opin. Cell Biol.,} {\bf 85}, 102269.

\bibitem{Harrington2023}
Harrington, M.~J., Mezzenga, R., and Miserez, A. (2024)
Fluid protein condensates for bio-inspired applications.
{\em Nat. Rev. Bioeng.,} {\bf 2}(3), 260--278.

\bibitem{Abyzov2022}
Abyzov, A., Blackledge, M., and Zweckstetter, M. (03, 2022)
Conformational Dynamics of Intrinsically Disordered Proteins Regulate Biomolecular Condensate Chemistry.
{\em Chem. Rev.,} {\bf 122}(6), 6719--6748.

\bibitem{Villegas2022}
Villegas, J.~A., Heidenreich, M., and Levy, E.~D. (2022)
Molecular and environmental determinants of biomolecular condensate formation.
{\em Nat. Chem. Biol.,} {\bf 18}(12), 1319--1329.

\bibitem{Scholl2021}
Scholl, D. and Deniz, A.~A. (2021)
Conformational freedom and topological confinement of proteins in biomolecular condensates.
{\em J. Mol. Biol.,} {\bf 434}(1), 167348.

\bibitem{Lohse2020}
Lohse, D. and Zhang, X. (2020)
Physicochemical hydrodynamics of droplets out of equilibrium.
{\em Nat. Rev. Phys.,} {\bf 2}, 426--443.

\bibitem{Andreotti2020}
Andreotti, B. and Snoeijer, J.~H. (2020)
Statics and dynamics of soft wetting.
{\em Annu. Rev. Fluid Mech.,} {\bf 52}, 285--308.

\bibitem{Andre2020}
Andr{\'e}, A. A.~M. and Spruijt, E. (2020)
Liquid--Liquid Phase Separation in Crowded Environments.
{\em Int. J. Mol. Sci.,} {\bf 21}(16).

\bibitem{Bracha2019}
Bracha, D., Walls, M.~T., and Brangwynne, C.~P. (Dec, 2019)
Probing and engineering liquid-phase organelles.
{\em Nat. Biotechnol.,} {\bf 37}(12), 1435--1445.

\bibitem{Nakashima2019}
Nakashima, K.~K., Vibhute, M.~A., and Spruijt, E. (2019)
Biomolecular chemistry in liquid phase separated compartments.
{\em Front. Mol. Biosci.,} {\bf 6}.

\bibitem{Cates2015a}
Cates, M.~E. and Tailleur, J. (2015)
Motility-induced phase separation.
{\em Annu. Rev. Condens. Matter Phys.,} {\bf 6}(1), 219--244.

\bibitem{Xu2014}
Xu, X., Ting, C.~L., Kusaka, I., and Wang, Z.-G. (2014)
Nucleation in Polymers and Soft Matter.
{\em Annu. Rev. Phys. Chem.,} {\bf 65}(1), 449--475.

\bibitem{Landfester2006}
Landfester, K. (2006)
Synthesis of Colloidal Particles in Miniemulsions.
{\em Annu. Rev. Mater. Res.,} {\bf 36}(1), 231--279.

\bibitem{Feric2013}
Feric, M. and Brangwynne, C.~P. (2013)
A nuclear F-actin scaffold stabilizes ribonucleoprotein droplets against gravity in large cells..
{\em Nat. Cell Biol.,} {\bf 15}(10), 1253--1259.

\bibitem{Hess2025}
Hess, N. and Joseph, J.~A. (2025)
Structured protein domains enter the spotlight: modulators of biomolecular condensate form and function.
{\em Trends Biochem. Sci.,} {\bf 50}(3), 206--223.

\bibitem{Holehouse2025}
Holehouse, A.~S. and Alberti, S. (2025/01/19, 2025)
Molecular determinants of condensate composition.
{\em Mol. Cell,} {\bf 85}(2), 290--308.

\bibitem{Pappu2023}
Pappu, R.~V., Cohen, S.~R., Dar, F., Farag, M., and Kar, M. (2023)
Phase Transitions of Associative Biomacromolecules.
{\em Chem. Rev.,} {\bf 123}(14), 8945--8987.

\bibitem{Choi2020a}
Choi, J.-M., Holehouse, A.~S., and Pappu, R.~V. (Jan, 2020)
Physical Principles Underlying the Complex Biology of Intracellular Phase Transitions.
{\em Annu. Rev. Biophys.,} {\bf 49}, 107--133.

\bibitem{Sherck2021}
Sherck, N., Shen, K., Nguyen, M., Yoo, B., K{\"o}hler, S., Speros, J.~C., Delaney, K.~T., Shell, M.~S., and Fredrickson, G.~H. (05, 2021)
Molecularly Informed Field Theories from Bottom-up Coarse-Graining.
{\em ACS Macro Lett.,} {\bf 10}(5), 576--583.

\bibitem{Julicher2024}
J\"{u}licher, F. and Weber, C.~A. (2024)
Droplet Physics and Intracellular Phase Separation.
{\em Annu. Rev. Condens. Matter Phys.,} {\bf 15}(1).

\bibitem{Cates2024}
Cates, M.~E. and Nardini, C. (2025)
Active phase separation: new phenomenology from non-equilibrium physics.
{\em Rep. Prog. Phys.,} {\bf 88}(5), 056601.

\bibitem{Jacobs2023}
Jacobs, W.~M. (06, 2023)
Theory and Simulation of Multiphase Coexistence in Biomolecular Mixtures.
{\em J. Chem. Theory Comput.,} {\bf 19}(12), 3429--3445.

\bibitem{Saar2023}
Saar, K.~L., Qian, D., Good, L.~L., Morgunov, A.~S., Collepardo-Guevara, R., Best, R.~B., and Knowles, T. P.~J. (2023)
Theoretical and Data-Driven Approaches for Biomolecular Condensates.
{\em Chem. Rev.,} {\bf 123}(14), 8988--9009.

\bibitem{Zwicker2022a}
Zwicker, D. (2022)
The intertwined physics of active chemical reactions and phase separation.
{\em Curr. Opin. Colloid Interface Sci.,} {\bf 61}, 101606.

\bibitem{Li2022b}
Li, L.-g. and Hou, Z. (2022)
Theoretical modelling of liquid--liquid phase separation: from particle-based to field-based simulation.
{\em Biophys. Rep.,} {\bf 8}(2), 55.

\bibitem{Ghosh2022}
Ghosh, K., Huihui, J., Phillips, M., and Haider, A. (2022)
Rules of Physical Mathematics Govern Intrinsically Disordered Proteins.
{\em Annu. Rev. Biophys.,} {\bf 51}(1).

\bibitem{Berry2018}
Berry, J., Brangwynne, C., and Haataja, M.~P. (Jan, 2018)
Physical Principles of Intracellular Organization via Active and Passive Phase Transitions.
{\em Rep. Prog. Phys.,} {\bf 81}, 046601.

\bibitem{Saarloos2024}
van Saarloos, W., Vitelli, V., and Zeravcic, Z. (2024)
Soft Matter: Concepts, Phenomena, and Applications,
Princeton University Press, .

\bibitem{Doi2013}
Doi, M. (2013)
Soft Matter Physics,
OUP Oxford, .

\bibitem{Onuki2007}
Onuki, A. (2007)
Phase transition dynamics,
Cambridge University Press,  1 edition.

\bibitem{Flory1942}
Flory, P.~I. (1942)
Thermodynamics of high polymer solutions.
{\em J. Chem. Phys.,} {\bf 10}(1), 51--61.

\bibitem{Huggins1941}
Huggins, M.~L. (1941)
Solutions of long chain compounds.
{\em J. Chem. Phys.,} {\bf 9}(5), 440--440.

\bibitem{Julicher2018}
J{\"u}licher, F., Grill, S.~W., and Salbreux, G. (Jul, 2018)
Hydrodynamic theory of active matter.
{\em Rep. Prog. Phys.,} {\bf 81}(7), 076601.

\bibitem{Zwicker2022}
Zwicker, D. and Laan, L. (2022)
Evolved interactions stabilize many coexisting phases in multicomponent liquids.
{\em Proc. Natl. Acad. Sci. USA,} {\bf 119}(28), e2201250119.

\bibitem{Mabillard2023}
Mabillard, J., Weber, C.~A., and J\"ulicher, F. (Jan, 2023)
Heat fluctuations in chemically active systems.
{\em Phys. Rev. E,} {\bf 107}, 014118.

\bibitem{Cahn1958}
Cahn, J.~W. and Hilliard, J.~E. (1958)
Free Energy of a Nonuniform System. I. Interfacial Free Energy.
{\em J. Chem. Phys.,} {\bf 28}(2), 258--267.

\bibitem{Hoyt1990}
Hoyt, J. (1990)
The continuum theory of nucleation in multicomponent systems.
{\em Acta Metall. Mater.,} {\bf 38}(8), 1405--1412.

\bibitem{Tang1991}
Tang, H. and Freed, K.~F. (1991)
Interfacial studies of incompressible binary blends.
{\em J. Chem. Phys.,} {\bf 94}(9), 6307--6322.

\bibitem{Li2022}
Li, A.~B., Miroshnik, L., Rummel, B.~D., Balakrishnan, G., Han, S.~M., and Sinno, T. (2022)
A unified theory of free energy functionals and applications to diffusion.
{\em Proc. Natl. Acad. Sci. USA,} {\bf 119}(23), e2203399119.

\bibitem{Mao2018}
Mao, S., Kuldinow, D., Haataja, M., and Kosmrlj, A. (2018)
Phase behavior and morphology of multicomponent liquid mixtures.
{\em Soft Matter,} {\bf 15}, 1297.

\bibitem{deGennes1980}
de~Gennes, P.~G. (1980)
Dynamics of fluctuations and spinodal decomposition in polymer blends.
{\em J. Chem. Phys.,} {\bf 72}(9), 4756--4763.

\bibitem{DeGroot2013}
De~Groot, S.~R. and Mazur, P. (2013)
Non-Equilibrium Thermodynamics,
Dover Publications Inc., .

\bibitem{Kramer1984}
Kramer, E.~J., Green, P., and Palmstr{\o}m, C.~J. (1984)
Interdiffusion and marker movements in concentrated polymer-polymer diffusion couples.
{\em Polymer,} {\bf 25}(4), 473--480.

\bibitem{Konig2021}
K{\"o}nig, B., Ronsin, O. J.~J., and Harting, J. (2021)
Two-dimensional Cahn--Hilliard simulations for coarsening kinetics of spinodal decomposition in binary mixtures.
{\em Phys. Chem. Chem. Phys.,} {\bf 23}, 24823--24833.

\bibitem{Bo2021}
Bo, S., Hubatsch, L., Bauermann, J., Weber, C.~A., and J\"ulicher, F. (Dec, 2021)
Stochastic dynamics of single molecules across phase boundaries.
{\em Phys. Rev. Research,} {\bf 3}, 043150.

\bibitem{VanVlimmeren1996}
{van Vlimmeren}, B. and Fraaije, J. (1996)
Calculation of noise distribution in mesoscopic dynamics models for phase separation of multicomponent complex fluids.
{\em Comput. Phys. Commun.,} {\bf 99}(1), 21--28.

\bibitem{Decker2011}
Decker, M., Jaensch, S., Pozniakovsky, A., Zinke, A., O'Connell, K.~F., Zachariae, W., Myers, E., and Hyman, A.~A. (2011)
Limiting Amounts of Centrosome Material Set Centrosome Size in C. elegans Embryos.
{\em Curr. Biol.,} {\bf 21}(15), 1259--1267.

\bibitem{Brangwynne2009}
Brangwynne, C.~P., Eckmann, C.~R., Courson, D.~S., Rybarska, A., Hoege, C., Gharakhani, J., J{\"u}licher, F., and Hyman, A.~A. (2009)
Germline P Granules Are Liquid Droplets That Localize by Controlled Dissolution/Condensation.
{\em Science,} {\bf 324}(5935), 1729--1732.

\bibitem{Hird1993}
Hird, S.~N. and White, J.~G. (1993)
Cortical and Cytoplasmic Flow Polarity in Early Embryonic Cells of \textit{Caenorhabditis elegans}.
{\em J. Cell Biol.,} {\bf 121}(6), 1343--1355.

\bibitem{Keating1998}
Keating, H.~H. and White, J.~G. (1998)
Centrosome dynamics in early embryos of Caenorhabditis elegans.
{\em J. Cell Sci.,} {\bf 111}, 3027--3033.

\bibitem{Mahen2011}
Mahen, R., Jeyasekharan, A.~D., Barry, N.~P., and Venkitaraman, A.~R. (2011)
Continuous polo-like kinase 1 activity regulates diffusion to maintain centrosome self-organization during mitosis.
{\em Proc. Natl. Acad. Sci. USA,} {\bf 108}, 9310--9315.

\bibitem{Griffin2011}
Griffin, E., Odde, D., and Seydoux, G. (2011)
Regulation of the MEX-5 gradient by a spatially segregated kinase/phosphatase cycle.
{\em Cell,} {\bf 146}(6), 955--68.

\bibitem{Landau1959_6}
Landau, L.~D. and Lifshits, E.~M. (1959)
Fluid Mechanics, Vol.~6 of Course of Theoretical Physics,
Pergamon Press, .

\bibitem{Zhou2021a}
Zhou, S. and Xie, Y.~M. (2021)
Numerical simulation of three-dimensional multicomponent Cahn--Hilliard systems.
{\em Int. J. Mech. Sci.,} {\bf 198}, 106349.

\bibitem{Zhu1999}
Zhu, J., Chen, L.-Q., Shen, J., and Tikare, V. (1999)
Coarsening kinetics from a variable-mobility Cahn-Hilliard equation: Application of a semi-implicit Fourier spectral method.
{\em Phys. Rev. E,} {\bf 60}(4), 3564.

\bibitem{Shrinivas2021}
Shrinivas, K. and Brenner, M.~P. (2021)
Phase separation in fluids with many interacting components.
{\em Proc. Natl. Acad. Sci. USA,} {\bf 118}(45).

\bibitem{Thampi2011}
Thampi, S.~P., Pagonabarraga, I., and Adhikari, R. (Oct, 2011)
Lattice-Boltzmann-Langevin simulations of binary mixtures.
{\em Phys. Rev. E,} {\bf 84}(4 Pt 2), 046709.

\bibitem{Ledesma-Aguilar2014}
Ledesma-Aguilar, R., Vella, D., and Yeomans, J.~M. (Nov, 2014)
Lattice-Boltzmann simulations of droplet evaporation.
{\em Soft Matter,} {\bf 10}(41), 8267--75.

\bibitem{Semprebon2016}
Semprebon, C., Kr\"uger, T., and Kusumaatmaja, H. (Mar, 2016)
Ternary free-energy lattice Boltzmann model with tunable surface tensions and contact angles.
{\em Phys. Rev. E,} {\bf 93}, 033305.

\bibitem{Gsell2022}
Gsell, S. and Merkel, M. (2022)
Phase separation dynamics in deformable droplets.
{\em Soft Matter,} {\bf 18}, 2672--2683.

\bibitem{GarciaOjalvo2012}
Garc{\'\i}a-Ojalvo, J. and Sancho, J.~M. (2012)
Noise in Spatially Extended Systems,
Springer New York, NY, .

\bibitem{Lord2014}
Lord, G.~J., Powell, C.~E., and Shardlow, T. (2014)
An Introduction to Computational Stochastic PDEs,
Cambridge Texts in Applied MathematicsCambridge University Press, .

\bibitem{Wodo2011}
Wodo, O. and Ganapathysubramanian, B. (2011)
Computationally efficient solution to the Cahn-Hilliard equation: Adaptive implicit time schemes, mesh sensitivity analysis and the 3D isoperimetric problem.
{\em J. Comput. Phys.,} {\bf 230}(15), 6037--6060.

\bibitem{Qian2022a}
Qian, D., Michaels, T. C.~T., and Knowles, T. P.~J. (08, 2022)
Analytical Solution to the Flory--Huggins Model.
{\em J. Phys. Chem. Lett.,} {\bf 13}(33), 7853--7860.

\bibitem{CahnHilliardBook2019}
Miranville, A. (2019)
The Cahn--Hilliard Equation: Recent Advances and Applications,
SIAM, .

\bibitem{Krapivsky2010}
Krapivsky, P.~L., Redner, S., and Ben-Naim, E. (2010)
A Kinetic View of Statistical Physics,
Cambridge University Press, .

\bibitem{Ji2023}
Ji, W., Hachmo, O., Barkai, N., and Amir, A. (sep, 2025)
Design principles of transcription factors with intrinsically disordered regions.
{\em eLife,} {\bf 14}, RP104956.

\bibitem{Michieletto2022}
Michieletto, D. and Marenda, M. (07, 2022)
Rheology and Viscoelasticity of Proteins and Nucleic Acids Condensates.
{\em JACS Au,} {\bf 2}(7), 1506--1521.

\bibitem{Yamaguchi2023}
Yamaguchi, Y. and Kawamura, H.
Introduction to Molecular-Scale Understanding of Surface Tension. (2023).

\bibitem{Dierkes1992}
Dierkes, U., Hildebrandt, S., K{\"u}ster, A., and Wohlrab, O. (1992)
Minimal surfaces I,
Springer, .

\bibitem{Ip1994}
Ip, S.~W. and Toguri, J.~M. (1994)
The Equivalency of Surface-Tension, Surface-Energy and Surface Free-Energy.
{\em J. Mater. Sci.,} {\bf 29}(3), 688--692.

\bibitem{Tolman1949}
Tolman, R.~C. (1949)
The effect of droplet size on surface tension.
{\em J. Chem. Phys.,} {\bf 17}(3), 333--337.

\bibitem{Vidal2020}
Vidal-Henriquez, E. and Zwicker, D. (2020)
Theory of droplet ripening in stiffness gradients.
{\em Soft Matter,} {\bf 16}, 5898--5905.

\bibitem{Thomson1872}
Thomson, W. (1872)
4. On the Equilibrium of Vapour at a Curved Surface of Liquid.
{\em Proc. R. Soc. Edinb.,} {\bf 7}, 63--68.

\bibitem{Kalikmanov2013}
Kalikmanov, V. (2013)
Nucleation theory.
{\em Lecture Notes in Physics,} {\bf 860}, 1--331.

\bibitem{Vidal2021}
Vidal-Henriquez, E. and Zwicker, D. (2021)
Cavitation controls droplet sizes in elastic media.
{\em Proc. Natl. Acad. Sci. USA,} {\bf 118}(40).

\bibitem{Wilken2024}
Wilken, S., Gutierrez, J., and Saleh, O.~A. (06, 2024)
{Nucleation dynamics of a model biomolecular liquid}.
{\em J. Chem. Phys.,} {\bf 160}(21), 214903.

\bibitem{Lifshitz1961}
Lifshitz, I.~M. and Slyozov, V.~V. (1961)
The kinetics of precipitation from supersaturated solid solutions.
{\em J. Phys. Chem. Solids,} {\bf 19}(1-2), 35--50.

\bibitem{Wagner1961}
Wagner, C. (1961)
{Theorie der Alterung von Niederschl{\"a}gen durch Uml{\"o}sen (Ostwald-Reifung)}.
{\em Z. Elektrochem.,} {\bf 65}, 581--591.

\bibitem{Zwicker2017}
Zwicker, D., Seyboldt, R., Weber, C.~A., Hyman, A.~A., and J{\"u}licher, F. (2017)
Growth and division of active droplets provides a model for protocells.
{\em Nat. Phys.,} {\bf 13}(4), 408--413.

\bibitem{Kulkarni2023}
Kulkarni, A., Vidal-Henriquez, E., and Zwicker, D. (2023)
Effective simulations of interacting active droplets.
{\em Sci. Rep.,} {\bf 13}(1), 733.

\bibitem{Weber2017}
Weber, C.~A., Lee, C.~F., and J{\"u}licher, F. (2017)
Droplet ripening in concentration gradients.
{\em New J. Phys.,} {\bf 19}(5), 053021.

\bibitem{Rosowski2019}
Rosowski, K.~A., Sai, T., Vidal-Henriquez, E., Zwicker, D., Style, R.~W., and Dufresne, E.~R. (Jan, 2020)
{Elastic ripening and inhibition of liquid-liquid phase separation}.
{\em Nat. Phys.,} {\bf 16}, 422--425.

\bibitem{Rosowski2020}
Rosowski, K.~A., Vidal-Henriquez, E., Zwicker, D., Style, R.~W., and Dufresne, E.~R. (2020)
Elastic stresses reverse Ostwald ripening.
{\em Soft Matter,} {\bf 16}, 5892.

\bibitem{Cochard2023}
Cochard, A., Safieddine, A., Combe, P., Benassy, M.-N., Weil, D., and Gueroui, Z. (2023)
Condensate functionalization with microtubule motors directs their nucleation in space and allows manipulating RNA localization.
{\em EMBO J.,} {\bf 42}(20), e114106.

\bibitem{Einstein1905}
Einstein, A. (1905)
{\"U}ber die von der molekularkinetischen Theorie der W{\"a}rme geforderte Bewegung von in ruhenden Fl{\"u}ssigkeiten suspendierten Teilchen.
{\em Ann. Phys.,} {\bf 322}, 549--560.

\bibitem{Garner2022}
Garner, R.~M., Molines, A.~T., Theriot, J.~A., and Chang, F. (2023)
Vast heterogeneity in cytoplasmic diffusion rates revealed by nanorheology and Doppelg{\"a}nger simulations.
{\em Biophys. J.,} {\bf 122}(5), 767--783.

\bibitem{Ostwald1897}
Ostwald, W. (1897)
Studien {\"u}ber die {Bildung} und {Umwandlung} fester {K{\"o}rper}.
{\em Z. Phys. Chem,} {\bf 22}(3), 289--330.

\bibitem{Voorhees1992}
Voorhees, P.~W. (1992)
Ostwald ripening of two-phase mixtures.
{\em Annu. Rev. Mater. Sci.,} {\bf 22}, 197--215.

\bibitem{Smoluchowski1918}
v.~Smoluchowski, M. (1918)
Versuch einer mathematischen Theorie der Koagulationskinetik kolloider L{\"o}sungen.
{\em Z. Phys. Chem.,} {\bf 92U}(1), 129--168.

\bibitem{Bray2003}
Bray, A.~J. (2003)
Coarsening dynamics of phase-separating systems.
{\em Phil. Trans. R. Soc. London A,} {\bf 361}(1805), 781--792.

\bibitem{vonHofe2025}
von Hofe, J., Abacousnac, J., Chen, M., Sasazawa, M., Jav{\'e}r~Kristiansen, I., Westrey, S., Grier, D.~G., and Saurabh, S. (07, 2025)
Multivalency Controls the Growth and Dynamics of a Biomolecular Condensate.
{\em J. Am. Chem. Soc.,} {\bf 147}(29), 25242--25253.

\bibitem{Webster1998}
Webster, A. and Cates, M.~E. (1998)
Stabilization of emulsions by trapped species.
{\em Langmuir,} {\bf 14}(8), 2068--2079.

\bibitem{Chen2023b}
Chen, S. and Wang, Z.-G. (Nov, 2023)
Charge Asymmetry Suppresses Coarsening Dynamics in Polyelectrolyte Complex Coacervation.
{\em Phys. Rev. Lett.,} {\bf 131}, 218201.

\bibitem{Lee2021}
Lee, D. S.~W., Wingreen, N.~S., and Brangwynne, C.~P. (2021)
Chromatin mechanics dictates subdiffusion and coarsening dynamics of embedded condensates.
{\em Nat. Phys.,} {\bf 17}, 531--538.

\bibitem{Zhu2001}
Zhu, J., Chen, L.-Q., and Shen, J. (oct, 2001)
Morphological evolution during phase separation and coarsening with strong inhomogeneous elasticity.
{\em Model. Simul. Mater. Sci. Eng.,} {\bf 9}(6), 499.

\bibitem{Tateno2021}
Tateno, M. and Tanaka, H. (2021)
Power-law coarsening in network-forming phase separation governed by mechanical relaxation.
{\em Nat. Commun.,} {\bf 12}(1), 912.

\bibitem{Siggia1979}
Siggia, E.~D. (1979)
Late stages of spinodal decomposition in binary mixtures.
{\em Phys. Rev. A,} {\bf 20}(2), 595--605.

\bibitem{Shimizu2015}
Shimizu, R. and Tanaka, H. (Jun, 2015)
A novel coarsening mechanism of droplets in immiscible fluid mixtures.
{\em Nat. Commun.,} {\bf 6}, 7407.

\bibitem{Wagner1998}
Wagner, A.~J. and Yeomans, J. (1998)
Breakdown of scale invariance in the coarsening of phase-separating binary fluids.
{\em Phys. Rev. Lett.,} {\bf 80}(7), 1429.

\bibitem{Wagner2001}
Wagner, A. and Cates, M.~E. (2001)
Phase ordering of two-dimensional symmetric binary fluids: A droplet scaling state.
{\em Europhys. Lett.,} {\bf 56}(4), 556--562.

\bibitem{Lee2023}
Lee, D. S.~W., Choi, C.-H., Sanders, D.~W., Beckers, L., Riback, J.~A., Brangwynne, C.~P., and Wingreen, N.~S. (2023)
Size distributions of intracellular condensates reflect competition between coalescence and nucleation.
{\em Nat. Phys.,} {\bf 19}, 586--596.

\bibitem{Banerjee2024}
Banerjee, D.~S., Chigumira, T., Lackner, R.~M., Kratz, J.~C., Chenoweth, D.~M., Banerjee, S., and Zhang, H. (2024)
Interplay of condensate material properties and chromatin heterogeneity governs nuclear condensate ripening.
{\em eLife,} {\bf 13}, RP101777.

\bibitem{Rossetto2024}
Rossetto, R., Wellecke, G., and Zwicker, D. (May, 2025)
Binding and dimerization control phase separation in a compartment.
{\em Phys. Rev. Res.,} {\bf 7}, 023145.

\bibitem{Rana2024}
Rana, U., Xu, K., Narayanan, A., Walls, M.~T., Panagiotopoulos, A.~Z., Avalos, J.~L., and Brangwynne, C.~P. (2024)
{Asymmetric oligomerization state and sequence patterning can tune multiphase condensate miscibility}.
{\em Nat. Chem.,} {\bf 16}, 1073--1082.

\bibitem{Bland2023}
Bland, T., Hirani, N., Briggs, D.~C., Rossetto, R., Ng, K., Taylor, I.~A., McDonald, N.~Q., Zwicker, D., and Goehring, N.~W. (2024)
Optimized PAR-2 RING dimerization mediates cooperative and selective membrane binding for robust cell polarity.
{\em EMBO J.,} {\bf 43}(15), 3214--3239.

\bibitem{Fritsch2021}
Fritsch, A.~W., Diaz-Delgadillo, A.~F., Adame-Arana, O., Hoege, C., Mittasch, M., Kreysing, M., Leaver, M., Hyman, A.~A., J{\"u}licher, F., and Weber, C.~A. (2021)
Local thermodynamics govern formation and dissolution of Caenorhabditis elegans P granule condensates.
{\em Proc. Natl. Acad. Sci. USA,} {\bf 118}(37).

\bibitem{Stormo2024}
Stormo, B.~M., McLaughlin, G.~A., Jalihal, A.~P., Frederick, L.~K., Cole, S.~J., Seim, I., Dietrich, F.~S., Chilkoti, A., and Gladfelter, A.~S. (2024)
Intrinsically disordered sequences can tune fungal growth and the cell cycle for specific temperatures.
{\em Curr. Biol.,} {\bf 34}(16), 3722--3734.e7.

\bibitem{Meyer2025}
Meyer, H.~M., Hotta, T., Malkovskiy, A.~V., Zheng, Y., and Ehrhardt, D.~W. (2025)
Manipulating condensation of thermo-sensitive SUF4 protein tunes flowering time in \textit{Arabidopsis thaliana}.
{\em Cell Rep.,} {\bf 44}(7).

\bibitem{adame2020liquid}
Adame-Arana, O., Weber, C.~A., Zaburdaev, V., Prost, J., and J{\"u}licher, F. (2020)
Liquid phase separation controlled by pH.
{\em Biophys. J.,} {\bf 119}(8), 1590--1605.

\bibitem{Jalihal2020}
Jalihal, A.~P., Pitchiaya, S., Xiao, L., Bawa, P., Jiang, X., Bedi, K., Parolia, A., Cieslik, M., Ljungman, M., Chinnaiyan, A.~M., and Walter, N.~G. (2020)
Multivalent Proteins Rapidly and Reversibly Phase-Separate upon Osmotic Cell Volume Change.
{\em Mol. Cell,} {\bf 79}(6), 978 -- 990.e5.

\bibitem{Zhang2022a}
Zhang, M., Zhu, C., Duan, Y., Liu, T., Liu, H., Su, C., and Lu, Y. (2022)
The intrinsically disordered region from PP2C phosphatases functions as a conserved CO2 sensor.
{\em Nat. Cell Biol.,} {\bf 24}, 1029--1037.

\bibitem{Agrawal2022}
Agrawal, A., Douglas, J.~F., Tirrell, M., and Karim, A. (2022)
Manipulation of coacervate droplets with an electric field.
{\em Proc. Natl. Acad. Sci. USA,} {\bf 119}(32), e2203483119.

\bibitem{Emenecker2021}
Emenecker, R.~J., Holehouse, A.~S., and Strader, L.~C. (2021)
Biological Phase Separation and Biomolecular Condensates in Plants.
{\em Annu. Rev. Plant Biol.,} {\bf 72}(1).

\bibitem{Keyport-Kik2024}
Keyport~Kik, S., Christopher, D., Glauninger, H., Hickernell, C.~W., Bard, J. A.~M., Lin, K.~M., Squires, A.~H., Ford, M., Sosnick, T.~R., and Drummond, D.~A. (2024)
An adaptive biomolecular condensation response is conserved across environmentally divergent species.
{\em Nat. Commun.,} {\bf 15}(1), 3127.

\bibitem{Cejkova2014}
{\v C}ejkov{\'a}, J., Nov{\'a}k, M., {\v S}t{\v e}p{\'a}nek, F., and Hanczyc, M.~M. (10, 2014)
Dynamics of Chemotactic Droplets in Salt Concentration Gradients.
{\em Langmuir,} {\bf 30}(40), 11937--11944.

\bibitem{Saha2016}
Saha, S., Weber, C.~A., Nousch, M., Adame-Arana, O., Hoege, C., Hein, M.~Y., Osborne-Nishimura, E., Mahamid, J., Jahnel, M., Jawerth, L., Pozniakovski, A., Eckmann, C.~R., J{\"u}licher, F., and Hyman, A.~A. (Sep, 2016)
Polar Positioning of Phase-Separated Liquid Compartments in Cells Regulated by an mRNA Competition Mechanism.
{\em Cell,} {\bf 166}(6), 1572--1584.e16.

\bibitem{Doan2024}
Doan, V.~S., Alshareedah, I., Singh, A., Banerjee, P.~R., and Shin, S. (2024)
Diffusiophoresis promotes phase separation and transport of biomolecular condensates.
{\em Nat. Commun.,} {\bf 15}(1), 7686.

\bibitem{Romano2025}
Romano, J., Golestanian, R., and Mahault, B.
Dynamics of Phase-Separated Interfaces in Inhomogenous and Driven Mixtures. (2025).

\bibitem{Kruger2018}
Kr{\"u}ger, S., Weber, C.~A., Sommer, J.-U., and J{\"u}licher, F. (2018)
Discontinuous switching of position of two coexisting phases.
{\em New J. Phys.,} {\bf 20}(7), 075009.

\bibitem{Hafner2023a}
H{\"a}fner, G. and M{\"u}ller, M. (2023)
Reaction-driven assembly: controlling changes in membrane topology by reaction cycles.
{\em Soft Matter,} {\bf 19}, 7281--7292.

\bibitem{Shim2022}
Shim, S. (04, 2022)
Diffusiophoresis, Diffusioosmosis, and Microfluidics: Surface-Flow-Driven Phenomena in the Presence of Flow.
{\em Chem. Rev.,} {\bf 122}(7), 6986--7009.

\bibitem{Haugerud2023}
Haugerud, I.~S., Jaiswal, P., and Weber, C.~A. (2024)
Nonequilibrium Wet--Dry Cycling Acts as a Catalyst for Chemical Reactions.
{\em J. Phys. Chem. B,} {\bf 128}(7), 1724--1736.

\bibitem{Tariq2024}
Tariq, D., Maurici, N., Bartholomai, B.~M., Chandrasekaran, S., Dunlap, J.~C., Bah, A., and Crane, B.~R. (mar, 2024)
Phosphorylation, disorder, and phase separation govern the behavior of Frequency in the fungal circadian clock.
{\em eLife,} {\bf 12}, RP90259.

\bibitem{Heltberg2022}
Heltberg, M.~S., Lucchetti, A., Hsieh, F.-S., {Minh Nguyen}, D.~P., Chen, S.-h., and Jensen, M.~H. (2022)
Enhanced DNA repair through droplet formation and p53 oscillations.
{\em Cell,} {\bf 185}(23), 4394--4408.e10.

\bibitem{Bartolucci2023a}
Bartolucci, G., Serrão, A.~C., Schwintek, P., Kühnlein, A., Rana, Y., Janto, P., Hofer, D., Mast, C.~B., Braun, D., and Weber, C.~A. (2023)
Sequence self-selection by cyclic phase separation.
{\em Proc. Natl. Acad. Sci. USA,} {\bf 120}(43), e2218876120.

\bibitem{Klosin2020}
Klosin, A., Oltsch, F., Harmon, T., Honigmann, A., J{{\"u}}licher, F., Hyman, A.~A., and Zechner, C. (Jan, 2020)
Phase separation provides a mechanism to reduce noise in cells.
{\em Science,} {\bf 367}(6476), 464--468.

\bibitem{Zechner2024}
Zechner, C. and J{\"u}licher, F. (2025)
Concentration buffering and noise reduction in non-equilibrium phase-separating systems.
{\em Cell Syst.,} {\bf 16}(2), 101168.

\bibitem{Devirie2021}
Deviri, D. and Safran, S.~A. (2021)
Physical theory of biological noise buffering by multicomponent phase separation.
{\em Proc. Natl. Acad. Sci. USA,} {\bf 118}(25).

\bibitem{Michaels2025}
Michaels, T., de~Monchaux-Irons, L., Han, N., Fr{\"u}hbauer, B., Jiang, J., Allain, F., Jagannathan, M., and Emmanouilidis, L.
Volume buffering in multi-component phase separation. (2025).

\bibitem{Cooper2000}
Cooper, G.~M. (2000)
The cell: A molecular approach,
Sunderland (MA): Sinauer Associates, .

\bibitem{Updike2009}
Updike, D.~L. and Strome, S. (12, 2009)
{A Genomewide RNAi Screen for Genes That Affect the Stability, Distribution and Function of P Granules in Caenorhabditis elegans}.
{\em Genetics,} {\bf 183}(4), 1397--1419.

\bibitem{Jain2016}
Jain, S., Wheeler, J.~R., Walters, R.~W., Agrawal, A., Barsic, A., and Parker, R. (2016)
ATPase-Modulated Stress Granules Contain a Diverse Proteome and Substructure.
{\em Cell,} {\bf 164}(3), 487--498.

\bibitem{Hubstenberger2017}
Hubstenberger, A., Courel, M., B{\'e}nard, M., Souquere, S., Ernoult-Lange, M., Chouaib, R., Yi, Z., Morlot, J.-B., Munier, A., Fradet, M., Daunesse, M., Bertrand, E., Pierron, G., Mozziconacci, J., Kress, M., and Weil, D. (2017)
P-Body Purification Reveals the Condensation of Repressed mRNA Regulons.
{\em Mol. Cell,} {\bf 68}(1), 144--157.e5.

\bibitem{Scholl2024}
Scholl, D., Boyd, T., Latham, A.~P., Salazar, A., Khan, A., Boeynaems, S., Holehouse, A.~S., Lander, G.~C., Sali, A., Park, D., Deniz, A.~A., and Lasker, K. (2024)
Cellular Function of a Biomolecular Condensate Is Determined by Its Ultrastructure.
{\em bioRxiv,}.

\bibitem{Tollervey2024}
Tollervey, F., Rios, M.~U., Zagoriy, E., Woodruff, J.~B., and Mahamid, J. (2025)
Molecular architectures of centrosomes in \textit{C. elegans} embryos visualized by cryo-electron tomography.
{\em Developmental Cell,} {\bf 60}(6), 885--900.e5.

\bibitem{Dollinger2025}
Dollinger, C., Potolitsyna, E., Martin, A.~G., Anand, A., Datar, G.~K., Schmit, J.~D., and Riback, J.~A. (2025)
Nanometer condensate organization in live cells derived from partitioning measurements.
{\em bioRxiv,}.

\bibitem{Currie2023}
Currie, S.~L., Xing, W., Muhlrad, D., Decker, C.~J., Parker, R., and Rosen, M.~K. (2023)
Quantitative reconstitution of yeast RNA processing bodies.
{\em Proc. Natl. Acad. Sci. USA,} {\bf 120}(14), e2214064120.

\bibitem{Arter2022}
Arter, W.~E., Qi, R., Erkamp, N.~A., Krainer, G., Didi, K., Welsh, T.~J., Acker, J., Nixon-Abell, J., Qamar, S., Guill{\'e}n-Boixet, J., Franzmann, T.~M., Kuster, D., Hyman, A.~A., Borodavka, A., George-Hyslop, P.~S., Alberti, S., and Knowles, T. P.~J. (2022)
Biomolecular condensate phase diagrams with a combinatorial microdroplet platform.
{\em Nat. Commun.,} {\bf 13}(1), 7845.

\bibitem{Qian2022}
Qian, D., Welsh, T.~J., Erkamp, N.~A., Qamar, S., Nixon-Abell, J., Krainer, G., St. George-Hyslop, P., Michaels, T. C.~T., and Knowles, T. P.~J. (Dec, 2022)
Tie-Line Analysis Reveals Interactions Driving Heteromolecular Condensate Formation.
{\em Phys. Rev. X,} {\bf 12}, 041038.

\bibitem{Ibrahim2024}
Ibrahim, K.~A., Naidu, A.~S., Miljkovic, H., Radenovic, A., and Yang, W. (2024)
Label-Free Techniques for Probing Biomolecular Condensates.
{\em ACS Nano,} {\bf 18}(16), 10738--10757.

\bibitem{McCall2020}
McCall, P.~M., Kim, K., Shevchenko, A., Ruer-Gru{\ss}, M., Peychl, J., Guck, J., Shevchenko, A., Hyman, A.~A., and Brugu{\'e}s, J. (2025)
A label-free method for measuring the composition of multicomponent biomolecular condensates.
{\em Nat. Chem.,}.

\bibitem{Dorner2024}
D{\"o}rner, K., Gut, M., Overwijn, D., Cao, F., Siketanc, M., Heinrich, S., Beuret, N., Sharpe, T., Lindorff-Larsen, K., and Maria, H. (2024)
Tag with Caution - How protein tagging influences the formation of condensates.
{\em bioRxiv,}.

\bibitem{Ning2020}
Ning, W., Guo, Y., Lin, S., Mei, B., Wu, Y., Jiang, P., Tan, X., Zhang, W., Chen, G., Peng, D., Chu, L., and Xue, Y. (Jan, 2020)
DrLLPS: a data resource of liquid-liquid phase separation in eukaryotes.
{\em Nucleic Acids Res,} {\bf 48}(D1), D288--D295.

\bibitem{GuillenBoixet2020}
Guill{\'e}n-Boixet, J., Kopach, A., Holehouse, A.~S., Wittmann, S., Jahnel, M., Schl{\"u}{\ss}ler, R., Kim, K., Trussina, I.~R., Wang, J., Mateju, D., Poser, I., Maharana, S., Ruer-Gru{\ss}, M., Richter, D., Zhang, X., Chang, Y.-T., Guck, J., Honigmann, A., Mahamid, J., Hyman, A.~A., Pappu, R.~V., Alberti, S., and Franzmann, T.~M. (2020)
RNA-Induced Conformational Switching and Clustering of G3BP Drive Stress Granule Assembly by Condensation.
{\em Cell,} {\bf 181}(2), 346--361.e17.

\bibitem{Joseph2020}
Joseph, J.~A., Espinosa, J.~R., Sanchez-Burgos, I., Garaizar, A., Frenkel, D., and Collepardo-Guevara, R. (2021)
Thermodynamics and kinetics of phase separation of protein-RNA mixtures by a minimal model.
{\em Biophys. J.,} {\bf 120}(7), 1219--1230.

\bibitem{Wheeler2018}
Wheeler, R.~J. and Hyman, A.~A. (05, 2018)
Controlling compartmentalization by non-membrane-bound organelles.
{\em Philos. Trans. R. Soc. B,} {\bf 373}(1747).

\bibitem{Banani2016}
Banani, S.~F., Rice, A.~M., Peeples, W.~B., Lin, Y., Jain, S., Parker, R., and Rosen, M.~K. (2016)
Compositional Control of Phase-Separated Cellular Bodies.
{\em Cell,} {\bf 166}(3), 651--663.

\bibitem{Kabalnov1987a}
Kabalnov, A.~S., Pertzov, A.~V., and Shchukin, E.~D. (00, 1987)
Ostwald ripening in two-component disperse phase systems: Application to emulsion stability.
{\em Colloids Surf.,} {\bf 24}(1), 19--32.

\bibitem{Thewes2023}
Thewes, F.~C., Kr\"uger, M., and Sollich, P. (Aug, 2023)
Composition Dependent Instabilities in Mixtures with Many Components.
{\em Phys. Rev. Lett.,} {\bf 131}, 058401.

\bibitem{Desouza2024}
de~Souza, J.~P. and Stone, H.~A. (07, 2024)
Exact analytical solution of the Flory--Huggins model and extensions to multicomponent systems.
{\em J. Chem. Phys.,} {\bf 161}(4), 044902.

\bibitem{Qian2023}
Qian, D., Ausserwoger, H., Sneideris, T., Farag, M., Pappu, R.~V., and Knowles, T. P.~J. (2024)
Dominance analysis to assess solute contributions to multicomponent phase equilibria.
{\em Proc. Natl. Acad. Sci. USA,} {\bf 121}(33), e2407453121.

\bibitem{Qian2023a}
Qian, D., Ausserwoger, H., Arter, W.~E., Scrutton, R.~M., Welsh, T.~J., Kartanas, T., Ermann, N., Qamar, S., Fischer, C.~M., Sneideris, T., St~George-Hyslop, P., Pappu, R.~V., and Knowles, T.~P. (Jun, 2025)
Molecular mechanisms of condensate modulation from energy-dominance analysis.
{\em Phys. Rev. Appl.,} {\bf 23}, 064017.

\bibitem{Nordenskiold2024}
Nordenski{\"o}ld, L., Shi, X., Korolev, N., Zhao, L., Zhai, Z., and Lindman, B. (2024)
Liquid-liquid phase separation (LLPS) in DNA and chromatin systems from the perspective of colloid physical chemistry.
{\em Adv. Colloid Interface Sci.,} {\bf 326}, 103133.

\bibitem{vonBulow2024}
von B{\"u}low, S., Tesei, G., Zaidi, F.~K., Mittag, T., and Lindorff-Larsen, K. (2025)
Prediction of phase-separation propensities of disordered proteins from sequence.
{\em Proc. Natl. Acad. Sci. USA,} {\bf 122}(13), e2417920122.

\bibitem{Semenov1998}
Semenov, A.~N. and Rubinstein, M. (02, 1998)
Thermoreversible Gelation in Solutions of Associative Polymers. 1. Statics.
{\em Macromolecules,} {\bf 31}(4), 1373--1385.

\bibitem{Rubinstein2001}
Rubinstein, M. and Semenov, A.~N. (02, 2001)
Dynamics of Entangled Solutions of Associating Polymers.
{\em Macromolecules,} {\bf 34}(4), 1058--1068.

\bibitem{Chang2017}
Chang, L.-W., Lytle, T.~K., Radhakrishna, M., Madinya, J.~J., V{\'e}lez, J., Sing, C.~E., and Perry, S.~L. (Nov, 2017)
Sequence and entropy-based control of complex coacervates.
{\em Nat. Commun.,} {\bf 8}(1), 1273.

\bibitem{Danielsen2023}
Danielsen, S. P.~O., Semenov, A.~N., and Rubinstein, M. (07, 2023)
Phase Separation and Gelation in Solutions and Blends of Heteroassociative Polymers.
{\em Macromolecules,} {\bf 56}(14), 5661--5677.

\bibitem{Harmon2017}
Harmon, T.~S., Holehouse, A.~S., Rosen, M.~K., and Pappu, R.~V. (nov, 2017)
Intrinsically disordered linkers determine the interplay between phase separation and gelation in multivalent proteins.
{\em eLife,} {\bf 6}, e30294.

\bibitem{GrandPre2023}
GrandPre, T., Zhang, Y., Pyo, A. G.~T., Weiner, B., Li, J.-L., Jonikas, M.~C., and Wingreen, N.~S. (Dec, 2023)
Impact of Linker Length on Biomolecular Condensate Formation.
{\em PRX Life,} {\bf 1}, 023013.

\bibitem{Adachi2024}
Adachi, K. and Kawaguchi, K. (Jul, 2024)
Predicting Heteropolymer Interactions: Demixing and Hypermixing of Disordered Protein Sequences.
{\em Phys. Rev. X,} {\bf 14}, 031011.

\bibitem{Maristany2025}
Maristany, M.~J., Gonzalez, A.~A., Espinosa, J.~R., Huertas, J., Collepardo-Guevara, R., and Joseph, J.~A. (feb, 2025)
Decoding phase separation of prion-like domains through data-driven scaling laws.
{\em eLife,} {\bf 13}, RP99068.

\bibitem{Sundaravadivelu-Devarajan2024}
Sundaravadivelu~Devarajan, D., Wang, J., Sza{\l}a-Mendyk, B., Rekhi, S., Nikoubashman, A., Kim, Y.~C., and Mittal, J. (2024)
Sequence-dependent material properties of biomolecular condensates and their relation to dilute phase conformations.
{\em Nat. Commun.,} {\bf 15}(1), 1912.

\bibitem{Rekhi2024}
Rekhi, S., Garcia, C.~G., Barai, M., Rizuan, A., Schuster, B.~S., Kiick, K.~L., and Mittal, J. (2024)
Expanding the molecular language of protein liquid--liquid phase separation.
{\em Nat. Chem.,} {\bf 16}, 1113--1124.

\bibitem{Biswas2025}
Biswas, S. and Potoyan, D.~A. (02, 2025)
Decoding biomolecular condensate dynamics: an energy landscape approach.
{\em PLoS Comput. Biol.,} {\bf 21}(2), 1--19.

\bibitem{Kar2022}
Kar, M., Dar, F., Welsh, T.~J., Vogel, L.~T., K{\"u}hnemuth, R., Majumdar, A., Krainer, G., Franzmann, T.~M., Alberti, S., Seidel, C. A.~M., Knowles, T. P.~J., Hyman, A.~A., and Pappu, R.~V. (2022)
Phase-separating RNA-binding proteins form heterogeneous distributions of clusters in subsaturated solutions.
{\em Proc. Natl. Acad. Sci. USA,} {\bf 119}(28), e2202222119.

\bibitem{Lan2023}
Lan, C., Kim, J., Ulferts, S., Aprile-Garcia, F., Weyrauch, S., Anandamurugan, A., Grosse, R., Sawarkar, R., Reinhardt, A., and Hugel, T. (2023)
Quantitative real-time in-cell imaging reveals heterogeneous clusters of proteins prior to condensation.
{\em Nat. Commun.,} {\bf 14}(1), 4831.

\bibitem{Gao2024}
Gao, G., Sumrall, E.~R., and Walter, N.~G. (2024)
Single molecule tracking reveals nanodomains in biomolecular condensates.
{\em bioRxiv,}.

\bibitem{Hofweber2019}
Hofweber, M. and Dormann, D. (2019)
Friend or foe---Post-translational modifications as regulators of phase separation and RNP granule dynamics.
{\em J. Biol. Chem.,} {\bf 294}(18), 7137--7150.

\bibitem{Soeding2019}
Soeding, J., Zwicker, D., Sohrabi-Jahromi, S., Boehning, M., and Kirschbaum, J. (2020)
Mechanisms of active regulation of biomolecular condensates.
{\em Trends Cell Biol.,} {\bf 30}(1), 4--14.

\bibitem{Shin2018}
Shin, Y., Chang, Y.-C., Lee, D.~S., Berry, J., Sanders, D.~W., Ronceray, P., Wingreen, N.~S., Haataja, M., and Brangwynne, C.~P. (2018)
Liquid nuclear condensates mechanically sense and restructure the genome.
{\em Cell,} {\bf 175}(6), 1481--1491.

\bibitem{Schmit2021}
Schmit, J.~D., Feric, M., and Dundr, M. (2021)
How Hierarchical Interactions Make Membraneless Organelles Tick Like Clockwork.
{\em Trends Biochem. Sci.,} {\bf 46}(7), 525--534.

\bibitem{Schmit2020}
Schmit, J.~D., Bouchard, J.~J., Martin, E.~W., and Mittag, T. (2020)
Protein Network Structure Enables Switching between Liquid and Gel States.
{\em J. Am. Chem. Soc.,} {\bf 142}(2), 874--883.

\bibitem{SanfeliuCerdan2025}
Sanfeliu-Cerd{\'a}n, N. and Krieg, M. (03, 2025)
The mechanobiology of biomolecular condensates.
{\em Biophys. Rev.,} {\bf 6}(1), 011310.

\bibitem{Woodruff2021}
Woodruff, J.~B. (2021)
The material state of centrosomes: lattice, liquid, or gel?.
{\em Curr. Opin. Struct. Biol.,} {\bf 66}, 139--147.

\bibitem{Paulin2025}
Paulin, O.~W., Garcia-Baucells, J., Zieger, L., Aland, S., Dammermann, A., and Zwicker, D.
Active viscoelastic condensates provide controllable mechanical anchor points. (2025).

\bibitem{Hester2023}
Hester, E.~W., Carney, S., Shah, V., Arnheim, A., Patel, B., Carlo, D.~D., and Bertozzi, A.~L. (2023)
Fluid dynamics alters liquid--liquid phase separation in confined aqueous two-phase systems.
{\em Proc. Natl. Acad. Sci. USA,} {\bf 120}(49), e2306467120.

\bibitem{Guo2015}
Guo, L. and Shorter, J. (Oct, 2015)
It's Raining Liquids: RNA Tunes Viscoelasticity and Dynamics of Membraneless Organelles.
{\em Mol. Cell,} {\bf 60}(2), 189--92.

\bibitem{Persson2020}
Persson, L.~B., Ambati, V.~S., and Brandman, O. (2020)
Cellular Control of Viscosity Counters Changes in Temperature and Energy Availability.
{\em Cell,} {\bf 183}(6), 1572--1585.e16.

\bibitem{Anderson1998}
Anderson, D.~M., McFadden, G.~B., and Wheeler, A.~A. (00, 1998)
Diffuse-Interface Methods in Fluid Mechanics.
{\em Annu. Rev. Fluid Mech.,} {\bf 30}(1), 139--165.

\bibitem{Jawerth2018}
Jawerth, L.~M., Ijavi, M., Ruer, M., Saha, S., Jahnel, M., Hyman, A.~A., J{{\"u}}licher, F., and Fischer-Friedrich, E. (2018)
Salt-dependent rheology and surface tension of protein condensates using optical traps.
{\em Phys. Rev. Lett.,} {\bf 121}(25), 258101.

\bibitem{jawerth2020protein}
Jawerth, L., Fischer-Friedrich, E., Saha, S., Wang, J., Franzmann, T., Zhang, X., Sachweh, J., Ruer, M., Ijavi, M., Saha, S., et al. (2020)
Protein condensates as aging Maxwell fluids.
{\em Science,} {\bf 370}(6522), 1317--1323.

\bibitem{Cheng2025}
Cheng, H.~H., Roggeveen, J.~V., Wang, H., Stone, H.~A., Shi, Z., and Brangwynne, C.~P. (2025)
Micropipette aspiration reveals differential RNA-dependent viscoelasticity of nucleolar subcompartments.
{\em Proc. Natl. Acad. Sci. USA,} {\bf 122}(22), e2407423122.

\bibitem{Tejedor2023}
Tejedor, A.~R., Collepardo-Guevara, R., Ram{\'\i}rez, J., and Espinosa, J.~R. (2023)
Time-Dependent Material Properties of Aging Biomolecular Condensates from Different Viscoelasticity Measurements in Molecular Dynamics Simulations.
{\em J. Phys. Chem. B,} {\bf 127}(20), 4441--4459.

\bibitem{Cohen2024}
Cohen, S.~R., Banerjee, P.~R., and Pappu, R.~V. (09, 2024)
{Direct computations of viscoelastic moduli of biomolecular condensates}.
{\em J. Chem. Phys.,} {\bf 161}(9), 095103.

\bibitem{tanaka2000viscoelastic}
Tanaka, H. (2000)
Viscoelastic phase separation.
{\em J. Phys. Condens. Matter,} {\bf 12}(15), R207.

\bibitem{tanaka2006viscoelastic}
Tanaka, H. and Araki, T. (2006)
Viscoelastic phase separation in soft matter: Numerical-simulation study on its physical mechanism.
{\em Chem. Eng. Sci.,} {\bf 61}(7), 2108--2141.

\bibitem{Tanaka2022}
Tanaka, H. (2022)
Viscoelastic phase separation in biological cells.
{\em Commun. Phys.,} {\bf 5}(1), 167.

\bibitem{Meng2023}
Meng, L. and Lin, J. (Feb, 2023)
Indissoluble biomolecular condensates via elasticity.
{\em Phys. Rev. Res.,} {\bf 5}, L012024.

\bibitem{Ghosh2021}
Ghosh, A., Kota, D., and Zhou, H.-X. (2021)
Shear relaxation governs fusion dynamics of biomolecular condensates.
{\em Nat. Commun.,} {\bf 12}(1), 5995.

\bibitem{Ceballos2022}
Ceballos, A.~V., A, J. A.~D., Preston, J.~M., Vairamon, C., Shen, C., Koder, R.~L., and Elbaum-Garfinkle, S. (2022)
Liquid to solid transition of elastin condensates.
{\em Proc. Natl. Acad. Sci. USA,} {\bf 119}(37), e2202240119.

\bibitem{patel2015liquid}
Patel, A., Lee, H.~O., Jawerth, L., Maharana, S., Jahnel, M., Hein, M.~Y., Stoynov, S., Mahamid, J., Saha, S., Franzmann, T.~M., et al. (2015)
A liquid-to-solid phase transition of the ALS protein FUS accelerated by disease mutation.
{\em Cell,} {\bf 162}(5), 1066--1077.

\bibitem{lin2015formation}
Lin, Y., Protter, D.~S., Rosen, M.~K., and Parker, R. (2015)
Formation and maturation of phase-separated liquid droplets by RNA-binding proteins.
{\em Molecular cell,} {\bf 60}(2), 208--219.

\bibitem{Takaki2023}
Takaki, R., Jawerth, L., Popovi{\'c}, M., and J{\"u}licher, F. (Aug, 2023)
Theory of Rheology and Aging of Protein Condensates.
{\em PRX Life,} {\bf 1}, 013006.

\bibitem{Roy2024}
Roy, H.~L. and Rios, P. D.~L.
Microscopic model for aging of biocondensates. (2024).

\bibitem{Garaizar2022}
Garaizar, A., Espinosa, J.~R., Joseph, J.~A., and Collepardo-Guevara, R. (2022)
Kinetic interplay between droplet maturation and coalescence modulates shape of aged protein condensates.
{\em Sci. Rep.,} {\bf 12}(1), 4390.

\bibitem{Garaizar2022a}
Garaizar, A., Espinosa, J.~R., Joseph, J.~A., Krainer, G., Shen, Y., Knowles, T.~P., and Collepardo-Guevara, R. (2022)
Aging can transform single-component protein condensates into multiphase architectures.
{\em Proc. Natl. Acad. Sci. USA,} {\bf 119}(26), e2119800119.

\bibitem{Lin2022}
Lin, J. (Apr, 2022)
Modeling the aging of protein condensates.
{\em Phys. Rev. Research,} {\bf 4}, L022012.

\bibitem{Woodruff2018}
Woodruff, J.~B., Hyman, A.~A., and Boke, E. (2018)
Organization and Function of Non-dynamic Biomolecular Condensates.
{\em Trends Biochem. Sci.,} {\bf 43}(2), 81--94.

\bibitem{Mittag2022}
Mittag, T. and Pappu, R.~V. (2022)
A conceptual framework for understanding phase separation and addressing open questions and challenges.
{\em Mol. Cell,} {\bf 82}(12), 2201--2214.

\bibitem{Zhang2024a}
Zhang, R., Mao, S., and Haataja, M.~P. (2024)
Chemically reactive and aging macromolecular mixtures I: Phase diagrams, spinodals, and gelation.
{\em J. Chem. Phys.,} {\bf 160}(24).

\bibitem{Zhang2024b}
Zhang, R., Mao, S., and Haataja, M.~P. (2024)
Chemically reactive and aging macromolecular mixtures II: Phase separation and coarsening.
{\em J Chem Phys,} {\bf 18}(161).

\bibitem{Kang2025}
Kang, W., Wu, Z., Huang, X., Qi, H., Wu, J., Wang, J., Li, J., Wu, S., Kang, B.-H., Li, B., Ma, J., and Xue, C. (2025)
Time-dependent catalytic activity in aging condensates.
{\em Nat. Commun.,} {\bf 16}(1), 6959.

\bibitem{Chen2024b}
Chen, F., Guo, W., and Shum, H.~C. (Sep, 2024)
Fractal-Dependent Growth of Solidlike Condensates.
{\em Phys. Rev. Lett.,} {\bf 133}, 118401.

\bibitem{Shen2023}
Shen, Y., Chen, A., Wang, W., Shen, Y., Ruggeri, F.~S., Aime, S., Wang, Z., Qamar, S., Espinosa, J.~R., Garaizar, A., George-Hyslop, P.~S., Collepardo-Guevara, R., Weitz, D.~A., Vigolo, D., and Knowles, T. P.~J. (2023)
The liquid-to-solid transition of FUS is promoted by the condensate surface.
{\em Proc. Natl. Acad. Sci. USA,} {\bf 120}(33), e2301366120.

\bibitem{Erkamp2023a}
Erkamp, N.~A., Sneideris, T., Ausserw{\"o}ger, H., Qian, D., Qamar, S., Nixon-Abell, J., St~George-Hyslop, P., Schmit, J.~D., Weitz, D.~A., and Knowles, T. P.~J. (2023)
Spatially non-uniform condensates emerge from dynamically arrested phase separation.
{\em Nat. Commun.,} {\bf 14}(1), 684.

\bibitem{Michaels2022}
Michaels, T. C.~T., Mahadevan, L., and Weber, C.~A. (Dec, 2022)
Enhanced potency of aggregation inhibitors mediated by liquid condensates.
{\em Phys. Rev. Res.,} {\bf 4}, 043173.

\bibitem{Bartolucci2023}
Bartolucci, G., Haugerud, I.~S., Michaels, T.~C., and Weber, C.~A. (2024)
The interplay between biomolecular assembly and phase separation.
{\em eLife,} {\bf 13}, RP93003.

\bibitem{Das2025}
Das, T., Zaidi, F.~K., Farag, M., Ruff, K.~M., Mahendran, T.~S., Singh, A., Gui, X., Messing, J., Taylor, J.~P., Banerjee, P.~R., Pappu, R.~V., and Mittag, T. (2025)
Tunable metastability of condensates reconciles their dual roles in amyloid fibril formation.
{\em Mol. Cell,} {\bf 85}(11), 2230--2245.e7.

\bibitem{Shen2020}
Shen, Y., Ruggeri, F.~S., Vigolo, D., Kamada, A., Qamar, S., Levin, A., Iserman, C., Alberti, S., George-Hyslop, P.~S., and Knowles, T. P.~J. (2020)
Biomolecular condensates undergo a generic shear-mediated liquid-to-solid transition.
{\em Nat. Nanotechnol.,} {\bf 15}(10), 841--847.

\bibitem{Hubatsch2024}
Hubatsch, L., Bo, S., Harmon, T.~S., Hyman, A.~A., Weber, C.~A., and J{\"u}licher, F. (September, 2025)
Transport kinetics across interfaces between coexisting liquid phases.
{\em eLife,}.

\bibitem{Zhang2024}
Zhang, Y., Pyo, A.~G., Kliegman, R., Jiang, Y., Brangwynne, C.~P., Stone, H.~A., and Wingreen, N.~S. (sep, 2024)
The exchange dynamics of biomolecular condensates.
{\em eLife,} {\bf 12}, RP91680.

\bibitem{Majee2024}
Majee, A., Weber, C.~A., and J\"ulicher, F. (Aug, 2024)
Charge separation at liquid interfaces.
{\em Phys. Rev. Res.,} {\bf 6}, 033138.

\bibitem{Folkmann2021}
Folkmann, A.~W., Putnam, A., Lee, C.~F., and Seydoux, G. (2021)
Regulation of biomolecular condensates by interfacial protein clusters.
{\em Science,} {\bf 373}(6560), 1218--1224.

\bibitem{Boeddeker2022}
B{\"o}ddeker, T.~J., Rosowski, K.~A., Berchtold, D., Emmanouilidis, L., Han, Y., Allain, F. H.~T., Style, R.~W., Pelkmans, L., and Dufresne, E.~R. (2022)
Non-specific adhesive forces between filaments and membraneless organelles.
{\em Nat. Phys,} {\bf 18}, 571--578.

\bibitem{Oh2025}
Oh, H.~J., Lee, Y., Hwang, H., Hong, K., Choi, H., Kang, J.~Y., and Jung, Y. (2025)
Size-controlled assembly of phase separated protein condensates with interfacial protein cages.
{\em Nat. Commun.,} {\bf 16}(1), 1009.

\bibitem{Favetta2025}
Favetta, B., Wang, H., Shi, Z., and Schuster, B.~S. (2025)
Amphiphilic Protein Surfactants Reduce the Interfacial Tension of Biomolecular Condensates.
{\em Langmuir,} {\bf 41}(35), 23827--23836.

\bibitem{Kelley2021}
Kelley, F.~M., Favetta, B., Regy, R.~M., Mittal, J., and Schuster, B.~S. (2021)
Amphiphilic proteins coassemble into multiphasic condensates and act as biomolecular surfactants.
{\em Proc. Natl. Acad. Sci. USA,} {\bf 118}(51), e2109967118.

\bibitem{Erkamp2025}
Erkamp, N.~A., Farag, M., Qiu, Y., Qian, D., Sneideris, T., Wu, T., Welsh, T.~J., Ausserw{\"o}ger, H., Krug, T.~J., Chauhan, G., Weitz, D.~A., Lew, M.~D., Knowles, T. P.~J., and Pappu, R.~V. (2025)
Differential interactions determine anisotropies at interfaces of RNA-based biomolecular condensates.
{\em Nat. Commun.,} {\bf 16}(1), 3463.

\bibitem{Binks2002}
Binks, B.~P. (2002)
Particles as surfactants---similarities and differences.
{\em Curr. Opin. Colloid Interface Sci.,} {\bf 7}(1), 21--41.

\bibitem{Aveyard2012}
Aveyard, R. (2012)
Can Janus particles give thermodynamically stable Pickering emulsions?.
{\em Soft Matter,} {\bf 8}, 5233--5240.

\bibitem{Golani2025}
Golani, G., Seal, M., Kar, M., Hyman, A.~A., Goldfarb, D., and Safran, S. (2025)
Mesoscale properties of protein clusters determine the size and nature of liquid-liquid phase separation (LLPS).
{\em Commun. Phys.,} {\bf 8}(1), 226.

\bibitem{HernandezArmendariz2023}
Hernandez-Armendariz, A., Sorichetti, V., Hayashi, Y., Koskova, Z., Brunner, A., Ellenberg, J., {\v S}ari{\'c}, A., and Cuylen-Haering, S. (2024)
A liquid-like coat mediates chromosome clustering during mitotic exit.
{\em Mol. Cell,} {\bf 84}(17), 3254--3270.e9.

\bibitem{Farag2022}
Farag, M., Cohen, S.~R., Borcherds, W.~M., Bremer, A., Mittag, T., and Pappu, R.~V. (2022)
Condensates formed by prion-like low-complexity domains have small-world network structures and interfaces defined by expanded conformations.
{\em Nat. Commun.,} {\bf 13}(1), 7722.

\bibitem{Maass2016}
Maass, C.~C., Kr{\"u}ger, C., Herminghaus, S., and Bahr, C. (2016)
Swimming Droplets.
{\em Annu. Rev. Condens. Matter Phys.,} {\bf 7}(Volume 7, 2016), 171--193.

\bibitem{JambonPuillet2023}
Jambon-Puillet, E., Testa, A., Lorenz, C., Style, R.~W., Rebane, A.~A., and Dufresne, E.~R. (2024)
Phase-separated droplets swim to their dissolution.
{\em Nat. Commun.,} {\bf 15}(3919).

\bibitem{Dai2024a}
Dai, Y., Wang, Z.-G., and Zare, R.~N. (2024)
Unlocking the electrochemical functions of biomolecular condensates.
{\em Nat. Chem. Biol.,} {\bf 20}(11), 1420--1433.

\bibitem{Zhou2018}
Zhou, H.-X. and Pang, X. (02, 2018)
Electrostatic Interactions in Protein Structure, Folding, Binding, and Condensation.
{\em Chem. Rev.,} {\bf 118}(4), 1691--1741.

\bibitem{Dai2024}
Dai, Y., Zhou, Z., Yu, W., Ma, Y., Kim, K., Rivera, N., Mohammed, J., Lantelme, E., Hsu-Kim, H., Chilkoti, A., and You, L. (2024)
Biomolecular condensates regulate cellular electrochemical equilibria.
{\em Cell,} {\bf 187}(21), 5951--5966.e18.

\bibitem{Pak2016}
Pak, C.~W., Kosno, M., Holehouse, A.~S., Padrick, S.~B., Mittal, A., Ali, R., Yunus, A.~A., Liu, D.~R., Pappu, R.~V., and Rosen, M.~K. (2016)
Sequence Determinants of Intracellular Phase Separation by Complex Coacervation of a Disordered Protein.
{\em Mol. Cell,} {\bf 63}(1), 72--85.

\bibitem{Wang2024b}
Wang, J., Chen, X., Chen, E.-Q., and Yang, S. (2024)
Interfacial Tensions of Polyelectrolyte Multiphase Coacervation.
{\em Macromolecules,} {\bf 57}(20), 9698--9710.

\bibitem{Yu2025}
Yu, W., Guo, X., Xia, Y., Ma, Y., Tong, Z., Yang, L., Song, X., Zare, R.~N., Hong, G., and Dai, Y. (2025)
Aging-dependent evolving electrochemical potentials of biomolecular condensates regulate their physicochemical activities.
{\em Nat. Chem.,} {\bf 17}, 756--766.

\bibitem{Posey2024}
Posey, A.~E., Bremer, A., Erkamp, N.~A., Pant, A., Knowles, T. P.~J., Dai, Y., Mittag, T., and Pappu, R.~V. (2024)
Biomolecular Condensates are Characterized by Interphase Electric Potentials.
{\em J. Am. Chem. Soc.,} {\bf 146}(41), 28268--28281.

\bibitem{Zhang2021d}
Zhang, P. and Wang, Z.-G. (12, 2021)
Interfacial Structure and Tension of Polyelectrolyte Complex Coacervates.
{\em Macromolecules,} {\bf 54}(23), 10994--11007.

\bibitem{Chen2022b}
Chen, S., Zhang, P., and Wang, Z.-G. (05, 2022)
Complexation between Oppositely Charged Polyelectrolytes in Dilute Solution: Effects of Charge Asymmetry.
{\em Macromolecules,} {\bf 55}(10), 3898--3909.

\bibitem{Li2022a}
Li, S.-F. and Muthukumar, M. (07, 2022)
Theory of Microphase Separation in Concentrated Solutions of Sequence-Specific Charged Heteropolymers.
{\em Macromolecules,} {\bf 55}(13), 5535--5549.

\bibitem{Luo2024a}
Luo, C., Hess, N., Aierken, D., Qiang, Y., Joseph, J.~A., and Zwicker, D. (09, 2025)
Theory of Condensate Size Control by Molecular Charge Asymmetry.
{\em ACS Macro Lett.,} {\bf 14}, 1484--1491.

\bibitem{Smokers2024b}
Smokers, I. B.~A., Lavagna, E., Freire, R. V.~M., Paloni, M., Voets, I.~K., Barducci, A., White, P.~B., Khajehpour, M., and Spruijt, E. (2025)
Selective Ion Binding and Uptake Shape the Microenvironment of Biomolecular Condensates.
{\em J. Am. Chem. Soc.,} {\bf 147}(29), 25692--25704.

\bibitem{Gurunian2024}
Gurunian, A., Lasker, K., and Deniz, A.~A. (2024)
Biomolecular Condensates can Induce Local Membrane Potentials.
{\em bioRxiv,}.

\bibitem{Deserno2001}
Deserno, M. (2001)
Rayleigh instability of charged droplets in the presence of counterions.
{\em Eur. Phys. J. E,} {\bf 6}(2), 163--168.

\bibitem{Rayleigh1882}
Rayleigh, L. (1882)
On the equilibrium of liquid conducting masses charged with electricity', Lond.
{\em Philos. Mag.,} {\bf 14}, 184--186.

\bibitem{Fredrickson2006}
Fredrickson, G. (2006)
The equilibrium theory of inhomogeneous polymers,
Oxford University Press, .

\bibitem{Dobrynin2005}
Dobrynin, A.~V. and Rubinstein, M. (2005)
Theory of polyelectrolytes in solutions and at surfaces.
{\em Prog. Polym. Sci.,} {\bf 30}(11), 1049--1118.

\bibitem{Muthukumar2017}
Muthukumar, M. (12, 2017)
50th Anniversary Perspective: A Perspective on Polyelectrolyte Solutions.
{\em Macromolecules,} {\bf 50}(24), 9528--9560.

\bibitem{Debye1923}
Debye, P. and H{{\"u}}ckel, E. (1923)
Zur Theorie der Elektrolyte. Gefrierpunktserniedrigung und verwandte Erscheinungen.
{\em Physikalische Zeitschrift,} {\bf 24}, 185--206.

\bibitem{Wright2007}
Wright, M. (2007)
An Introduction to Aqueous Electrolyte Solutions,
Wiley, .

\bibitem{Ahn2024}
Ahn, S.~Y. and Obermeyer, A.~C. (2024)
Selectivity of Complex Coacervation in Multiprotein Mixtures.
{\em JACS Au,} {\bf 4}(10), 3800--3812.

\bibitem{Sing2017}
Sing, C.~E. (Jan, 2017)
Development of the modern theory of polymeric complex coacervation.
{\em Adv Colloid Interface Sci,} {\bf 239}, 2--16.

\bibitem{Obermeyer2016}
Obermeyer, A.~C., Mills, C.~E., Dong, X.-H., Flores, R.~J., and Olsen, B.~D. (2016)
Complex coacervation of supercharged proteins with polyelectrolytes.
{\em Soft Matter,} {\bf 12}, 3570--3581.

\bibitem{Bungenberg1929}
Bungenberg~de Jong, H. and Kruyt, H. (1929)
Coacervation (partial miscibility in colloid systems).
In \emph{Proc. K. Ned. Akad. Wet}
Vol.~32,  pp. 849--856.

\bibitem{Oparin1952}
Oparin, A.~I. (1938)
The Origin of Life,
Dover Publications, Inc., New York,
Translated by Sergius Morgulis.

\bibitem{Brangwynne2012}
Brangwynne, C.~P. and Hyman, A.~A. (2012)
In Retrospect: The Origin of Life.
{\em Nature,} {\bf 491}(7425), 524--525.

\bibitem{Overbeek1957}
Overbeek, J.~T. and Voorn, M.~J. (May, 1957)
Phase separation in polyelectrolyte solutions; theory of complex coacervation.
{\em J. Cell. Physiol. Suppl.,} {\bf 49}(Suppl 1), 7--22; discussion, 22--6.

\bibitem{Kumari2022}
Kumari, S., Dwivedi, S., and Podgornik, R. (06, 2022)
On the nature of screening in Voorn--Overbeek type theories.
{\em J. Chem. Phys.,} {\bf 156}(24), 244901.

\bibitem{Adar2017}
Adar, R.~M., Markovich, T., and Andelman, D. (05, 2017)
{Bjerrum pairs in ionic solutions: A Poisson-Boltzmann approach}.
{\em J. Chem. Phys.,} {\bf 146}(19), 194904.

\bibitem{Theillet2014}
Theillet, F.-X., Binolfi, A., Frembgen-Kesner, T., Hingorani, K., Sarkar, M., Kyne, C., Li, C., Crowley, P.~B., Gierasch, L., Pielak, G.~J., et al. (2014)
Physicochemical properties of cells and their effects on intrinsically disordered proteins (IDPs).
{\em Chem. Rev.,} {\bf 114}(13), 6661--6714.

\bibitem{Li2018b}
Li, L., Srivastava, S., Andreev, M., Marciel, A.~B., de~Pablo, J.~J., and Tirrell, M.~V. (04, 2018)
Phase Behavior and Salt Partitioning in Polyelectrolyte Complex Coacervates.
{\em Macromolecules,} {\bf 51}(8), 2988--2995.

\bibitem{Dobrynin1995}
{Andrey V. Dobrynin} and {Michael Rubinstein} (1995)
Flory Theory of a Polyampholyte Chain.
{\em J. Phys. II France,} {\bf 5}(5), 677--695.

\bibitem{Rumyantsev2025}
Rumyantsev, A.~M. and Johner, A. (2025)
Electrostatically Stabilized Microstructures: From Clusters to Necklaces to Bulk Microphases.
{\em ACS Macro Lett.,} {\bf 14}(4), 472--483.

\bibitem{Sear2003}
Sear, R.~P. and Cuesta, J. (2003)
Instabilities in complex mixtures with a large number of components..
{\em Phys. Rev. Lett.,} {\bf 91}(24), 245701--245701/4.

\bibitem{Sear2005}
Sear, R. (2005)
The cytoplasm of living cells: A functional mixture of thousands of components.
{\em J. Phys.: Condens. Matter,} {\bf 17}(45), S3587--S3595.

\bibitem{Sear2008}
Sear, R.~P. (2008)
Phase separation of equilibrium polymers of proteins in living cells.
{\em Faraday Discuss.,} {\bf 139}, 21--34.

\bibitem{Jacobs2013}
Jacobs, W.~M. and Frenkel, D. (2013)
Predicting phase behavior in multicomponent mixtures..
{\em J. Chem. Phys.,} {\bf 139}(2), 024108.

\bibitem{Jacobs2017}
Jacobs, W.~M. and Frenkel, D. (Feb, 2017)
Phase Transitions in Biological Systems with Many Components.
{\em Biophys. J.,} {\bf 112}(4), 683--691.

\bibitem{Binous2021}
Binous, H. and Bellagi, A. (2021)
Calculation of ternary liquid-liquid equilibrium data using arc-length continuation.
{\em Eng. Rep.,} {\bf 3}(2), e12296.

\bibitem{Lee1992}
Lee, D.~D., Choy, J.~H., and Lee, J.~K. (1992)
Computer generation of binary and ternary phase diagrams via a convex hull method.
{\em J. Phase Equilibria,} {\bf 13}(4), 365--372.

\bibitem{Wolff2011}
Wolff, J., Marques, C.~M., and Thalmann, F. (Mar, 2011)
Thermodynamic Approach to Phase Coexistence in Ternary Phospholipid-Cholesterol Mixtures.
{\em Phys. Rev. Lett.,} {\bf 106}, 128104.

\bibitem{Dhamankar2024}
Dhamankar, S., Jiang, S., and Webb, M.~A. (2025)
Accelerating multicomponent phase-coexistence calculations with physics-informed neural networks.
{\em Mol. Syst. Des. Eng.,} {\bf 10}, 89--101.

\bibitem{Livan2018}
Livan, G., Novaes, M., and Vivo, P. (2018)
Introduction to random matrices theory and practice,
Springer, .

\bibitem{Qiang2024}
Qiang, Y., Luo, C., and Zwicker, D. (Oct, 2025)
Scaling laws for phase coexistence in multicomponent mixtures.
{\em Phys. Rev. Res.,} {\bf 7}, 043008.

\bibitem{Rostam2023}
Rostam, N., Ghosh, S., Chow, C. F.~W., Hadarovich, A., Landerer, C., Ghosh, R., Moon, H., Hersemann, L., Mitrea, D.~M., Klein, I.~A., Hyman, A.~A., and Toth-Petroczy, A. (Apr, 2023)
CD-CODE: crowdsourcing condensate database and encyclopedia.
{\em Nat. Methods,} {\bf 20}(673--676).

\bibitem{Chaderjian2025}
Chaderjian, A.~S., Wilken, S., and Saleh, O.~A.
Diverse, Distinct, and Densely Packed DNA Droplets. (2025).

\bibitem{DeLaCruz2024}
De~La~Cruz, N., Pradhan, P., Veettil, R.~T., Conti, B.~A., Oppikofer, M., and Sabari, B.~R. (2024)
Disorder-mediated interactions target proteins to specific condensates.
{\em Mol. Cell,} {\bf 84}(18), 3497--3512.e9.

\bibitem{Chen2023a}
Chen, F. and Jacobs, W.~M. (2024)
Emergence of multiphase condensates from a limited set of chemical building blocks.
{\em J. Chem. Theory Comput.,} {\bf 20}(15).

\bibitem{Graf2022}
Graf, I.~R. and Machta, B.~B. (Aug, 2022)
Thermodynamic stability and critical points in multicomponent mixtures with structured interactions.
{\em Phys. Rev. Research,} {\bf 4}, 033144.

\bibitem{Carugno2022}
Carugno, G., Neri, I., and Vivo, P. (jul, 2022)
Instabilities of complex fluids with partially structured and partially random interactions.
{\em Phys. Biol.,} {\bf 19}(5), 056001.

\bibitem{Kilgore2025}
Kilgore, H.~R., Chinn, I., Mikhael, P.~G., Mitnikov, I., Dongen, C.~V., Zylberberg, G., Afeyan, L., Banani, S.~F., Wilson-Hawken, S., Lee, T.~I., Barzilay, R., and Young, R.~A. (2025)
Protein codes promote selective subcellular compartmentalization.
{\em Science,} {\bf 387}(6738), 1095--1101.

\bibitem{Chen2023}
Chen, F. and Jacobs, W.~M. (06, 2023)
{Programmable phase behavior in fluids with designable interactions}.
{\em J. Chem. Phys.,} {\bf 158}(21), 214118.

\bibitem{Teixeira2023}
Teixeira, R.~B., Carugno, G., Neri, I., and Sartori, P. (2024)
Liquid Hopfield model: Retrieval and localization in multicomponent liquid mixtures.
{\em Proc. Natl. Acad. Sci. USA,} {\bf 121}(48), e2320504121.

\bibitem{Quinodoz2024}
Quinodoz, S.~A., Jiang, L., Abu-Alfa, A.~A., Comi, T.~J., Zhao, H., Yu, Q., Wiesner, L.~W., Botello, J.~F., Donlic, A., Soehalim, E., Bhat, P., Zorbas, C., Wacheul, L., Ko{\v s}mrlj, A., Lafontaine, D. L.~J., Klinge, S., and Brangwynne, C.~P. (2025)
Mapping and engineering RNA-driven architecture of the multiphase nucleolus.
{\em Nature,} {\bf 644}(8076), 557--566.

\bibitem{Riback2020}
Riback, J.~A., Zhu, L., Ferrolino, M.~C., Tolbert, M., Mitrea, D.~M., Sanders, D.~W., Wei, M.-T., Kriwacki, R.~W., and Brangwynne, C.~P. (2020)
Composition-dependent thermodynamics of intracellular phase separation.
{\em Nature,} {\bf 581}, 209--214.

\bibitem{Luo2024}
Luo, C., Qiang, Y., and Zwicker, D. (Jul, 2024)
Beyond pairwise: Higher-order physical interactions affect phase separation in multicomponent liquids.
{\em Phys. Rev. Res.,} {\bf 6}, 033002.

\bibitem{Galvanetto2024}
Galvanetto, N., Ivanovi{\'c}, M.~T., Grosso, S. A.~D., Chowdhury, A., Sottini, A., Nettels, D., Best, R.~B., and Schuler, B. (2025)
Material properties of biomolecular condensates emerge from nanoscale dynamics.
{\em Proc. Natl. Acad. Sci. USA,} {\bf 122}(23), e2424135122.

\bibitem{Snead2025}
Snead, W.~T., Skillicorn, M.~K., Shrinivas, K., and Gladfelter, A.~S. (2025)
Immiscible proteins compete for RNA binding to order condensate layers.
{\em Proc. Natl. Acad. Sci. USA,} {\bf 122}(32), e2504778122.

\bibitem{Yan2025a}
Yan, X., Kuster, D., Mohanty, P., Nijssen, J., Pombo-Garc{\'\i}a, K., {Garcia Morato}, J., Rizuan, A., Franzmann, T.~M., Sergeeva, A., Ly, A.~M., Liu, F., Passos, P.~M., George, L., Wang, S.-H., Shenoy, J., Danielson, H.~L., Ozguney, B., Honigmann, A., Ayala, Y.~M., Fawzi, N.~L., Dickson, D.~W., Rossoll, W., Mittal, J., Alberti, S., and Hyman, A.~A. (2025)
Intra-condensate demixing of TDP-43 inside stress granules generates pathological aggregates.
{\em Cell,} {\bf 188}(15), 4123--4140.e18.

\bibitem{Li2024}
Li, T. and Jacobs, W.~M. (Jun, 2024)
Predicting the Morphology of Multiphase Biomolecular Condensates from Protein Interaction Networks.
{\em PRX Life,} {\bf 2}, 023013.

\bibitem{Yo2021}
Yu, H., Lu, S., Gasior, K., Singh, D., Vazquez-Sanchez, S., Tapia, O., Toprani, D., Beccari, M.~S., Yates, J.~R., Cruz, S.~D., Newby, J.~M., Lafarga, M., Gladfelter, A.~S., Villa, E., and Cleveland, D.~W. (2021)
HSP70 chaperones RNA-free TDP-43 into anisotropic intranuclear liquid spherical shells.
{\em Science,} {\bf 371}(6529), eabb4309.

\bibitem{Mao2020}
Mao, S., Chakraverti-Wuerthwein, M.~S., Gaudio, H., and Ko\ifmmode~\check{s}\else \v{s}\fi{}mrlj, A. (Nov, 2020)
Designing the Morphology of Separated Phases in Multicomponent Liquid Mixtures.
{\em Phys. Rev. Lett.,} {\bf 125}, 218003.

\bibitem{Pyo2023}
Pyo, A. G.~T., Zhang, Y., and Wingreen, N.~S. (2023)
Proximity to criticality predicts surface properties of biomolecular condensates.
{\em Proc. Natl. Acad. Sci. USA,} {\bf 120}(23), e2220014120.

\bibitem{Gouveia2022}
Gouveia, B., Kim, Y., Shaevitz, J.~W., Petry, S., Stone, H.~A., and Brangwynne, C.~P. (2022)
Capillary forces generated by biomolecular condensates.
{\em Nature,} {\bf 609}(7926), 255--264.

\bibitem{Cahn1977}
Cahn, J.~W. (1977)
Critical point wetting.
{\em J. Chem. Phys.,} {\bf 66}(8), 3667--3672.

\bibitem{Bonn2009}
Bonn, D., Eggers, J., Indekeu, J., Meunier, J., and Rolley, E. (May, 2009)
Wetting and spreading.
{\em Rev. Mod. Phys.,} {\bf 81}, 739--805.

\bibitem{deGennes1985}
de~Gennes, P.~G. (Jul, 1985)
Wetting: statics and dynamics.
{\em Rev. Mod. Phys.,} {\bf 57}, 827--863.

\bibitem{deGennes2003}
De~Gennes, P.-G., Brochard-Wyart, F., and Qu{\'e}r{\'e}, D. (2003)
Capillarity and wetting phenomena: drops, bubbles, pearls, waves,
Springer Science \& Business Media, .

\bibitem{Zhao2021}
Zhao, X., Bartolucci, G., Honigmann, A., J{\"u}licher, F., and Weber, C.~A. (dec, 2021)
Thermodynamics of wetting, prewetting and surface phase transitions with surface binding.
{\em New J. Phys.,} {\bf 23}(12), 123003.

\bibitem{cahn1977critical}
Cahn, J.~W. (1977)
Critical point wetting.
{\em J. Chem. Phys.,} {\bf 66}(8), 3667--3672.

\bibitem{Zhao2024}
Zhao, X., Liese, S., Honigmann, A., J{\"u}licher, F., and Weber, C.~A. (oct, 2024)
Theory of wetting dynamics with surface binding.
{\em New J. Phys.,} {\bf 26}(10), 103025.

\bibitem{style2013universal}
Style, R.~W., Boltyanskiy, R., Che, Y., Wettlaufer, J., Wilen, L.~A., and Dufresne, E.~R. (2013)
Universal deformation of soft substrates near a contact line and the direct measurement of solid surface stresses.
{\em Phys. Rev. Lett.,} {\bf 110}(6), 066103.

\bibitem{Lu2025}
Lu, T., Liese, S., Visser, B.~S., van Haren, M. H.~I., Lipi{\'n}ski, W.~P., Huck, W. T.~S., Weber, C.~A., and Spruijt, E. (2025)
Controlling Multiphase Coacervate Wetting and Self-Organization by Interfacial Proteins.
{\em J. Am. Chem. Soc.,} {\bf 147}(26), 22622--22633.

\bibitem{Feric2016}
Feric, M., Vaidya, N., Harmon, T.~S., Mitrea, D.~M., Zhu, L., Richardson, T.~M., Kriwacki, R.~W., Pappu, R.~V., and Brangwynne, C.~P. (2016)
Coexisting Liquid Phases Underlie Nucleolar Subcompartments.
{\em Cell,} {\bf 165}(7), 1686--1697.

\bibitem{volmer1939kinetik}
Volmer, M. et al. (1939)
Kinetik der phasenbildung,
T. Steinkopff, .

\bibitem{turnbull1950kinetics}
Turnbull, D. (1950)
Kinetics of heterogeneous nucleation.
{\em J. Chem. Phys.,} {\bf 18}(2), 198--203.

\bibitem{fletcher1958size}
Fletcher, N.~H. (1958)
Size effect in heterogeneous nucleation.
{\em J. Chem. Phys.,} {\bf 29}(3), 572--576.

\bibitem{Shevtsov2011}
Shevtsov, S.~P. and Dundr, M. (2011)
Nucleation of nuclear bodies by RNA.
{\em Nat. Cell Biol.,} {\bf 13}(2), 167--173.

\bibitem{Shimobayashi2021}
Shimobayashi, S.~F., Ronceray, P., Sanders, D.~W., Haataja, M.~P., and Brangwynne, C.~P. (2021)
Nucleation landscape of biomolecular condensates.
{\em Nature,} {\bf 599}(7885), 503--506.

\bibitem{Schede2023}
Schede, H.~H., Natarajan, P., Chakraborty, A.~K., and Shrinivas, K. (2023)
A model for organization and regulation of nuclear condensates by gene activity.
{\em Nat. Commun.,} {\bf 14}(1), 4152.

\bibitem{Visser2024a}
Visser, B.~S., van Haren, M. H.~I., Lipi{\'n}ski, W.~P., van Leijenhorst-Groener, K.~A., Claessens, M.~M., Queir{\'o}s, M. V.~A., Ramos, C. H.~I., Eeftens, J., and Spruijt, E. (2024)
Controlling interfacial protein adsorption, desorption and aggregation in biomolecular condensates.
{\em bioRxiv,}.

\bibitem{Broedersz2014}
Broedersz, C.~P. and MacKintosh, F.~C. (2014)
Modeling semiflexible polymer networks.
{\em Rev. Mod. Phys.,} {\bf 86}(3), 995.

\bibitem{setru2021hydrodynamic}
Setru, S.~U., Gouveia, B., Alfaro-Aco, R., Shaevitz, J.~W., Stone, H.~A., and Petry, S. (2021)
A hydrodynamic instability drives protein droplet formation on microtubules to nucleate branches.
{\em Nat. Phys.,} {\bf 17}(4), 493--498.

\bibitem{Jijumon2022}
Jijumon, A.~S., Bodakuntla, S., Genova, M., Bangera, M., Sackett, V., Besse, L., Maksut, F., Henriot, V., Magiera, M.~M., Sirajuddin, M., and Janke, C. (2022)
Lysate-based pipeline to characterize microtubule-associated proteins uncovers unique microtubule behaviours.
{\em Nat. Cell Biol.,} {\bf 24}(2), 253--267.

\bibitem{Hernandez-Vega2017}
Hern{\'a}ndez-Vega, A., Braun, M., Scharrel, L., Jahnel, M., Wegmann, S., Hyman, B.~T., Alberti, S., Diez, S., and Hyman, A.~A. (Sep, 2017)
Local Nucleation of Microtubule Bundles through Tubulin Concentration into a Condensed Tau Phase.
{\em Cell Rep.,} {\bf 20}(10), 2304--2312.

\bibitem{Feigeles2025}
Feigeles, C.~A., Brasovs, A., Puchalski, A., Laukat, O., Kornev, K.~G., and Weirich, K.~L. (2025)
Protein condensates induce biopolymer filament bundling and network remodeling via capillary interactions.
{\em Soft Matter,} {\bf 21}, 7872--7880.

\bibitem{Strom2024}
Strom, A.~R., Kim, Y., Zhao, H., Chang, Y.-C., Orlovsky, N.~D., Ko{\v s}mrlj, A., Storm, C., and Brangwynne, C.~P. (2024)
Condensate interfacial forces reposition DNA loci and probe chromatin viscoelasticity.
{\em Cell,} {\bf 187}(19), 5282--5297.e20.

\bibitem{Boddeker2023}
B\"oddeker, T.~J., Rusch, A., Leeners, K., Murrell, M.~P., and Dufresne, E.~R. (Dec, 2023)
Actin and Microtubules Position Stress Granules.
{\em PRX Life,} {\bf 1}, 023010.

\bibitem{Quail2021}
Quail, T., Golfier, S., Elsner, M., Ishihara, K., Murugesan, V., Renger, R., J{{\"u}}licher, F., and Brugu{\'e}s, J. (2021)
Force generation by protein--DNA co-condensation.
{\em Nat. Phys.,} {\bf 17}, 1007--1012.

\bibitem{Takaki2025}
Takaki, R., Savich, Y., Brugu\'es, J., and J\"ulicher, F. (Mar, 2025)
Active Loop Extrusion Guides DNA-Protein Condensation.
{\em Phys. Rev. Lett.,} {\bf 134}, 128401.

\bibitem{botterbusch2021interactions}
Botterbusch, S. and Baumgart, T. (2021)
Interactions between Phase-Separated Liquids and Membrane Surfaces.
{\em Appl. Sci.,} {\bf 11}(3).

\bibitem{Lipowsky2023}
Lipowsky, R. (2023)
Remodeling of Biomembranes and Vesicles by Adhesion of Condensate Droplets.
{\em Membranes,} {\bf 13}(2).

\bibitem{Mondal2023}
Mondal, S. and Baumgart, T. (2023)
Membrane reshaping by protein condensates.
{\em Biochim. Biophys. Acta Biomembr.,} {\bf 1865}(3), 184121.

\bibitem{Mangiarotti2025}
Mangiarotti, A., Sabri, E., Schmidt, K.~V., Hoffmann, C., Milovanovic, D., Lipowsky, R., and Dimova, R. (2025)
Lipid packing and cholesterol content regulate membrane wetting and remodeling by biomolecular condensates.
{\em Nat. Commun.,} {\bf 16}(1), 2756.

\bibitem{Lipowsky2025}
Lipowsky, R. (2025)
Complex remodeling of biomembranes and vesicles by condensate droplets.
{\em Soft Matter,} pp.~--.

\bibitem{GrandPre2024a}
GrandPre, T., Pyo, A. G.~T., and Wingreen, N.~S. (2024)
Membrane wetting by biomolecular condensates is facilitated by mobile tethers.
{\em bioRxiv,}.

\bibitem{Snead2022}
Snead, W.~T., Jalihal, A.~P., Gerbich, T.~M., Seim, I., Hu, Z., and Gladfelter, A.~S. (2022)
Membrane surfaces regulate assembly of ribonucleoprotein condensates.
{\em Nat. Cell Biol.,} {\bf 24}, 461--470.

\bibitem{Liu2023c}
Liu, Z., Yethiraj, A., and Cui, Q. (2023)
Sensitive and selective polymer condensation at membrane surface driven by positive co-operativity.
{\em Proc. Natl. Acad. Sci. USA,} {\bf 120}(15), e2212516120.

\bibitem{Agudo2021}
Agudo-Canalejo, J., Schultz, S.~W., Chino, H., Migliano, S.~M., Saito, C., Koyama-Honda, I., Stenmark, H., Brech, A., May, A.~I., Mizushima, N., and Knorr, R.~L. (2021)
Wetting regulates autophagy of phase-separated compartments and the cytosol.
{\em Nature,} {\bf 591}(7848), 142--146.

\bibitem{Winter2024}
Winter, A., Liu, Y., Ziepke, A., Dadunashvili, G., and Frey, E. (2025)
Phase separation on deformable membranes: Interplay of mechanical coupling and dynamic surface geometry.
{\em Phys. Rev. E,} {\bf 111}, 044405.

\bibitem{Bergeron2021endocytic}
Bergeron-Sandoval, L.-P., Kumar, S., Heris, H.~K., Chang, C. L.~A., Cornell, C.~E., Keller, S.~L., Fran{\c c}ois, P., Hendricks, A.~G., Ehrlicher, A.~J., Pappu, R.~V., and Michnick, S.~W. (2021)
Endocytic proteins with prion-like domains form viscoelastic condensates that enable membrane remodeling.
{\em Proc. Natl. Acad. Sci. USA,} {\bf 118}(50).

\bibitem{Kozak2022}
Kozak, M. and Kaksonen, M. (apr, 2022)
Condensation of Ede1 promotes the initiation of endocytosis.
{\em eLife,} {\bf 11}, e72865.

\bibitem{Nair2025}
Nair, K.~S., Radhakrishnan, S., and Bajaj, H. (2025)
Dynamic Duos: Coacervate-Lipid Membrane Interactions in Regulating Membrane Transformation and Condensate Size.
{\em Small,} {\bf 21}(20), 2501470.

\bibitem{Mokbel2023}
Mokbel, M., Mokbel, D., Liese, S., Weber, C., and Aland, S. (2024)
A Simulation Method for the Wetting Dynamics of Liquid Droplets on Deformable Membranes.
{\em SIAM J. Sci. Comput.,} {\bf 46}(6), B806--B829.

\bibitem{Wang2024a}
Wang, Y., Li, S., Mokbel, M., May, A.~I., Liang, Z., Zeng, Y., Wang, W., Zhang, H., Yu, F., Sporbeck, K., Jiang, L., Aland, S., Agudo-Canalejo, J., Knorr, R.~L., and Fang, X. (2024)
Biomolecular condensates mediate bending and scission of endosome membranes.
{\em Nature,} {\bf 634}, 1204--1210.

\bibitem{kusumaatmajae2021wetting}
Kusumaatmaja, H., May, A.~I., Feeney, M., McKenna, J.~F., Mizushima, N., Frigerio, L., and Knorr, R.~L. (2021)
Wetting of phase-separated droplets on plant vacuole membranes leads to a competition between tonoplast budding and nanotube formation.
{\em Proc. Natl. Acad. Sci. USA,} {\bf 118}(36).

\bibitem{Pike2009}
Pike, L.~J. (2009)
The challenge of lipid rafts.
{\em Journal of Lipid Research,} {\bf 50}, S323--S328.

\bibitem{Veatch2005}
Veatch, S.~L. and Keller, S.~L. (2005)
Seeing spots: Complex phase behavior in simple membranes.
{\em Biochim. Biophys. Acta Mol. Cell Res.,} {\bf 1746}(3), 172--185.

\bibitem{Shaw2021}
Shaw, T.~R., Ghosh, S., and Veatch, S.~L. (2021)
Critical Phenomena in Plasma Membrane Organization and Function.
{\em Annu. Rev. Phys. Chem.,} {\bf 72}(Volume 72, 2021), 51--72.

\bibitem{Litschel2023}
Litschel, T., Kelley, C.~F., Cheng, X., Babl, L., Mizuno, N., Case, L.~B., and Schwille, P. (2024)
Membrane-induced 2D phase separation of the focal adhesion protein talin.
{\em Nat. Commun.,} {\bf 15}(1), 4986.

\bibitem{Rouches2021}
Rouches, M., Veatch, S.~L., and Machta, B.~B. (2021)
Surface densities prewet a near-critical membrane.
{\em Proc. Natl. Acad. Sci. USA,} {\bf 118}(40), e2103401118.

\bibitem{Wang2022a}
Wang, H.-Y., Chan, S.~H., Dey, S., Castello-Serrano, I., Ditlev, J.~A., Rosen, M.~K., Levental, K.~R., and Levental, I. (2023)
Coupling of protein condensates to ordered lipid domains determines functional membrane organization.
{\em Sci. Adv,} {\bf 9}(17).

\bibitem{Lee2023a}
Lee, Y., Park, S., Yuan, F., Hayden, C.~C., Wang, L., Lafer, E.~M., Choi, S.~Q., and Stachowiak, J.~C. (2023)
Transmembrane coupling of liquid-like protein condensates.
{\em Nat. Commun.,} {\bf 14}(1), 8015.

\bibitem{Caballero2023}
Caballero, N., Kruse, K., and Giamarchi, T. (Jul, 2023)
Phase separation on surfaces in the presence of matter exchange.
{\em Phys. Rev. E,} {\bf 108}, L012801.

\bibitem{Zhang2021b}
Zhang, Y., Lee, D. S.~W., Meir, Y., Brangwynne, C.~P., and Wingreen, N.~S. (Jun, 2021)
Mechanical Frustration of Phase Separation in the Cell Nucleus by Chromatin.
{\em Phys. Rev. Lett.,} {\bf 126}, 258102.

\bibitem{Qi2021}
Qi, Y. and Zhang, B. (2021)
Chromatin network retards nucleoli coalescence.
{\em Nat. Commun.,} {\bf 12}(1), 6824.

\bibitem{Williams2020}
Williams, J.~F., Surovtsev, I.~V., Schreiner, S.~M., Chen, Z., Raiymbek, G., Nguyen, H., Hu, Y., Biteen, J.~S., Mochrie, S.~G., Ragunathan, K., and King, M.~C. (Phase sep. / Drops, Elasticity, 2024)
The condensation of HP1$\alpha$/Swi6 imparts nuclear stiffness.
{\em Cell Rep.,} {\bf 43}(7), 114373.

\bibitem{Style2015a}
Style, R.~W., Wettlaufer, J.~S., and Dufresne, E.~R. (Jan, 2015)
Surface tension and the mechanics of liquid inclusions in compliant solids.
{\em Soft Matter,} {\bf 11}(4), 672--9.

\bibitem{Ronceray2022}
Ronceray, P., Mao, S., Kosmrlj, A., and Haataja, M.~P. (2022)
Liquid demixing in elastic networks: Cavitation, permeation, or size selection?.
{\em Europhys. Lett.,} {\bf 137}(6), 67001.

\bibitem{Style2018}
Style, R.~W., Sai, T., Fanelli, N., Ijavi, M., Smith-Mannschott, K., Xu, Q., Wilen, L.~A., and Dufresne, E.~R. (2018)
Liquid-liquid phase separation in an elastic network.
{\em Phys. Rev. X,} {\bf 8}(1), 011028.

\bibitem{Fernandez-Rico2021}
Fern{\'a}ndez-Rico, C., Sai, T., Sicher, A., Style, R.~W., and Dufresne, E.~R. (2021)
Putting the Squeeze on Phase Separation.
{\em JACS Au,} {\bf 2}(1), 66--73.

\bibitem{Wei2020}
Wei, X., Zhou, J., Wang, Y., and Meng, F. (2020)
Modeling Elastically Mediated Liquid-Liquid Phase Separation.
{\em Phys. Rev. Lett.,} {\bf 125}(26), 268001.

\bibitem{Kothari2020}
Kothari, M. and Cohen, T. (2020)
Effect of elasticity on phase separation in heterogeneous systems.
{\em J. Mech. Phys. Solids,} {\bf 145}, 104153.

\bibitem{Kim2020}
Kim, J.~Y., Liu, Z., Weon, B.~M., Cohen, T., Hui, C.-Y., Dufresne, E.~R., and Style, R.~W. (2020)
Extreme cavity expansion in soft solids: Damage without fracture.
{\em Science advances,} {\bf 6}(13), eaaz0418.

\bibitem{Liu2023b}
Liu, J.~X., Haataja, M.~P., Ko{\v s}mrlj, A., Datta, S.~S., Arnold, C.~B., and Priestley, R.~D. (2023)
Liquid--liquid phase separation within fibrillar networks.
{\em Nat. Commun.,} {\bf 14}(1), 6085.

\bibitem{Paulin2022}
Paulin, O.~W., Morrow, L.~C., Hennessy, M.~G., and MacMinn, C.~W. (2022)
Fluid--fluid phase separation in a soft porous medium.
{\em J. Mech. Phys. Solids,} {\bf 164}, 104892.

\bibitem{Paulin2023}
Paulin, O.~W.
Formation and collapse of non-wetting cavities in soft porous media
PhD thesis University of Oxford (2023).

\bibitem{Furthauer2022}
F{\"u}rthauer, S. and Shelley, M.~J. (2022)
How Cross-Link Numbers Shape the Large-Scale Physics of Cytoskeletal Materials.
{\em Annu. Rev. Condens. Matter Phys.,} {\bf 13}(1), 365--384.

\bibitem{Qiang2024a}
Qiang, Y., Luo, C., and Zwicker, D. (Apr, 2024)
Nonlocal Elasticity Yields Equilibrium Patterns in Phase Separating Systems.
{\em Phys. Rev. X,} {\bf 14}, 021009.

\bibitem{Mannattil2024}
Mannattil, M., Diamant, H., and Andelman, D. (Sep, 2025)
Theory of Microphase Separation in Elastomers.
{\em Phys. Rev. Lett.,} {\bf 135}, 108101.

\bibitem{Oudich2025}
Oudich, H., Carrara, P., and {De Lorenzis}, L. (2026)
Phase-field modeling of elastic microphase separation.
{\em J. Mech. Phys. Solids,} {\bf 206}, 106380.

\bibitem{Paulin2025a}
{Paulin}, O.~W., {Qiang}, Y., and {Zwicker}, D. (August, 2025)
{Dynamics of phase separation in non-local elastic networks}.
{\em arXiv e-prints,} p. arXiv:2508.09829.

\bibitem{Fernandez-Rico2023}
Fern{\'a}ndez-Rico, C., Schreiber, S., Oudich, H., Lorenz, C., Sicher, A., Sai, T., Bauernfeind, V., Heyden, S., Carrara, P., Lorenzis, L.~D., Style, R.~W., and Dufresne, E.~R. (2024)
Elastic microphase separation produces robust bicontinuous materials.
{\em Nat. Mater.,} {\bf 23}, 124--130.

\bibitem{Kothari2022}
Kothari, M. and Cohen, T. (2022)
The crucial role of elasticity in regulating liquid--liquid phase separation in cells.
{\em Biomech. Model. Mechanobiol.,} {\bf 22}(2), 645--654.

\bibitem{Curk2023}
Curk, T. and Luijten, E. (2023)
Phase separation and ripening in a viscoelastic gel.
{\em Proc. Natl. Acad. Sci. USA,} {\bf 120}(32), e2304655120.

\bibitem{Prost2015}
Prost, J., J{\"u}licher, F., and Joanny, J.-F. (2015)
Active gel physics.
{\em Nat. Phys.,} {\bf 11}(2), 111--117.

\bibitem{Tayar2023}
Tayar, A.~M., Caballero, F., Anderberg, T., Saleh, O.~A., Cristina~Marchetti, M., and Dogic, Z. (2023)
Controlling liquid--liquid phase behaviour with an active fluid.
{\em Nat. Mater.,} {\bf 22}(11), 1401--1408.

\bibitem{Bodinilefranc2025}
Bodini-Lefranc, Q., Schindelwig, J., Weidinger, D., Engleder, L., and F{\"u}rthauer, S.
Arrested coarsening, oscillations, and memory from a conserved phase separating nucleator in a self-straining cytoskeletal network. (2025).

\bibitem{rai2018kinase}
Rai, A.~K., Chen, J.-X., Selbach, M., and Pelkmans, L. (July, 2018)
Kinase-controlled phase transition of membraneless organelles in mitosis.
{\em Nature,} {\bf 559}(7713), 211--216.

\bibitem{Zwicker2014}
Zwicker, D., Decker, M., Jaensch, S., Hyman, A.~A., and J{\"u}licher, F. (2014)
Centrosomes are autocatalytic droplets of pericentriolar material organized by centrioles.
{\em Proc. Natl. Acad. Sci. USA,} {\bf 111}(26).

\bibitem{Hondele2019}
Hondele, M., Sachdev, R., Heinrich, S., Wang, J., Vallotton, P., Fontoura, B. M.~A., and Weis, K. (Sep, 2019)
DEAD-box ATPases are global regulators of phase-separated organelles.
{\em Nature,} {\bf 573}(7772), 144--148.

\bibitem{Atkin2010}
Atkin, P. and de~Paula, J. (2010)
Atkins' physical chemistry,
W. H. Freeman and Company,  9th edition.

\bibitem{JulicherAjdari1997}
J\"ulicher, F., Ajdari, A., and Prost, J. (Oct, 1997)
Modeling molecular motors.
{\em Rev. Mod. Phys.,} {\bf 69}, 1269--1282.

\bibitem{Hanggi1990}
H{{\"a}}nggi, P., Talkner, P., and Borkovec, M. (1990)
Reaction-rate theory: fifty years after Kramers.
{\em Rev. Mod. Phys.,} {\bf 62}(2), 251.

\bibitem{Pagonabarraga1997}
Pagonabarraga, I., P{\'e}rez-Madrid, A., and Rub{\'\i}, J. (1997)
Fluctuating hydrodynamics approach to chemical reactions.
{\em Phys. A: Stat. Mech. Appl.,} {\bf 237}(1), 205--219.

\bibitem{Bazant2013}
Bazant, M.~Z. (May, 2013)
Theory of chemical kinetics and charge transfer based on nonequilibrium thermodynamics.
{\em Acc. Chem. Res.,} {\bf 46}(5), 1144--60.

\bibitem{Kirschbaum2021}
Kirschbaum, J. and Zwicker, D. (2021)
Controlling biomolecular condensates via chemical reactions.
{\em J. R. Soc. Interface,} {\bf 18}(179), 20210255.

\bibitem{qian2007phosphorylation}
Qian, H. (2007)
Phosphorylation Energy Hypothesis: Open Chemical Systems and Their Biological Functions.
{\em Annu. Rev. Phys. Chem.,} {\bf 58}(1), 113--142.

\bibitem{Avanzini2021}
Avanzini, F., Penocchio, E., Falasco, G., and Esposito, M. (2021)
Nonequilibrium thermodynamics of non-ideal chemical reaction networks.
{\em J. Chem. Phys.,} {\bf 154}(9), 094114.

\bibitem{Aslyamov2023}
Aslyamov, T., Avanzini, F., Fodor, E., and Esposito, M. (Sep, 2023)
Nonideal Reaction-Diffusion Systems: Multiple Routes to Instability.
{\em Phys. Rev. Lett.,} {\bf 131}, 138301.

\bibitem{Avanzini2024}
Avanzini, F., Aslyamov, T., Fodor, {\'E}., and Esposito, M. (11, 2024)
{Nonequilibrium thermodynamics of non-ideal reaction--diffusion systems: Implications for active self-organization}.
{\em J. Chem. Phys.,} {\bf 161}(17), 174108.

\bibitem{Vrugt2024}
te~Vrugt, M. and Wittkowski, R. (2025)
Metareview: a survey of active matter reviews.
{\em Eur. Phys. J. E,} {\bf 48}(2), 12.

\bibitem{Bauermann2022}
Bauermann, J., Weber, C.~A., and J{\"u}licher, F. (2022)
Energy and Matter Supply for Active Droplets.
{\em Annalen der Physik,} {\bf 534}(9), 2200132.

\bibitem{Zwicker2024}
Zwicker, D.
Chemically active droplets. (2024).

\bibitem{Ziethen2023}
Ziethen, N., Kirschbaum, J., and Zwicker, D. (Jun, 2023)
Nucleation of Chemically Active Droplets.
{\em Phys. Rev. Lett.,} {\bf 130}, 248201.

\bibitem{Zwicker2015}
Zwicker, D., Hyman, A.~A., and J{\"u}licher, F. (2015)
Suppression of Ostwald ripening in active emulsions.
{\em Phys. Rev. E,} {\bf 92}(1), 012317.

\bibitem{Tena-solsona2019}
Tena-Solsona, M., Janssen, J., Wanzke, C., Schnitter, F., Park, H., Rie{\ss}, B., Gibbs, J.~M., Weber, C.~A., and Boekhoven, J. (2021)
Accelerated Ripening in Chemically Fueled Emulsions.
{\em ChemSystemsChem,} {\bf 3}(2), e2000034.

\bibitem{Bauermann2023}
Bauermann, J., Bartolucci, G., Boekhoven, J., Weber, C.~A., and J\"ulicher, F. (Dec, 2023)
Formation of liquid shells in active droplet systems.
{\em Phys. Rev. Res.,} {\bf 5}, 043246.

\bibitem{Demarchi2023}
Demarchi, L., Goychuk, A., Maryshev, I., and Frey, E. (Mar, 2023)
Enzyme-Enriched Condensates Show Self-Propulsion, Positioning, and Coexistence.
{\em Phys. Rev. Lett.,} {\bf 130}, 128401.

\bibitem{Zwicker2018b}
Zwicker, D., Baumgart, J., Redemann, S., M{\"u}ller-Reichert, T., Hyman, A.~A., and J{\"u}licher, F. (2018)
Positioning of particles in active droplets.
{\em Phys. Rev. Lett.,} {\bf 121}(15), 158102.

\bibitem{Wurtz2018}
Wurtz, J.~D. and Lee, C.~F. (2018)
Chemical-reaction-controlled phase separated drops: formation, size selection, and coarsening.
{\em Phys. Rev. Lett.,} {\bf 120}(7), 078102.

\bibitem{Sastre2024}
Sastre, J., Thatte, A., Bergmann, A.~M., Stasi, M., Tena-Solsona, M., Weber, C.~A., and Boekhoven, J. (2025)
Size control and oscillations of active droplets in synthetic cells.
{\em Nat. Commun.,} {\bf 16}(1), 2003.

\bibitem{bauermann2024}
Bauermann, J., Bartolucci, G., Boekhoven, J., J{\"u}licher, F., and Weber, C.~A.
Critical transition between intensive and extensive active droplets. (2024).

\bibitem{Oono1988}
Oono, Y. and Bahiana, M. (Aug, 1988)
2/3-Power Law for Copolymer Lamellar Thickness Implies a 1/3 -Power Law for Spinodal Decomposition.
{\em Phys. Rev. Lett.,} {\bf 61}, 1109--1111.

\bibitem{Glotzer1994}
Glotzer, S.~C., Stauffer, D., and Jan, N. (1994)
Monte Carlo simulations of phase separation in chemically reactive binary mixtures.
{\em Phys. Rev. Lett.,} {\bf 72}(26), 4109--4112.

\bibitem{Glotzer1994a}
Glotzer, S.~C. and Coniglio, A. (1994)
Self-consistent solution of phase separation with competing interactions.
{\em Phys. Rev. E,} {\bf 50}(5), 4241--4244.

\bibitem{Christensen1996}
Christensen, J.~J., Elder, K., and Fogedby, H.~C. (1996)
Phase segregation dynamics of a chemically reactive binary mixture.
{\em Phys. Rev. E,} {\bf 54}(3), R2212--R2215.

\bibitem{Muratov2002}
Muratov, C.~B. (2002)
Theory of domain patterns in systems with long-range interactions of Coulomb type.
{\em Phys. Rev. E,} {\bf 66}(6 Pt 2), 066108.

\bibitem{Liese2023}
Liese, S., Zhao, X., Weber, C.~A., and J{\"u}licher, F. (2025)
Chemically active wetting.
{\em Proc. Natl. Acad. Sci. USA,} {\bf 122}(15), e2403083122.

\bibitem{Ziethen2024}
Ziethen, N. and Zwicker, D. (06, 2024)
Heterogeneous nucleation and growth of sessile chemically active droplets.
{\em J. Chem. Phys.,} {\bf 160}(22), 224901.

\bibitem{Ohta1986}
Ohta, T. and Kawasaki, K. (1986)
Equilibrium morphology of block copolymer melts.
{\em Macromolecules,} {\bf 19}, 2621--2632.

\bibitem{Liu1989}
Liu, F. and Goldenfeld, N. (1989)
Dynamics of phase separation in block copolymer melts.
{\em Phys. Rev. A,} {\bf 39}(9), 4805--4810.

\bibitem{Seul1995}
Seul, M. and Andelman, D. (1995)
Domain Shapes and Patterns: The Phenomenology of Modulated Phases.
{\em Science,} {\bf 267}(5197), 476--483.

\bibitem{Huberman1976}
Huberman, B.~A. (1976)
Striations in chemical reactions.
{\em J. Chem. Phys.,} {\bf 65}(5), 2013--2019.

\bibitem{Puri1994}
Puri, S. and Frisch, H. (1994)
Segregation dynamics of binary mixtures with simple chemical reactions.
{\em J. Phys. A,} {\bf 27}, 6027--6038.

\bibitem{Glotzer1995}
Glotzer, S.~C., Di~Marzio, E.~A., and Muthukumar, M. (1995)
Reaction-controlled morphology of phase-separating mixtures.
{\em Phys. Rev. Lett.,} {\bf 74}(11), 2034--2037.

\bibitem{Carati1997}
Carati, D. and Lefever, R. (1997)
Chemical freezing of phase separation in immiscible binary mixtures.
{\em Phys. Rev. E,} {\bf 56}(3 B), 3127--3136.

\bibitem{Tran-Cong1996}
Tran-Cong, Q. and Harada, A. (1996)
Reaction-induced ordering phenomena in binary polymer mixtures.
{\em Phys. Rev. Lett.,} {\bf 76}, 1162--1165.

\bibitem{Motoyama1996}
Motoyama, M. (1996)
Morphology of Binary Mixtures Which Undergo Phase Separation during Chemical Reactions.
{\em J. Phys. Soc. Jpn.,} {\bf 65}(7), 1894--1897.

\bibitem{Motoyama1997}
Motoyama, M. and Ohta, T. (1997)
Morphology of Phase-Separating Binary Mixtures with Chemical Reaction.
{\em J. Phys. Soc. Jpn.,} {\bf 66}(9), 2715--2725.

\bibitem{Toxvaerd1996}
Toxvaerd, S. (1996)
Molecular dynamics simulations of phase separation in chemically reactive binary mixtures.
{\em Phys. Rev. E,} {\bf 53}(4), 3710--3716.

\bibitem{TranCongMiyata2017}
Tran-Cong-Miyata, Q. and Nakanishi, H. (2017)
Phase separation of polymer mixtures driven by photochemical reactions: current status and perspectives.
{\em Polym. Int.,} {\bf 66}(2), 213--222.

\bibitem{Tran-Cong-Miyata2011}
Tran-Cong-Miyata, Q., Kinohira, T., Van-Pham, D.-T., Hirose, A., Norisuye, T., and Nakanishi, H. (00, 2011)
Phase separation of polymer mixtures driven by photochemical reactions: Complexity and fascination.
{\em Curr. Opin. Solid State Mater. Sci.,} {\bf 15}(6), 254--261.

\bibitem{Cho2023a}
Cho, Y. and Jacobs, W.~M. (Mar, 2023)
Tuning Nucleation Kinetics via Nonequilibrium Chemical Reactions.
{\em Phys. Rev. Lett.,} {\bf 130}, 128203.

\bibitem{Mullins1963}
Mullins, W.~W. and Sekerka, R.~F. (1963)
Morphological stability of a particle growing by diffusion or heat flow.
{\em J. Appl. Phys.,} {\bf 34}(2), 323--329.

\bibitem{Golestanian2017}
Golestanian, R. (2017)
Origin of life: Division for multiplication.
{\em Nat. Phys.,} {\bf 13}(4), 323--324.

\bibitem{Fries2025a}
Fries, J., Berthin, R., Luo, C., Jardat, M., Zwicker, D., Dahirel, V., and Illien, P.
Chemically active droplets in crowded environments. (2025).

\bibitem{Zippo2025}
Zippo, E., Dormann, D., Speck, T., and Stelzl, L.~S. (2025)
Molecular simulations of enzymatic phosphorylation of disordered proteins and their condensates.
{\em Nat. Commun.,} {\bf 16}(1), 4649.

\bibitem{Case2019a}
Case, L.~B., Ditlev, J.~A., and Rosen, M.~K. (2019)
Regulation of Transmembrane Signaling by Phase Separation.
{\em Annu. Rev. Biophys.,} {\bf 48}(Volume 48, 2019), 465--494.

\bibitem{Goychuk2024}
Goychuk, A., Demarchi, L., Maryshev, I., and Frey, E. (Jul, 2024)
Self-consistent sharp interface theory of active condensate dynamics.
{\em Phys. Rev. Res.,} {\bf 6}, 033082.

\bibitem{Golestanian2009}
Golestanian, R. (May, 2009)
Anomalous Diffusion of Symmetric and Asymmetric Active Colloids.
{\em Phys. Rev. Lett.,} {\bf 102}, 188305.

\bibitem{Rasshofer2025}
Ra{\ss}hofer, F., Bauer, S., Ziepke, A., Maryshev, I., and Frey, E.
Capillary wave formation in conserved active emulsions. (2025).

\bibitem{Qiang2025a}
{Qiang}, Y., {Luo}, C., and {Zwicker}, D. (August, 2025)
{Self-propulsion via non-transitive phase coexistence in chemically active mixtures}.
{\em arXiv e-prints,} p. arXiv:2508.09816.

\bibitem{Laha2024}
Laha, S., Bauermann, J., J\"ulicher, F., Michaels, T. C.~T., and Weber, C.~A. (Nov, 2024)
Chemical reactions regulated by phase-separated condensates.
{\em Phys. Rev. Res.,} {\bf 6}, 043092.

\bibitem{Menou2023}
Menou, L., Luo, C., and Zwicker, D. (2023)
Physical interactions in non-ideal fluids promote Turing patterns.
{\em J. R. Soc. Interface,} {\bf 20}(204), 20230244.

\bibitem{Tran-Cong1999a}
Tran-Cong, Q., Kawai, J., Nishikawa, Y., and Jinnai, H. (1999)
Phase separation with multiple length scales in polymer mixtures induced by autocatalytic reactions.
{\em Phys. Rev. E,} {\bf 60}(2 Pt A), R1150--3.

\bibitem{Luo2023}
Luo, C. and Zwicker, D. (Sep, 2023)
Influence of physical interactions on spatiotemporal patterns.
{\em Phys. Rev. E,} {\bf 108}, 034206.

\bibitem{Coupe2024}
Coupe, S.~T. and Fakhri, N. (2024)
Nonequilibrium phases of a biomolecular condensate facilitated by enzyme activity.
{\em bioRxiv,}.

\bibitem{Haugerud2025}
Haugerud, I.~S., Vuijk, H.~D., Boekhoven, J., and Weber, C.~A.
Excitability and oscillations of active droplets. (2025).

\bibitem{Miangolarra2022}
Miangolarra, A.~M. and Castellana, M. (2022)
On Non-ideal Chemical-Reaction Networks and Phase Separation.
{\em J. Stat. Phys.,} {\bf 190}(1), 23.

\bibitem{Sang2022}
Sang, D., Shu, T., Pantoja, C.~F., {Ib{\'a}{\\textasciitilde n}ez de Opakua}, A., Zweckstetter, M., and Holt, L.~J. (2022)
Condensed-phase signaling can expand kinase specificity and respond to macromolecular crowding.
{\em Mol. Cell,} {\bf 82}(19), 3693--3711.e10.

\bibitem{Bauermann2025}
Bauermann, J., Bartolucci, G., Weber, C.~A., and J{\"u}licher, F.
The droplet size distribution and its dynamics in chemically active emulsions. (2025).

\bibitem{Bergmann2023}
Bergmann, A.~M., Bauermann, J., Bartolucci, G., Donau, C., Stasi, M., Holtmannsp{\"o}tter, A.-L., J{\"u}licher, F., Weber, C.~A., and Boekhoven, J. (2023)
Liquid spherical shells are a non-equilibrium steady state of active droplets.
{\em Nat. Commun.,} {\bf 14}(1), 6552.

\bibitem{Fries2024}
Fries, J., Diaz, J., Jardat, M., Pagonabarraga, I., Illien, P., and Dahirel, V. (2025)
Active droplets controlled by enzymatic reactions.
{\em J. R. Soc. Interface,} {\bf 22}(226), 20240803.

\bibitem{Banani2024}
Banani, S.~F., Goychuk, A., Natarajan, P., Zheng, M.~M., Dall{\textquoteright}Agnese, G., Henninger, J.~E., Kardar, M., Young, R.~A., and Chakraborty, A.~K. (2024)
Active RNA synthesis patterns nuclear condensates.
{\em bioRxiv,}.

\bibitem{MartinezCalvo2024}
Mart\'{\i}nez-Calvo, A., Zhou, J., Zhang, Y., and Wingreen, N.~S. (Aug, 2025)
Sticky Enzymes: Increased Metabolic Efficiency via Substrate-Dependent Enzyme Clustering.
{\em PRX Life,} {\bf 3}, 033011.

\bibitem{wurtz2018stress}
Wurtz, J.~D. and Lee, C.~F. (2018)
Stress granule formation via ATP depletion-triggered phase separation.
{\em New J. Phys.,} {\bf 20}(4), 045008.

\bibitem{Morgan2021}
Morgan, C., Fozard, J.~A., Hartley, M., Henderson, I.~R., Bomblies, K., and Howard, M. (2021)
Diffusion-mediated HEI10 coarsening can explain meiotic crossover positioning in Arabidopsis.
{\em Nat. Commun.,} {\bf 12}(1), 4674.

\bibitem{Zhang2021c}
Zhang, L., Stauffer, W., Liu, C., Shao, H., Abuzahriyeh, N., Jiang, R., Zwicker, D., Liu, X., Yao, X., and Dernburg, A.~F. (2025)
Crossover patterning through condensation and coarsening of pro-crossover factors.
{\em Nat. Cell Biol.,}.

\bibitem{Durand2022}
Durand, S., Lian, Q., Jing, J., Ernst, M., Grelon, M., Zwicker, D., and Mercier, R. (2022)
Joint control of meiotic crossover patterning by the synaptonemal complex and HEI10 dosage.
{\em Nat. Commun.,} {\bf 13}(1), 5999.

\bibitem{Girard2023}
Girard, C., Zwicker, D., and Mercier, R. (2023)
{The regulation of meiotic crossover distribution: a coarse solution to a century-old mystery?}.
{\em Biochem. Soc. Trans.,} {\bf 51}(3), 1179--1190.

\bibitem{Ernst2024}
Ernst, M., Mercier, R., and Zwicker, D. (2024)
Interference Length reveals regularity of crossover placement across species.
{\em Nat. Commun.,} {\bf 15}(8973).

\bibitem{Chauhan2025}
Chauhan, G., Wilkinson, E.~G., Yuan, Y., Cohen, S.~R., Onishi, M., Pappu, R.~V., and Strader, L.~C. (2025)
Active transport enables protein condensation in cells.
{\em Sci. Adv.,} {\bf 11}(21), eadv7875.

\bibitem{Marchetti2013}
Marchetti, M.~C., Joanny, J.~F., Ramaswamy, S., Liverpool, T.~B., Prost, J., Rao, M., and Simha, R.~A. (2013)
Hydrodynamics of soft active matter.
{\em Rev. Mod. Phys.,} {\bf 85}(3), 1143--1189.

\bibitem{Ramaswamy2010}
Ramaswamy, S. (2010)
The Mechanics and Statistics of Active Matter.
{\em Annu. Rev. Condens. Matter Phys.,} {\bf 1}(Volume 1, 2010), 323--345.

\bibitem{Franzmann2018a}
Franzmann, T.~M., Jahnel, M., Pozniakovsky, A., Mahamid, J., Holehouse, A.~S., N{\"u}ske, E., Richter, D., Baumeister, W., Grill, S.~W., Pappu, R.~V., Hyman, A.~A., and Alberti, S. (2018)
Phase separation of a yeast prion protein promotes cellular fitness.
{\em Science,} {\bf 359}(6371), eaao5654.

\bibitem{Dine2018}
Dine, E., Gil, A.~A., Uribe, G., Brangwynne, C.~P., and Toettcher, J.~E. (06, 2018)
Protein Phase Separation Provides Long-Term Memory of Transient Spatial Stimuli.
{\em Cell Syst.,} {\bf 6}(6), 655--663.e5.

\bibitem{Badia2021}
Badia, M. and Bolognesi, B. (2021)
Assembling the right type of switch: Protein condensation to signal cell death.
{\em Curr. Opin. Cell Biol.,} {\bf 69}, 55--61.

\bibitem{Zhang2024c}
Zhang, R., Yang, W., Zhang, R., Rijal, S., Youssef, A., Zheng, W., and Tian, X. (2024)
Phase Separation to Resolve Growth-Related Circuit Failures.
{\em bioRxiv,}.

\bibitem{Bussi2023}
Bussi, C., Mangiarotti, A., Vanhille-Campos, C., Aylan, B., Pellegrino, E., Athanasiadi, N., Fearns, A., Rodgers, A., Franzmann, T.~M., {\v S}ari{\'c}, A., Dimova, R., and Gutierrez, M.~G. (2023)
Stress granules plug and stabilize damaged endolysosomal membranes.
{\em Nature,} {\bf 623}, 1062--1069.

\bibitem{Ji2025}
Ji, B.-T., Pan, H.-T., Qian, Z.-G., and Xia, X.-X. (2025)
Programming biological communication between distinct membraneless compartments.
{\em Nat. Chem. Biol.,} {\bf 21}, 1110--1117.

\bibitem{Ye2025}
Ye, S., Benhamou~Goldfajn, N., So, C.~L., Inoue, T., and Cai, D. (2025)
Rainbow Nucleus Charts Dynamic Interactome of Membrane-less Organelles.
{\em bioRxiv,}.

\bibitem{Modi2023}
Modi, N., Chen, S., Adjei, I. N.~A., Franco, B.~L., Bishop, K. J.~M., and Obermeyer, A.~C. (2023)
Designing negative feedback loops in enzymatic coacervate droplets.
{\em Chem. Sci.,} {\bf 14}, 4735--4744.

\bibitem{Zentner2025}
Zentner, A., Halingstad, E.~V., Chalk, C., Brenner, M.~P., Murugan, A., Winfree, E., and Shrinivas, K.
Information processing driven by multicomponent surface condensates. (2025).

\bibitem{Murugan2025}
Murugan, A., Zwicker, D., Lorenz, C., and Dufresne, E.~R.
Could Living Cells Use Phase Transitions to Process Information?. (2025).

\bibitem{Hilbert2025}
Hilbert, L., Gadzekpo, A., Vecchio, S.~L., Wellh{\"a}usser, M., Tschurikow, X., Prizak, R., Becker, B., Burghart, S., and Oprzeska-Zingrebe, E.~A. (2025)
Chromatin-associated condensates as an inspiration for the system architecture of future DNA computers.
{\em Ann. NY Acad. Sci.,}.

\bibitem{Spruijt2023}
Spruijt, E. (2023)
Open questions on liquid--liquid phase separation.
{\em Commun. Chem.,} {\bf 6}(1), 23.

\bibitem{Zwicker2023}
Zwicker, D. (2023)
Droplets Come to Life.
{\em Physics,} {\bf 16}(45).

\bibitem{Wenisch2025}
Wenisch, M., Li, Y., Braun, M.~G., Eylert, L., Sp{\"a}th, F., Poprawa, S.~M., Rieger, B., Synatschke, C.~V., Niederholtmeyer, H., and Boekhoven, J. (2025)
Toward synthetic life---Emergence, growth, creation of offspring, decay, and rescue of fuel-dependent synthetic cells.
{\em Chem,} p. 102578.

\bibitem{Hallatschek2023}
Hallatschek, O., Datta, S.~S., Drescher, K., Dunkel, J., Elgeti, J., Waclaw, B., and Wingreen, N.~S. (2023)
Proliferating active matter.
{\em Nat. Rev. Phys.,} {\bf 5}, 407--419.

\bibitem{Goldbeter2018}
Goldbeter, A. (2018)
Dissipative structures in biological systems: bistability, oscillations, spatial patterns and waves.
{\em Philos. Trans. R. Soc. A,} {\bf 376}(2124), 20170376.

\bibitem{Julicher2009}
J{\"u}licher, F. and Prost, J. (2009)
Generic theory of colloidal transport.
{\em Eur. Phys. J. E,} {\bf 29}, 27--36.

\bibitem{Kirkwood1949}
Kirkwood, J.~G. and Buff, F.~P. (03, 1949)
{The Statistical Mechanical Theory of Surface Tension}.
{\em J. Chem. Phys.,} {\bf 17}(3), 338--343.

\bibitem{Joanny1978}
Joanny, J.-F. and Leibler, L. (1978)
Interface in molten polymer mixtures near the consolute point.
{\em J. Phys. France,} {\bf 39}(9), 951--953.

\bibitem{Mangiarotti2023}
Mangiarotti, A., Chen, N., Zhao, Z., Lipowsky, R., and Dimova, R. (2023)
Wetting and complex remodeling of membranes by biomolecular condensates.
{\em Nat. Commun.,} {\bf 14}(1), 2809.

\bibitem{Bressloff2020}
Bressloff, P.~C. (2020)
Active suppression of Ostwald ripening: Beyond mean-field theory.
{\em Phys. Rev. E,} {\bf 101}(4), 042804.

\bibitem{Bressloff2020a}
Bressloff, P.~C. (2020)
Two-dimensional droplet ripening in a concentration gradient.
{\em J. Phys. A,} {\bf 53}(36), 365002.

\bibitem{Saffman1975}
Saffman, P.~G. and Delbr{\"u}ck, M. (1975)
Brownian motion in biological membranes..
{\em Proc. Natl. Acad. Sci. USA,} {\bf 72}(8), 3111--3113.

\end{thebibliography}

\end{document}